\documentclass[aps]{revtex4}
\usepackage{amsfonts}
\usepackage{amsmath}
\usepackage{amssymb,epsf}

\begin{document}

\title{Extended phase space of AdS Black Holes \\
in Einstein-Gauss-Bonnet gravity with a quadratic nonlinear
electrodynamics}
\author{S. H. Hendi$^{1,2}$\footnote{
email address: hendi@shirazu.ac.ir}, S. Panahiyan$^{1}$\footnote{
email address: sh.panahiyan@gmail.com} and M.
Momennia$^{1}$\footnote{ email address: momennia1988@gmail.com}}
\affiliation{$^1$ Physics Department and Biruni Observatory,
College of Sciences, Shiraz
University, Shiraz 71454, Iran\\
$^{2}$ Research Institute for Astronomy and Astrophysics of
Maragha (RIAAM), P.O. Box 55134-441, Maragha, Iran}

\begin{abstract}
In this paper, we consider quadratic Maxwell invariant as a
correction to the Maxwell theory and study thermodynamic behavior
of the black holes in Einstein and Gauss-Bonnet gravities. We
consider cosmological constant as a thermodynamic pressure to
extend phase space. Next, we obtain critical values in case of
variation of nonlinearity and Gauss-Bonnet parameters. Although
the general thermodynamical behavior of the black hole solutions
is the same as usual Van der Waals system, we show that in special
case of the nonlinear electromagnetic field, there will be a
turning point for the phase diagrams and usual Van der Waals is
not observed. This theory of nonlinear electromagnetic field
provides two critical horizon radii. We show that this unusual
behavior of phase diagrams is due to existence of second critical
horizon radius. It will be pointed out that the power of the
gravity and nonlinearity of the matter field modify the critical
values. We generalize the study by considering the effects of
dimensionality on critical values and make comparisons between our
models with their special sub classes. In addition, we examine the
possibility of the existence of the reentrant phase transitions
through two different methods.
\end{abstract}

\maketitle

\section{introduction}

It is well known that black hole behaves as a thermodynamic system
and it can be interpreted with a physical temperature and an
entropy \cite{1}. Finding the connection between the laws of black
hole mechanics and the laws of ordinary thermodynamics is one of
the remarkable achievements of theoretical physics during the last
forty years. Therefore, it is very natural to study various
thermodynamic aspects of black holes, such as thermal stability,
phase transition and black hole evaporation. Studying phase
transition in black holes would be one of the fascinating topics
in this regard because this phenomena plays an important role in
order to explore thermodynamic properties of various systems near
the critical points. The first attempt to investigate the phase
transition of the black holes has been done by Hut and Davies
\cite{2}.

In general, there are four different approaches for studying the
phase transition of the black holes, theoretically. First, the
cosmological constant, $\Lambda$, is considered to be pressure of
the system, in which its related conjugate quantity will be
volume. By this consideration, the critical behavior can be
studied through phase diagrams. Second, studying phase transition
in black holes by using the Clausius-Clapeyron-Ehrenfest's
equations \cite{3}. Considering the analogy between the
thermodynamic state variables and various black hole parameters
($V\leftrightarrow Q$ and $P\leftrightarrow -\Phi$), puts us in a
position to write down the Ehrenfest's equations for the black
holes \cite{4} and study their phase transition. Third, an
alternative approach to investigate the phase transition was
suggested by Ruppeiner in 1979 \cite{5} in which proposed a
geometrical way to study thermodynamical phase transitions.
Fourth, investigating the phase transitions of black holes through
the canonical ensemble by calculating heat capacity \cite{6}.

In this paper, we are going to investigate the phase transition of
black hole solutions in the asymptotically AdS spacetime by
considering the first mentioned approach (the cosmological
constant as a pressure of the system) in both Einstein and
Gauss-Bonnet (GB) gravities. In context of AdS/CFT correspondence,
it was proposed that variation of $\Lambda $ corresponds to
variation of the number of the colors on boundary of the
Yang-Mills theory with chemical potential interpretation
\cite{ADSCFT,9}. There are two main reasons to investigate the
asymptotically AdS black holes. First reason is that the AdS/CFT
correspondence attracts attentions to the physics of
asymptotically AdS black holes in recent years; the main focus is
on understanding strongly coupled thermal field theories living on
the AdS boundary. Even from a bulk perspective such black holes
and their thermodynamics which exhibits various phase transitions,
are quite interesting. Second, the behavior of the black hole
phase transition in the asymptotically AdS spacetime is like the
Van der Waals liquid/gas \cite{Chamblin,Goldenfeld}, and also, is
different from those of black holes in the flat space
\cite{7,Vahidinia}. Studying thermodynamic behavior of black holes
in an asymptotically AdS spacetime has been done first by Hawking
and Page in 1983 \cite{8}. After that, the critical behaviors of
the black holes by including the cosmological constant as a
thermodynamic pressure have been investigated in
\cite{Chamblin,9}. In this approach, the black hole mass $M$ is
considered as the Enthalpy of the system. Studying phase
transition of
black holes with Einstein gravity has been done in many literatures \cite%
{KubiznakMann}. In addition, the critical behavior of charged
AdS-GB black holes has been investigated in
\cite{10,GBMaxwell,maximal pressure}.

One the other hand, in electrodynamic point of view, the
self-energy of a point-like charge has a divergency at the origin.
In order to remove this singularity, Born and Infeld introduced an
interesting kind of nonlinear electrodynamics (NED) in 1934
\cite{11}. Coupling of NED with the gravity was first done by
Hoffmann \cite{12}. The effects of Born-Infeld NED coupled to the
gravitational field have been studied in various contexts such as
superconductors \cite{13}, wormholes \cite{14,15}, static black holes \cite%
{16} and rotating black objects \cite{17}. In addition, another motivation
for considering Born-Infeld NED comes from the fact that it naturally arises
in the low-energy limit of the heterotic string theory \cite{18}. Recently,
two different Born-Infeld types of NED have been introduced by Soleng \cite%
{19} and Hendi \cite{20}. The Soleng Lagrangian has a logarithmic form and,
like Born-Infeld theory, removes divergency of the electric field while the
Lagrangian proposed by Hendi has an exponential form and does not cancel the
divergency of the electric field but its singularity is much weaker than
that in Maxwell theory. Investigation of black object solutions coupled to
these two nonlinear fields has been done in \cite{15,21}. The Lagrangian of
mentioned Born-Infeld type nonlinear theories, for weak nonlinearity, can be
written with the following form
\begin{equation}
\mathcal{L}(\mathcal{F})=-\mathcal{F}+\beta \mathcal{F}^{2}+O\left( \beta
^{2}\right) ,  \label{Lagrangian}
\end{equation}%
where $\beta $ is proportional to the inverse value of square
nonlinearity parameter in Born-Infeld-type theories, so it gets
just positive values. In Eq. (\ref{Lagrangian}),
$\mathcal{F}=F_{\mu \nu }F^{\mu \nu }$ is the Maxwell invariant,
$F_{\mu \nu }=\partial _{\mu }A_{\nu }-\partial _{\nu }A_{\mu }$
is the electromagnetic field tensor and $A_{\mu } $ is the gauge
potential. In addition, $\beta $ denotes nonlinearity
parameter which is small. For $\beta \longrightarrow 0$, $\mathcal{L}(%
\mathcal{F})$ reduces to the standard Maxwell Lagrangian, $\mathcal{L}%
_{Maxwell}(\mathcal{F})=-\mathcal{F}$, as it should be. In this paper, we
take into account the Eq. (\ref{Lagrangian}) as a NED source coupled to the
Einstein and GB gravities and investigate the effects of nonlinearity on the
properties of the phase transition.

It is worthwhile to mention the motivations for considering the
NED Lagrangian and specially Lagrangian (\ref{Lagrangian}).
Nonlinear field theories are of interest to different branches of
mathematical physics because most physical systems are inherently
nonlinear in the nature. The main reason to consider NED comes
from the fact that these theories are considerably richer than the
Maxwell field and in special case they reduce to the linear
Maxwell theory. Various limitations of the Maxwell theory, such as
description of the self-interaction of virtual electron-positron
pairs \cite{22} and the radiation propagation inside specific
materials \cite{Lorenci}, motivate one to consider NED \cite{22}.
Besides, NED improves the basic concept of gravitational redshift
and its dependency of any background magnetic field as compared to
the well-established method introduced by standard general
relativity. In addition, it was recently shown that NED
objects can remove both of the big bang and black hole singularities \cite%
{AyonBeato1,AyonBeato2,Corda}. Moreover, from astrophysical point
of view, one finds that the effects of NED become indeed quite
important in superstrongly magnetized compact objects, such as
pulsars and particular neutron stars (also the so-called magnetars
and strange quark magnetars) \cite{Mosquera}. Also, since the
gravitational redshift of magnetized compact objects is connected
to the mass--radius relation of the objects, it is important to
note that NED affects the mass--radius relation of the objects. It
is worthwhile to mention that one can find regular black hole
solutions of the Einstein field equations coupled to a suitable NED \cite%
{AyonBeato1,AyonBeato2}. In addition, an interesting property
which is common to all the NED models is that these models satisfy
the zeroth and first laws of black hole mechanics. The appropriate
world-volume dynamics on a curved $D3$-brane may provide a
plausible frame-work at Planck scale by incorporating the
Einstein-NED. At this point, elimination of
strong intrinsic curvature in the regime by the strong nonlinearity in the $%
U(1)$ gauge theory is remarkable \cite{AyonBeato2,Gopakumar}. From
the point of view of AdS/CFT correspondence in hydrodynamic
models, it has been shown that, unlike gravitational correction,
higher-derivative terms for abelian fields in the form of NED do
not affect the ratio of shear viscosity to entropy density
\cite{Brigante}. Motivated by the recent results mentioned above
and the fact that we accepted NED as a generalization of the
Maxwell theory, it is natural to apply NED theories for charged
objects such as black holes.

Now, we focus on motivations of considering the nonlinear term of the
electromagnetic field perturbatively. Although various theories of NED have
been created with different primitive motivations, only for the weak
nonlinearity (Eq. (\ref{Lagrangian})), they contain physical and
experimental importances. As we know, using the Maxwell theory in various
branches leads to near accurate or acceptable consequences. So, in
transition from the Maxwell theory to NED, the logical decision is to
consider the effects of weak nonlinearity variations, not strong ones. This
means that, one can expect to obtain precise physical results with
experimental agreements, provided one regards the nonlinearity as a
correction to the Maxwell field. On the other hand, several reasonable
papers have been published by considering Eq. (\ref{Lagrangian}) as an
effective Lagrangian of electrodynamics \cite%
{22,23,HendiMomen,ThreeMag,HendiEPJC,HendiPanah,HendiIJMPD}. Heisenberg and
Euler have shown that quantum corrections lead to nonlinear properties of
vacuum \cite{22}. Also, it was proved that in the low energy limit of
heterotic string theory, a quartic correction of the Maxwell field strength
tensor appears \cite{23}. So it is natural to consider Eq. (\ref{Lagrangian}%
) as an effective and suitable Lagrangian of electrodynamics instead of the
Maxwell one. Investigating the effects of nonlinearity parameter of Eq. (\ref%
{Lagrangian}) coupled to the Einstein, GB and third order Lovelock
gravities have been done in \cite{HendiMomen,ThreeMag},
\cite{HendiEPJC,HendiPanah} and \cite{HendiIJMPD}, respectively.
In this paper, we are dealing with stringy corrected
electrodynamics, and therefore, obtained results are applicable in
context of string theory. In order to separate the valid domains
of the stringy corrected and classical Born-Infeld theories, we
will present a limiting point (vertical line in the figures).

Motivations for considering GB gravity can be found in literature.
For example, we refer the reader to the interesting nontrivial
causal structure of GB gravity with superluminal graviton modes
\cite{SL}. Finally, we should note that although most of
thermodynamic works in black hole physics related to
asymptotically AdS solutions, there were some attempts to
investigate thermodynamical behavior of dS black holes \cite{dS}.

The outline of our paper is as follows. Section II is devoted to
introduction to Einstein and GB black hole solutions and their conserved
quantities. Next, we extend the phase space by considering cosmological
constant as thermodynamic pressure and calculate critical values and then we
plot diagrams for different cases. We give a detailed discussion regarding
diagrams, their physical interpretations, and the effects of both nonlinear
electromagnetic and gravitational parameters. We finish our paper with some
closing remarks.

\section{Field equations and conserved quantities}

In order to study phase transition of Einstein-Gauss-Bonnet gravity in
presence of a generalized nonlinear electromagnetic field, one can employ
the following Lagrangian
\begin{equation}
\mathcal{L}_{\mathrm{tot}}=\mathcal{L}_{EN}-2\Lambda +\alpha \mathcal{L}%
_{GB}+\mathcal{L}(\mathcal{F}),  \label{fieldLagrangian}
\end{equation}%
where the Lagrangian of Einstein gravity is the Ricci scalar, $\mathcal{L}%
_{EN}=\mathcal{R}$, and $\Lambda $ is the negative cosmological
constant. In third term of Eq. (\ref{fieldLagrangian}), $\alpha $
is the GB coefficient with dimension (Length)$^{2}$ and
$\mathcal{L}_{GB}$ is the Lagrangian of GB gravity with following
form
\begin{equation}
\mathcal{L}_{GB}=R_{abcd}R^{abcd}-4R_{ab}R^{ab}+\mathcal{R}^{2}.  \label{LGB}
\end{equation}

Using variational method, we obtain the following field equations
\begin{equation}
G_{ab}^{E}+\Lambda g_{ab}+\alpha G_{ab}^{GB}=\frac{1}{2}g_{ab}\mathcal{L}(%
\mathcal{F})-2\mathcal{L}_{\mathcal{F}}F_{ac}F_{b}^{c},  \label{Feq1}
\end{equation}%
\begin{equation}
\partial _{a}\left( \sqrt{-g}\mathcal{L}_{\mathcal{F}}F^{ab}\right) =0,
\label{Feq2}
\end{equation}%
where $G_{ab}^{E}$ is the Einstein tensor, $G_{ab}^{GB}=2\left(
R_{acde}R_{b}^{cde}-2R_{acbd}R^{cd}-2R_{ac}R_{b}^{c}+\mathcal{R}%
R_{ab}\right) -1/2\mathcal{L}_{GB}g_{ab}$ and $L_{\mathcal{F}}=d\mathcal{L}(%
\mathcal{F})/d\mathcal{F}$.

Now, we are interested in studying topological black holes and their phase
diagrams, therefore, we employ the following static metric
\begin{equation}
ds^{2}=-f(r)dt^{2}+\frac{dr^{2}}{f(r)}+r^{2}d\Omega _{n-1}^{2},
\label{metric}
\end{equation}%
in which%
\begin{equation}
d\Omega _{n-1}^{2}=\left\{
\begin{array}{cc}
d\theta _{1}^{2}+\sum\limits_{i=2}^{n-1}\prod\limits_{j=1}^{i-1}\sin
^{2}\theta _{j}d\theta _{i}^{2} & k=1 \\
d\theta _{1}^{2}+\sinh ^{2}\theta _{1}d\theta _{2}^{2}+\sinh ^{2}\theta
_{1}\sum\limits_{i=3}^{n-1}\prod\limits_{j=2}^{i-1}\sin ^{2}\theta
_{j}d\theta _{i}^{2} & k=-1 \\
\sum\limits_{i=1}^{n-1}d\theta _{i}^{2} & k=0%
\end{array}%
\right. ,  \label{dOmega}
\end{equation}%
with volume $\omega _{n-1}$.

We use Eq. (\ref{Feq2}) and mentioned metric to obtain radial
electromagnetic field tensor as \cite{HendiMomen,HendiPanah}
\begin{equation}
F_{tr}=\frac{q}{r^{n-1}}-\frac{4q^{3}\beta }{r^{3n-3}}+O(\beta ^{2}).
\label{Ftr}
\end{equation}

In order to find Einstein solutions one can use two methods: one, by putting
$\alpha =0$ and using mentioned field equations. Second approach is
obtaining GB metric function through use of field equations and series
expanding it for small values of GB parameter. In order to give more
specific details, we use the second approach. Therefore, in case of GB
gravity, one can obtain metric function in form of \cite{HendiPanah}
\begin{equation}
f(r)=k+\frac{r^{2}}{2\alpha ^{\prime }}\left( 1-\sqrt{\Psi (r)}\right) ,
\label{f(r)}
\end{equation}%
with
\begin{equation}
\Psi (r)=1+\frac{8\alpha ^{\prime }}{n(n-1)}\left( \Lambda +\frac{n(n-1)m}{%
2r^{n}}-\frac{nq^{2}}{(n-2)r^{2n-2}}+\frac{2nq^{4}\beta }{r^{4n-4}(3n-4)}%
\right) +O\left( \beta ^{2}\right) ,  \label{Psi(r)}
\end{equation}%
where $m$ is an integration constant that is related to mass and $\alpha
^{\prime }=(n-2)(n-3)\alpha $. It is evident that for case of small values
of nonlinearity, the metric function will lead to the GB-Maxwell gravity. As
for Einstein gravity, series expanding of GB metric function for small
values of $\alpha ^{\prime }$ will lead to
\begin{equation}
f\left( r\right) =f_{EN}-\frac{4q^{4}}{\left( n-1\right) \left( 3n-4\right)
r^{4n-6}}\beta +\frac{f_{EN}^{2}}{r^{2}}\alpha ^{\prime }+O\left( \alpha
^{\prime }\beta ,\alpha ^{\prime 2},\beta ^{2}\right) ,  \label{f(r)expand}
\end{equation}%
where the metric function of Einstein-Maxwell gravity is
\begin{equation}
f_{EN}=k-\frac{2\Lambda r^{2}}{n\left( n-1\right) }-\frac{m}{r^{n-2}}+\frac{%
2q^{2}}{\left( n-1\right) \left( n-2\right) r^{2n-4}}.  \label{f(r)EM}
\end{equation}

Next step is devoted to calculating conserved quantities. In general, for
both Einstein and GB gravities one can find total mass of black hole in form
of \cite{HendiMomen,HendiPanah}
\begin{equation}
M=\frac{\omega _{n-1}\left( n-1\right) m}{16\pi }.  \label{Mass}
\end{equation}

It is notable that, although the form of total mass in GB and
Einstein gravities seems to be the same, its value is different
for Einstein and GB branches ($k\neq 0$). In other words, the
geometrical mass for GB gravity will be
\begin{equation}
m_{GB}=k\left( k\alpha ^{\prime }+r_{+}^{2}\right) r_{+}^{n-4}-\frac{%
2r_{+}^{n}\Lambda }{n\left( n-1\right) }+\frac{2q^{2}}{\left( n-1\right)
\left( n-2\right) r_{+}^{n-2}}-\frac{4q^{4}\beta }{\left( n-1\right) \left(
3n-4\right) r_{+}^{3n-4}}+O\left( \beta ^{2}\right) ,  \label{GB mass}
\end{equation}%
where $r_{+}$\ satisfies $f(r=r_{+})=0$, and Eq. (\ref{GB mass})
reduces to geometrical mass for Einstein gravity in case of
$\alpha ^{\prime }=0$.

Previously, it was seen that obtained metric functions are
representing black holes with essential singularity located at
$r=0$. Geometrical properties of the solutions were investigated
in \cite{HendiMomen,HendiPanah} and it was shown that these
solutions can be interpreted as asymptotically AdS black holes.
Therefore, by using the definition of surface gravity and its
relation with Hawking temperature we find temperature of these two
black holes as \cite{HendiPanah}
\begin{equation}
T=\frac{k(n-1)(n-2)r_{+}^{4n-6}\left( 1+\frac{(n-4)\alpha ^{\prime }}{%
(n-2)r_{+}^{2}}\right) -2r_{+}^{4n-4}\Lambda -2r_{+}^{2n-2}q^{2}+4q^{4}\beta
}{4\pi (n-1)r_{+}^{4n-5}\left( 1+\frac{2k\alpha ^{\prime }}{r_{+}^{2}}%
\right) }+O\left( \beta ^{2}\right) ,  \label{T}
\end{equation}%
where in order to find temperature of Einstein gravity, it is sufficient to
set $\alpha ^{\prime }=0$ \cite{HendiMomen}.

Due to the fact that solutions are asymptotically AdS, in order to
find entropy of these two gravities, one can use Gibbs-Duhem
relation. Therefore, we obtain the following relation
\cite{HendiPanah}
\begin{equation}
S=\frac{V_{n-1}}{4}\left( 1+\frac{2\left( n-1\right) \alpha ^{\prime }}{%
(n-3)r_{+}^{2}}k\right) r_{+}^{n-1},  \label{S}
\end{equation}%
where in order to find entropy related to Einstein gravity, one
should set GB parameter to zero \cite{HendiMomen}.

Generally, as one can see, the topological structure of spacetime
modifies the amount of contribution of GB parameter, for the GB
gravity. Although the presence of GB gravity in flat case is
evident in metric function, regarding to conserved and
thermodynamic quantities no contribution of GB gravity was seen
and obtained values are same as Einstein gravity. Considering this
fact and equation of Gibbs free energy, the thermodynamic behavior
and phase diagrams of GB and Einstein gravities in case of flat
horizon are same. In addition, it was shown that
\cite{HendiMomen,HendiPanah} obtained conserved and thermodynamic
quantities satisfy the first law of thermodynamics with the
following form
\begin{equation}
dM=TdS+\Phi dQ,  \label{First1}
\end{equation}%
where $Q=\frac{q}{4\pi }$ and $\Phi =\int F_{tr}dr|_{r=r_{+}}$.

\section{Extended phase space and phase diagrams}

In order to investigate the phase structure of the solutions, we employ the
approach in which the cosmological constant is a thermodynamic variable
corresponding to thermodynamical pressure with the following relation
\begin{equation}
P=-\frac{\Lambda }{8\pi }.  \label{P}
\end{equation}

This consideration could be justified due to the fact that in
quantum context, fundamental fixed parameters could vary and they
are not fixed. In the absence of cosmological constant, a
sourceless solution of the Einstein's equation is Minkowski
spacetime where the isometry transformations are governed by
Poincare group. In the presence of cosmological constant,
Minkowski is no longer a valid solution and it is replaced by the
(anti-)de Sitter spacetime with (anti-)de Sitter group
description. Considering that we are now employing the (anti-)de
Sitter group for describing kinematics, the ordinary notions of
energy and momentum, as well as the relationship between them and
the causal structure of spacetime will be modified \cite{Cosm}.
Therefore, we expect to see the effects of this distortion on
spacetime of black holes and its corresponding thermodynamical
values which is evident from calculated thermodynamical values. As
one can see the conjugating thermodynamic variable to this
assumption (cosmological constant as pressure) will be volume
where in literature the derived volume for different types of
black
holes are same as that for the topology of the spacetime \cite%
{KubiznakMann,10}. In order to calculate the volume of these thermodynamical
systems, we use the following relation
\begin{equation}
V=\left( \frac{\partial H}{\partial P}\right) _{S,Q}.  \label{V}
\end{equation}

In addition, it was proven that the Smarr formula should be
extended to Lovelock gravity as well as nonlinear theories of
electrodynamics \cite {KubiznakMann,GBMaxwell,Smarr}. Scaling
argument was used to derive an extension of the first law and its
related modified Smarr relation that includes variations in the
cosmological constant, Lovelock coefficient, and also the
nonlinearity parameter \cite{KubiznakMann,GBMaxwell,Smarr}. In our
case, perturbative Lovelock gravity with NED, $M$ can be a
function of entropy, pressure, charge, Lovelock parameter,
nonlinearity coupling coefficient. Regarding the previous section,
we find that those thermodynamic quantities satisfy the following
differential form,
\begin{equation}
dM=TdS+\Phi dQ+VdP+\mathcal{A}d\alpha +\mathcal{B}d\beta ,  \label{First2}
\end{equation}
where
\begin{eqnarray*}
\mathcal{A} &=&\left( \frac{dM}{d\alpha }\right) _{S,Q,P,\beta }=\frac{%
(n-1)k^{2}r_{+}^{n-4}}{16\pi }, \\
\mathcal{B} &=&\left( \frac{dM}{d\beta }\right) _{S,Q,P,\alpha }=-\frac{q^{4}%
}{4(3n-4)\pi r_{+}^{3n-4}}.
\end{eqnarray*}
Moreover, scaling argument helps us to obtain the generalized
Smarr relation for our black hole solutions in the extended phase
space
\begin{equation}
M=\frac{n-1}{n-2}TS+\Phi dQ-\frac{2}{n-2}PV-\frac{2(n-1)}{(n-2)(n-3)}%
\mathcal{A}\alpha +\frac{2}{n-2}\mathcal{B}d\beta .  \label{Smarr}
\end{equation}
We should note that the obtained relations is valid for
perturbative Lovelock gravity with NED. Hereafter, we treat the
cosmological constant as a thermodynamical variable, while
Gauss-Bonnet coefficient and nonlinearity parameter of NED as two
constants. With doing so the total finite mass of the black hole
will play the role of Enthalpy and the corresponding Gibbs free
energy will be in form of
\begin{equation}
G=H-TS=M-TS.  \label{G}
\end{equation}

The obtained volume for our considered cases is
\begin{equation}
V=\frac{\omega _{n-1}{r_{+}}^{n}}{n},  \label{V2}
\end{equation}%
which is consistent with topological structure of spherical symmetric
spacetime. This result is consistent with what was derived previously \cite%
{KubiznakMann,10} and shows the fact that although considering GB gravity
modifies the metric function and some conserved quantities of the black
hole, it does not change the volume of the black hole. In other words, the
volume of the black hole is solely dependant on the cosmological constant
(pressure). Due to relation between volume and radius of the black hole, we
use horizon radius (specific volume) in order to investigate the critical
behavior of these systems \cite{KubiznakMann,10}.

Next step will be calculating critical values. In order to do so,
we use the method in which critical values are obtained through
the use of $P-r_{+}$ diagrams. Since the critical point is an
inflection point on the critical isotherm $P-r_{+}$ diagram, we
use the following relations to obtain the proper equations for
critical quantities
\begin{equation}
\left( \frac{\partial P}{\partial r_{+}}\right) _{T}=\left( \frac{\partial
^{2}P}{\partial r_{+}^{2}}\right) _{T}=0.  \label{inflection}
\end{equation}

It will be constructive to give a short description regarding
different phase diagrams and the information they contain before
presenting tables and phase diagrams. $G-T$ diagrams are
representing energy level of different states that phase
transition takes place between them. The characteristic swallow
tail that is seen in these diagrams shows the process that we know
as phase transition. It also gives interesting information
regarding temperature of critical points. For $T-r_{+}$ diagrams,
it contains information regarding critical temperature and horizon
radius in which phase transition takes place. Also, it gives some
insight about single state regions which in our case is
small/large black holes. It also helps us to understand the
effects of different parameters on critical temperature and
horizon radius, and whether by changing value of a parameter,
system needs more or less energy in order to have phase
transition. If one is interested in studying
conductor/superconductor transition that these nonlinear
electromagnetic fields are representing, studying these diagrams
will give more information regarding to conductivity and
superconductivity regions. Studying $P-r_{+}$ diagrams gives us
information regarding the behavior of pressure as a function of
horizon radius, and critical pressure and horizon radius of phase
transition. Finally, using the fact that the free energy,
temperature, and the pressure of the system are constant during
the phase transition, one can plot the coexistence curve of two
phases. One of the reasons for studying these diagrams is the
similarity between phase structure of black holes and the Van der
Waals thermodynamical systems. Here, we have used the geometric
units and investigate thermodynamic behavior and critical point of
the solutions, qualitatively.

Using Eq. (\ref{inflection}) one can find $T_{c}$ in one of the
equations and replace it in other equation which leads to the
following relations for calculating critical horizon radius
\begin{equation}
\left\{
\begin{array}{cc}
k\left( n-2\right) r_{+}^{4}+8\left( 4n-5\right) \beta
q^{4}r_{+}^{10-4n}-2\left( 2n-3\right) q^{2}r_{+}^{8-2n}=0, &
\begin{array}{c}
Einstein \\
\end{array}
\\
&  \\
\begin{array}{c}
\left( 12n-48\right) \alpha ^{\prime 2}k^{3}-12\alpha ^{\prime
}k^{2}r_{+}^{2}+\left\{ 4\left( 4n-5\right) \beta q^{2}r_{+}^{10-4n}-\left(
2n-3\right) r_{+}^{8-2n}\right\} 2q^{2}\vspace{0.23cm} \\
+\left[ \left\{ 4\left( 4n-7\right) \beta q^{2}r_{+}^{8-4n}-\left(
2n-5\right) r_{+}^{6-2n}\right\} 12q^{2}\alpha ^{\prime }+\left( n-2\right)
r_{+}^{4}\right] k=0,%
\end{array}
& GB%
\end{array}%
\right. .  \label{rc}
\end{equation}

\begin{figure}[tbp]
$%
\begin{array}{ccc}
\epsfxsize=5cm \epsffile{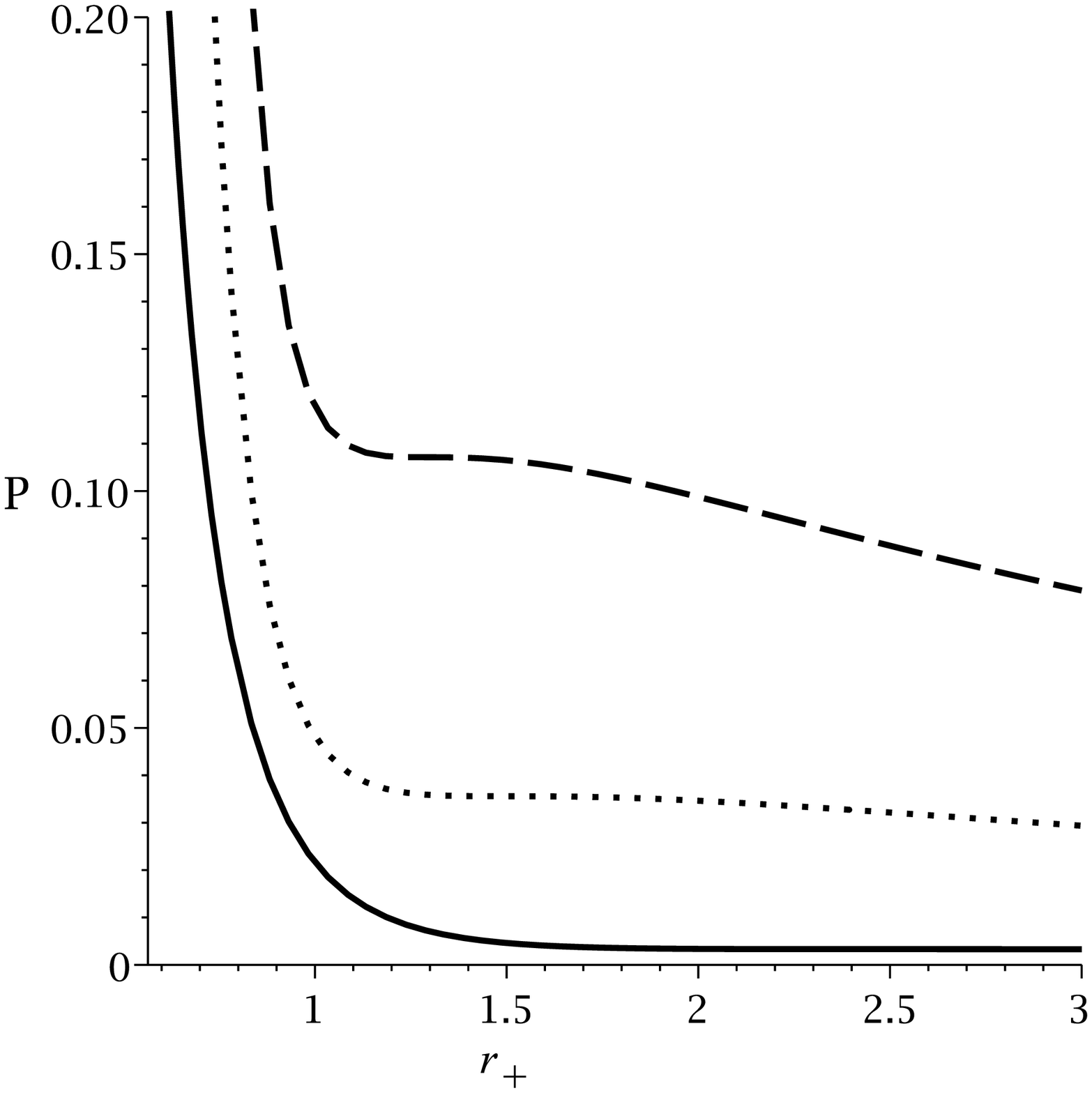} & \epsfxsize=5cm %
\epsffile{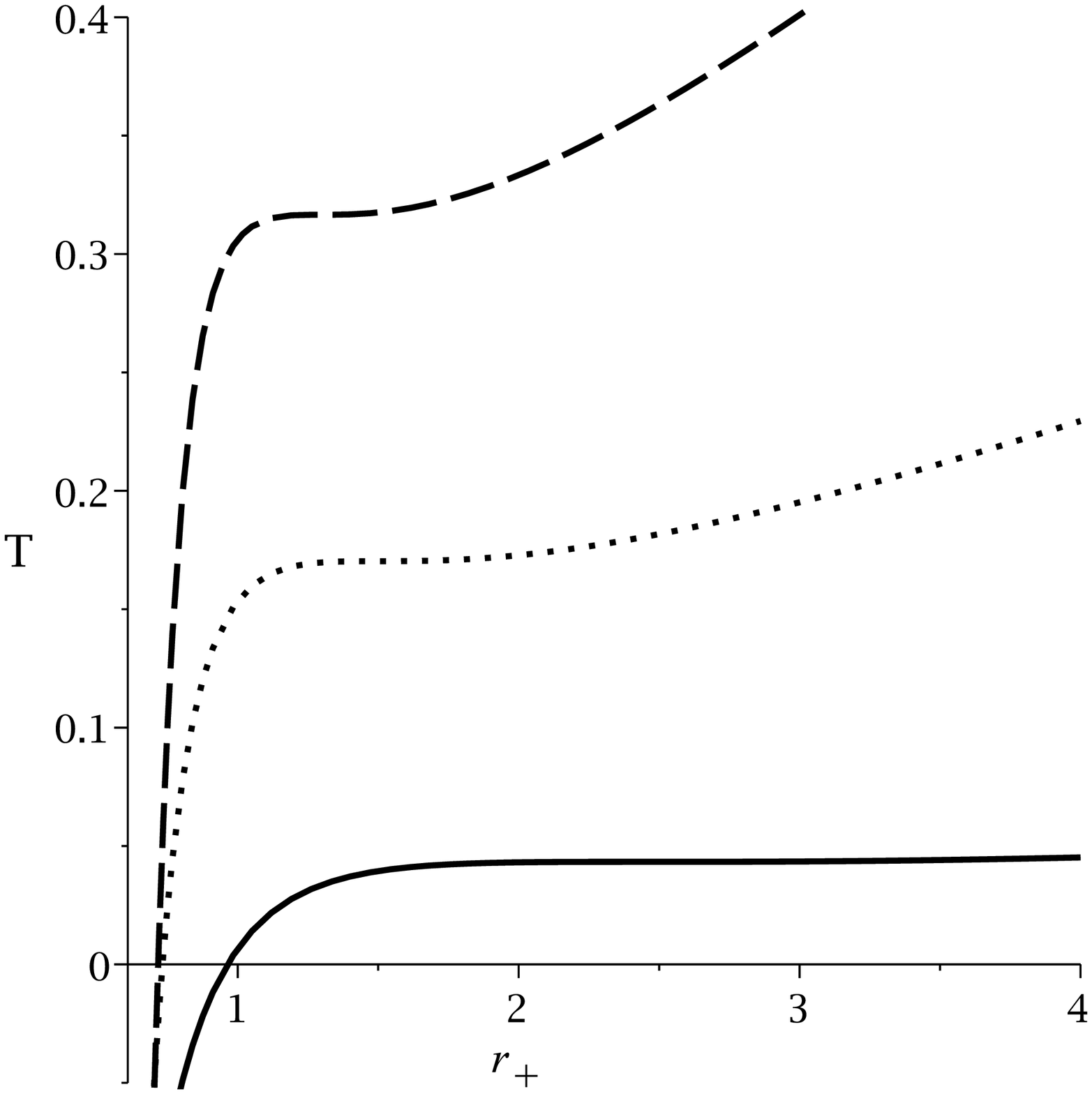} & \epsfxsize=5cm \epsffile{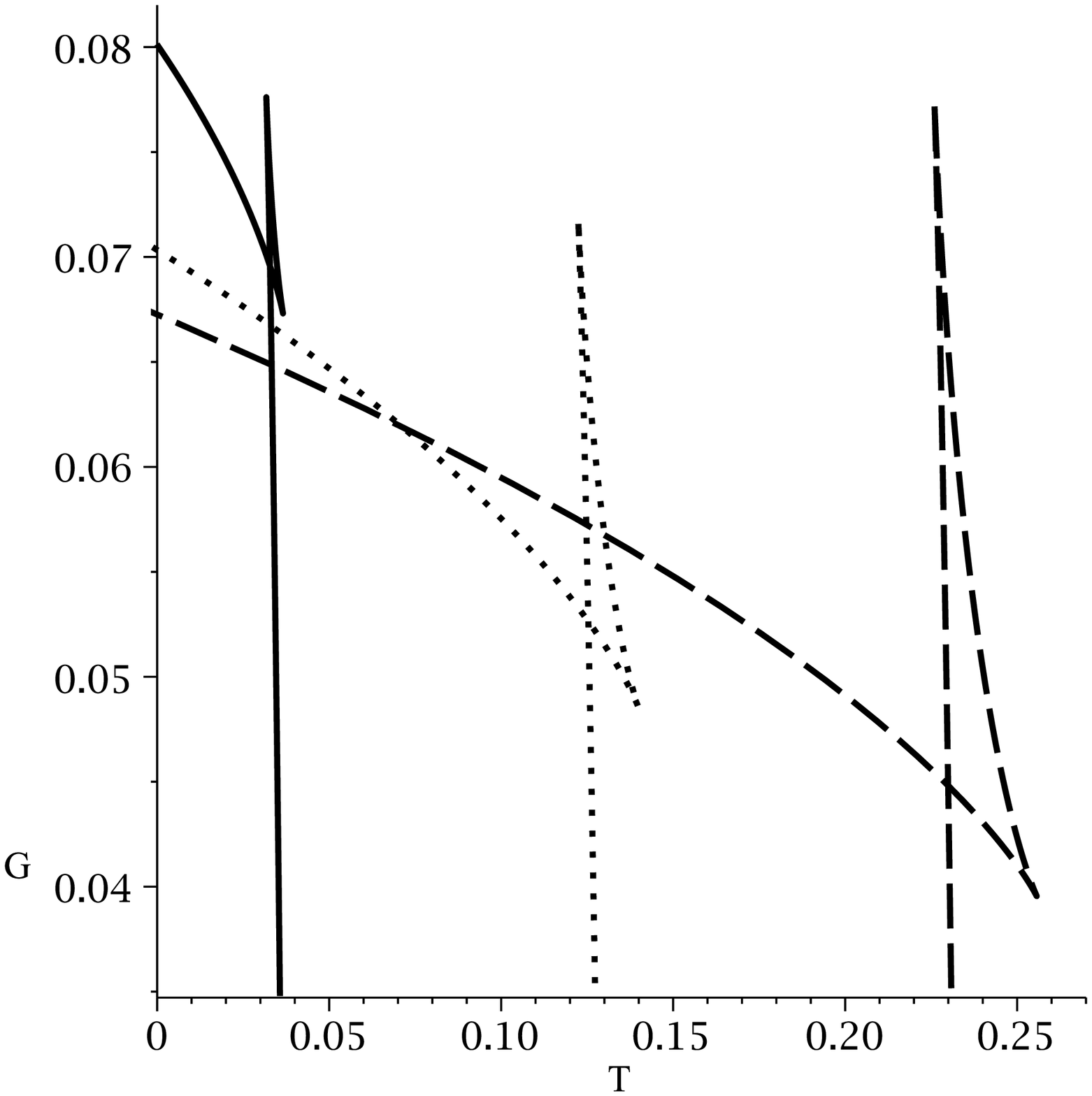}%
\end{array}
$%
\caption{\textbf{\emph{Maxwell solutions for Einstein gravity:}} $P-r_{+}$
for $T=T_{c}$ (Left), $T-r_{+}$ for $P=P_{c}$ (Middle) and $G-T$ for $%
P=0.5P_{c}$ (Right) diagrams for $k=1$, $q=1$, $n=3$ (continuous line), $n=4$
(dotted line) and $n=5$ (dashed line).}
\label{EMbeta0dim}
\end{figure}

As one can see, due to complexity of the obtained relation for
critical horizon radius, it is not possible to find critical
horizon radius analytically. Therefore, we employ the numerical
method in order to calculate critical quantities and study the
effects of variation of parameters in case of spherical horizon
($k=1$). The numerical calculations show that for cases of flat
($k=0$) and hyperbolic ($k=-1$) horizons, the critical pressure
and critical temperature for both Einstein and GB gravities are
negative, so like GB-Maxwell black holes \cite{GBMaxwell}, the
phase transition does not take place. Here, we present various
tables in order to study the effects of different parameters
on critical values. Next, by using the information of these tables, we plot $%
P-r_{+}$, $T-r_{+}$ and $G-T$ diagrams for Einstein and GB gravities in the
presence of nonlinear corrected Maxwell field (Figs. \ref{EMbeta0dim} -- \ref%
{GBn456}). It is notable that following results for critical
pressure and temperature are obtained by using larger critical
horizon radius. In order for higher orders of corrections to be
small enough and do not acquire values higher than Maxwell term,
we have plotted a vertical line which represents the limit for
different cases.

It is notable to mention that in order to have a well-defined
vacuum solution with $m=q=0$, the pressure $P$ has to satisfy the
following constraint \cite{GBMaxwell,maximal pressure}%
\begin{equation}
0\leq \frac{64\pi \alpha ^{\prime }P}{n(n-1)}\leq 1,  \label{Pmax1}
\end{equation}%
which puts a large bound for the pressure as maximal pressure
\begin{equation}
P\leq P_{\max }=\frac{n(n-1)}{64\pi \alpha ^{\prime }}.  \label{Pmax2}
\end{equation}

It means that only for sufficiently small pressures, the solution
Eq. (\ref{f(r)}) possesses an asymptotic AdS region. However, in
this paper, we choose suitable parameters for pressure to be
smaller than maximal pressure everywhere.
\begin{figure}[tbp]
$%
\begin{array}{ccc}
\epsfxsize=5cm \epsffile{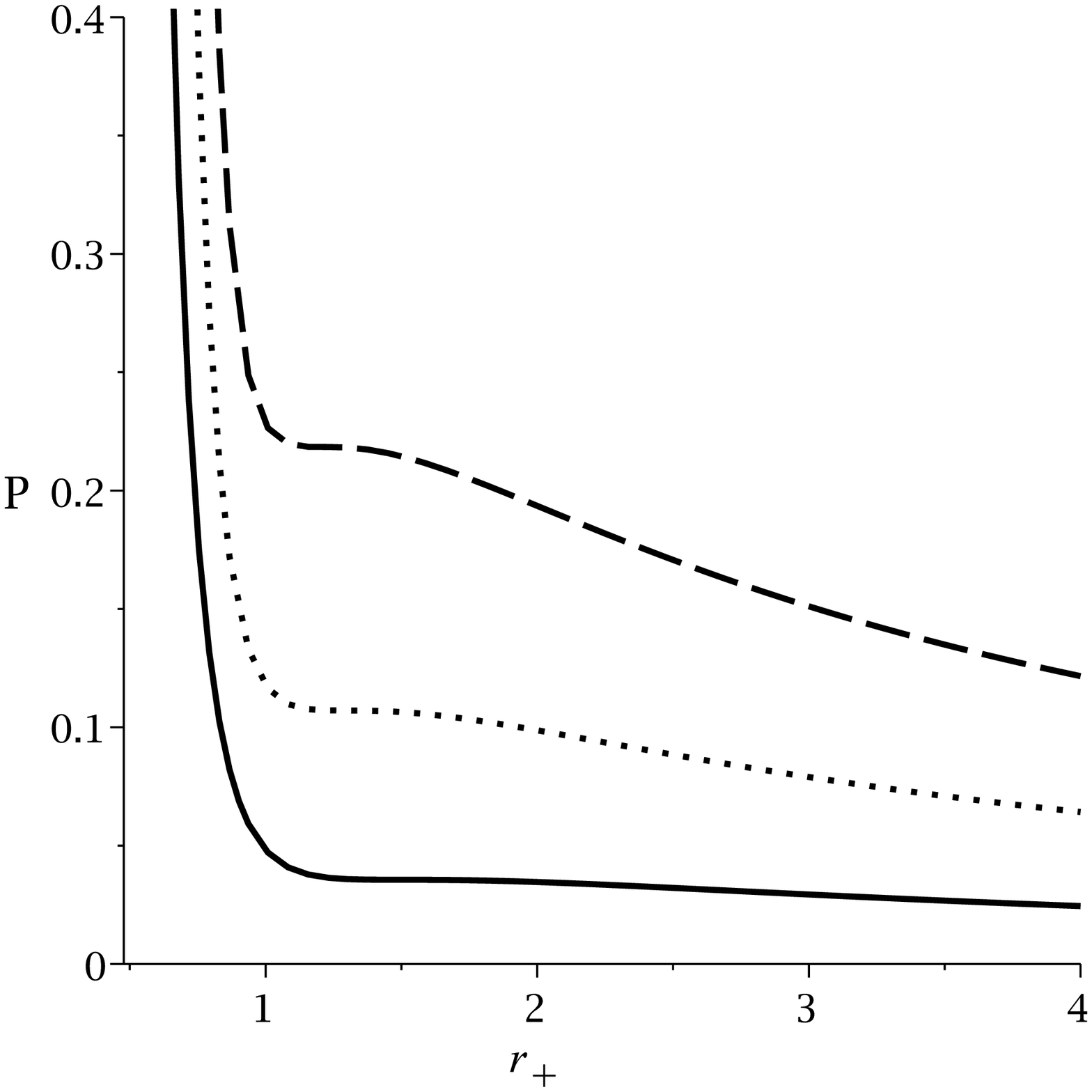} & \epsfxsize=5cm %
\epsffile{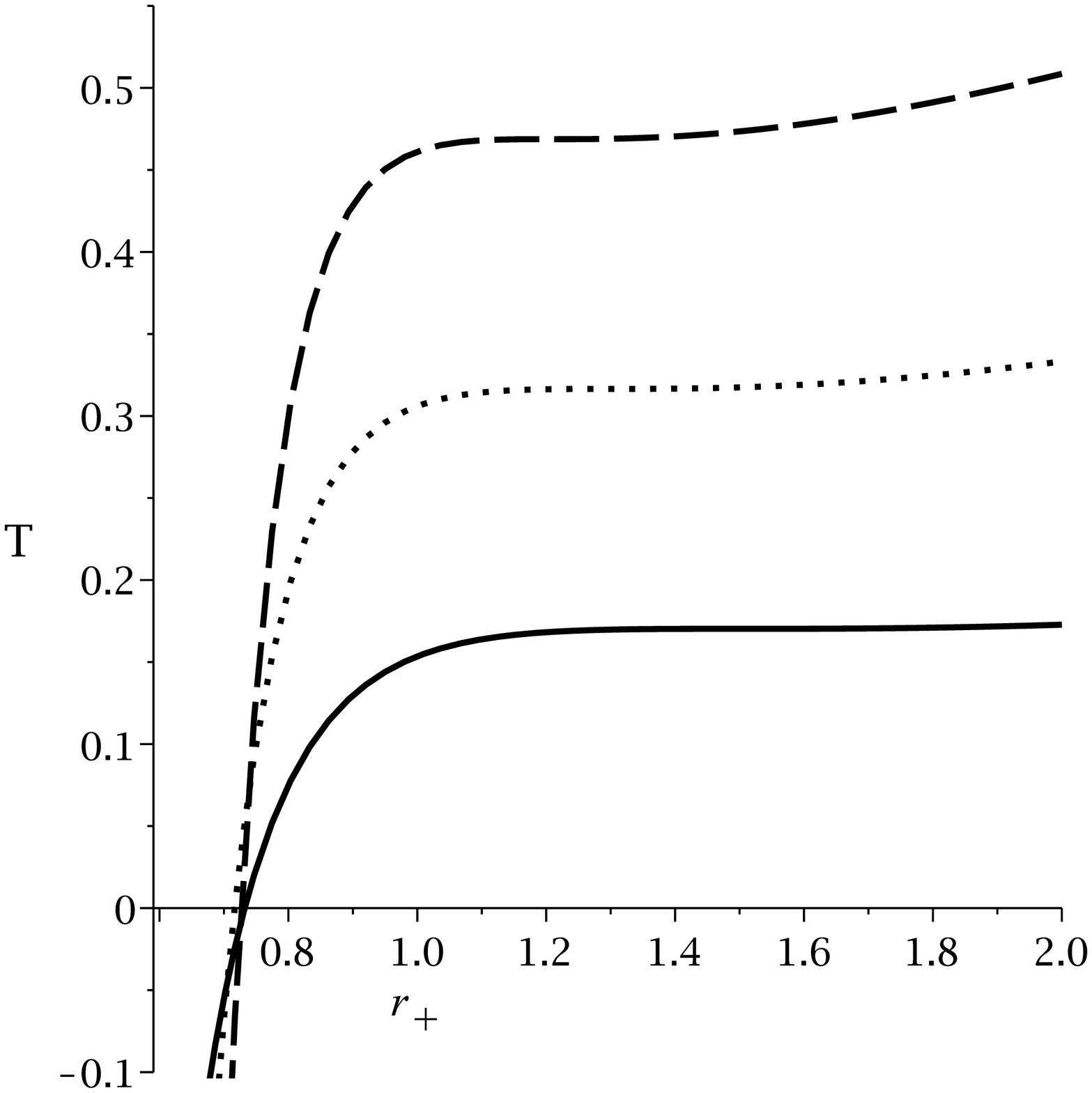} & \epsfxsize=5cm \epsffile{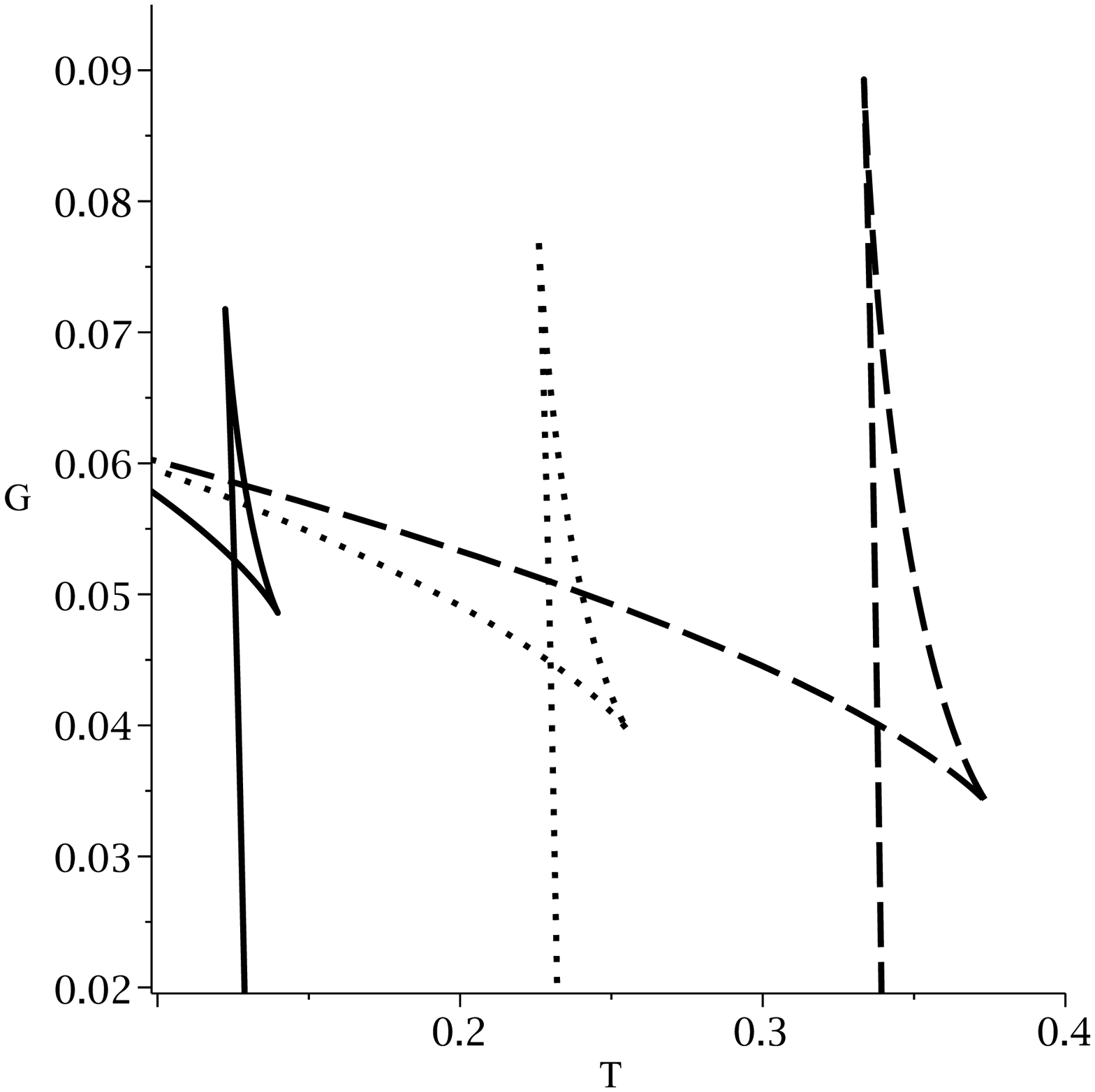}%
\end{array}
$%
\caption{\textbf{\emph{Maxwell solutions for GB gravity:}} $P-r_{+}$ for $%
T=T_{c}$ (Left), $T-r_{+}$ for $P=P_{c}$ (Middle) and $G-T$ for $P=0.5P_{c}$
(Right) diagrams for $k=1$, $q=1$, $\protect\alpha ^{\prime }=10^{-4}$, $n=4$
(continuous line), $n=5$ (dotted line) and $n=6$ (dashed line).}
\label{GBbeta0dim}
\end{figure}
\begin{figure}[tbp]
$%
\begin{array}{ccc}
\epsfxsize=5cm \epsffile{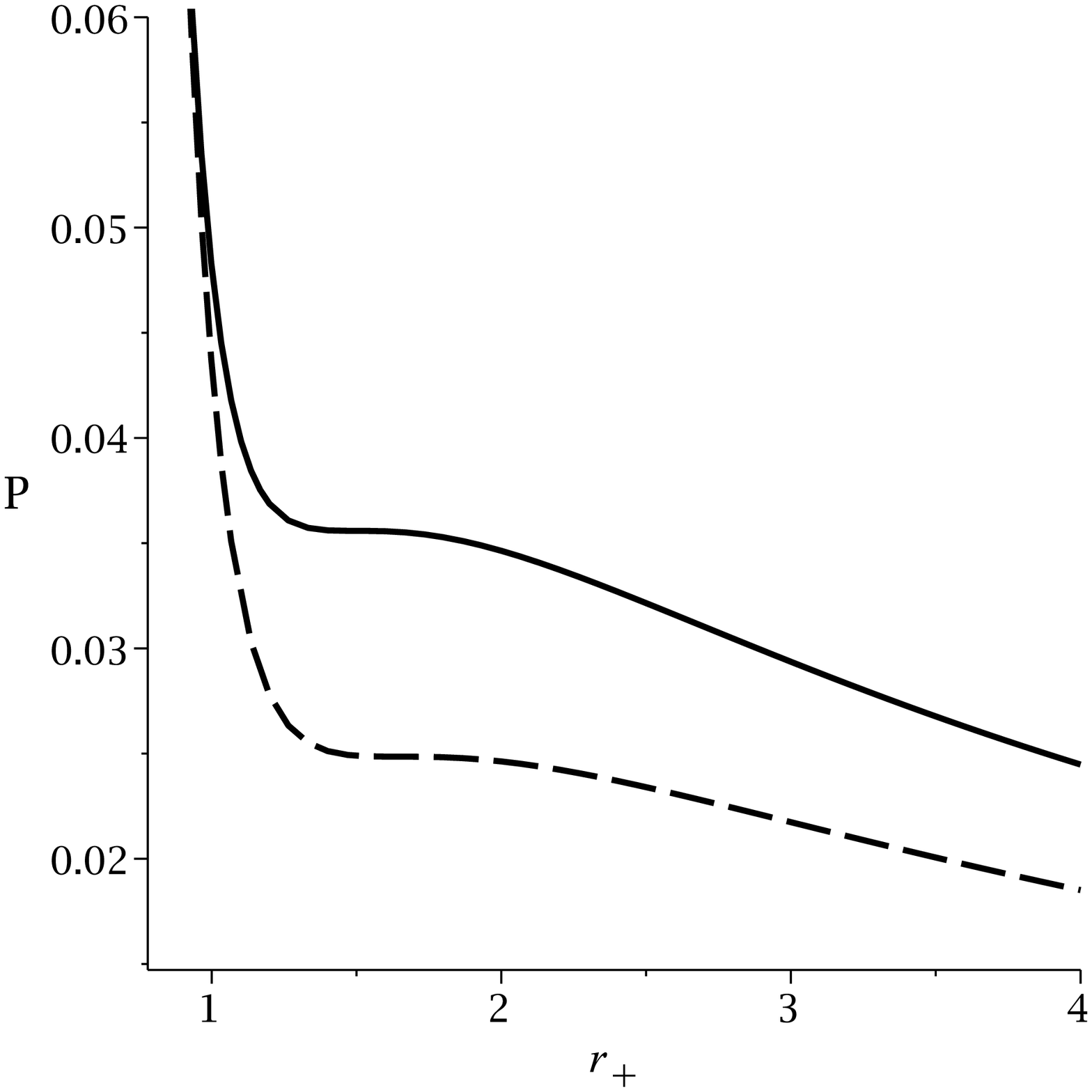} & \epsfxsize=5cm %
\epsffile{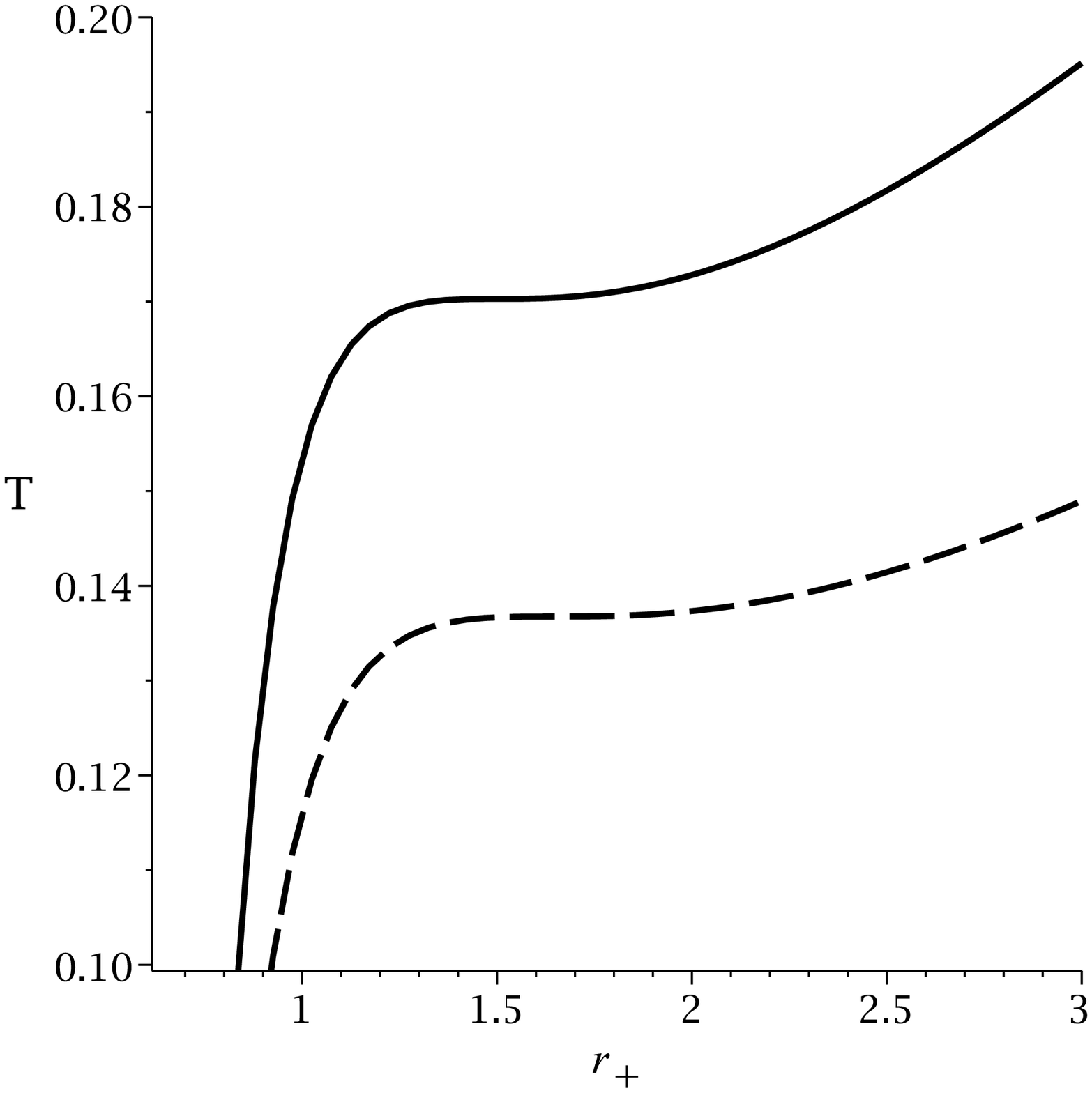} & \epsfxsize=5cm %
\epsffile{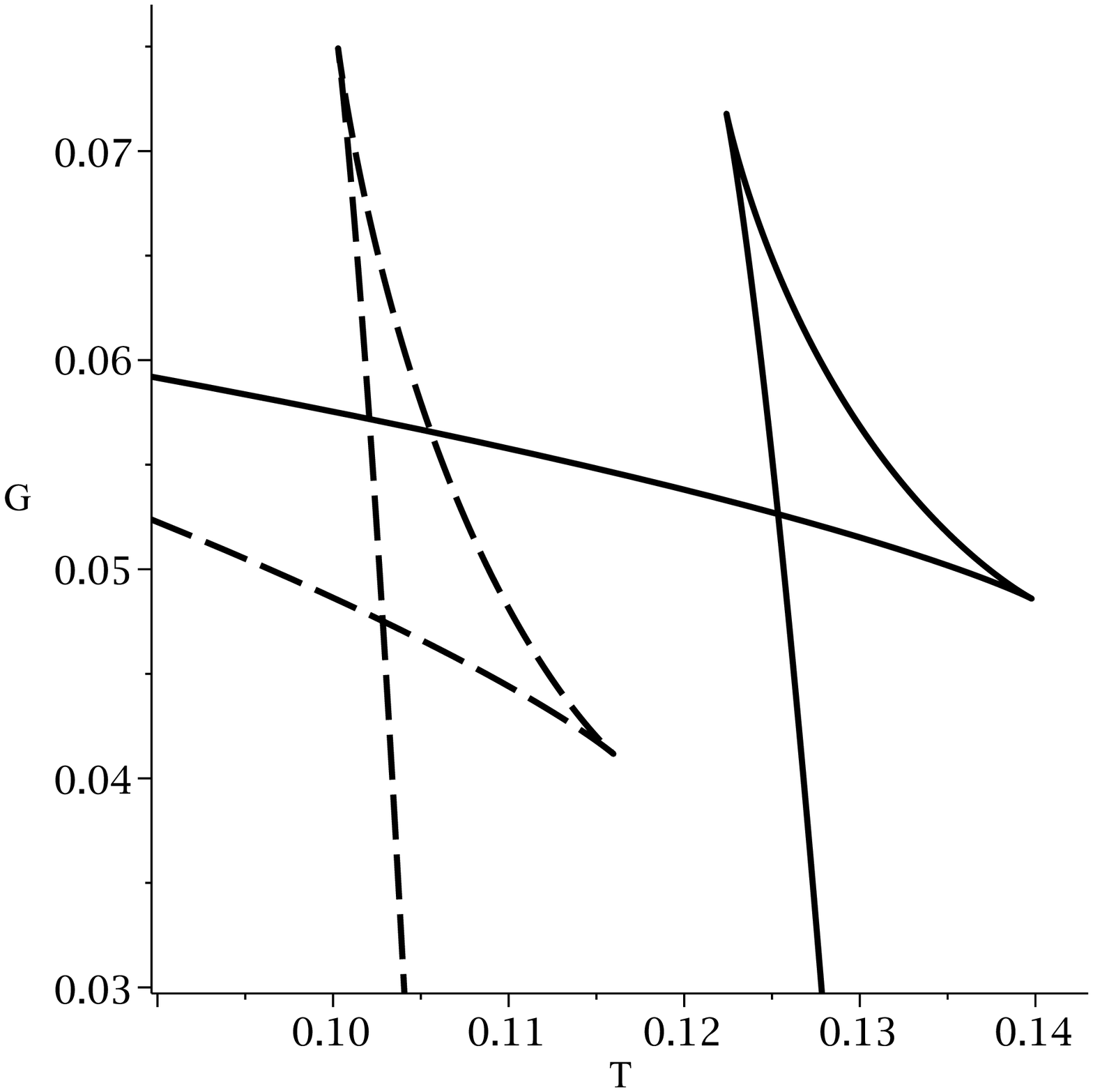}%
\end{array}
$%
\caption{\textbf{\emph{Maxwell solutions for Einstein and GB gravities:}} $%
P-r_{+}$ for $T=T_{c}$ (Left), $T-r_{+}$ for $P=P_{c}$ (Middle) and $G-T$
for $P=0.5P_{c}$ (Right) diagrams for $k=1$, $n=4$, $q=1$, $\protect\alpha %
^{\prime }=0.1$ for Einstein (continuous line) and GB (dashed line)
gravities.}
\label{comparEandGBbeta0n4}
\end{figure}
\begin{figure}[tbp]
$%
\begin{array}{ccc}
\epsfxsize=5cm \epsffile{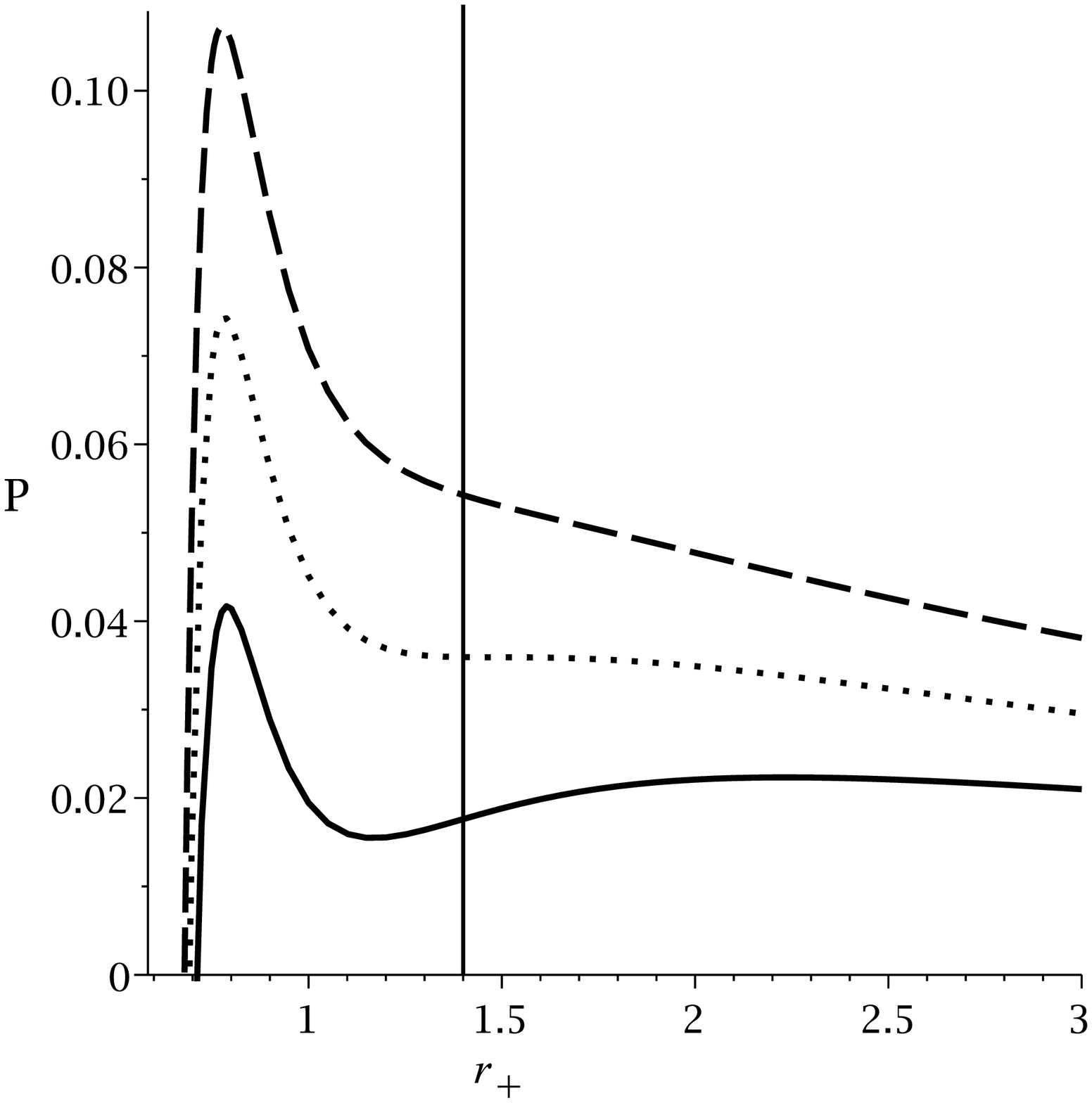} & \epsfxsize=5cm %
\epsffile{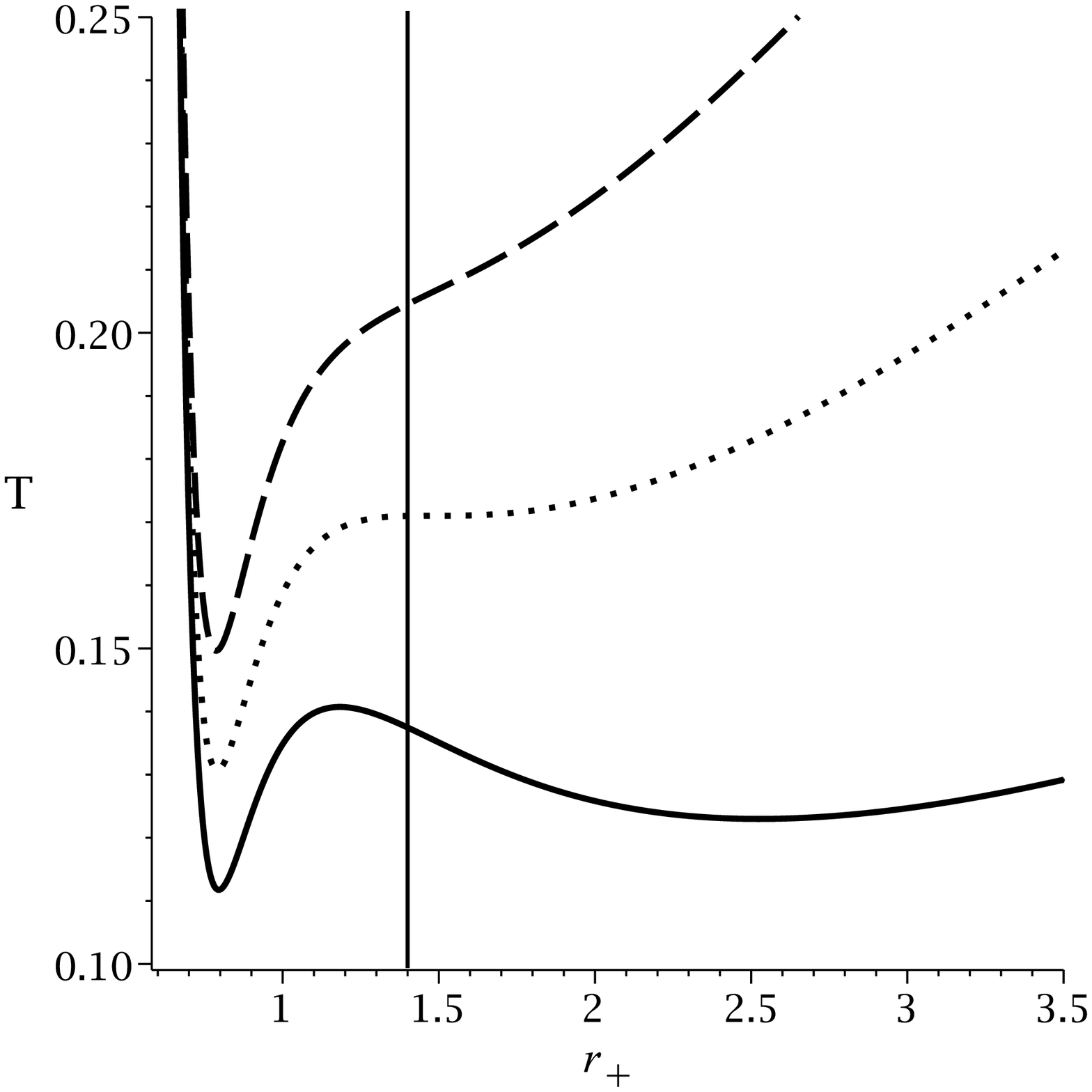} & \epsfxsize=5cm \epsffile{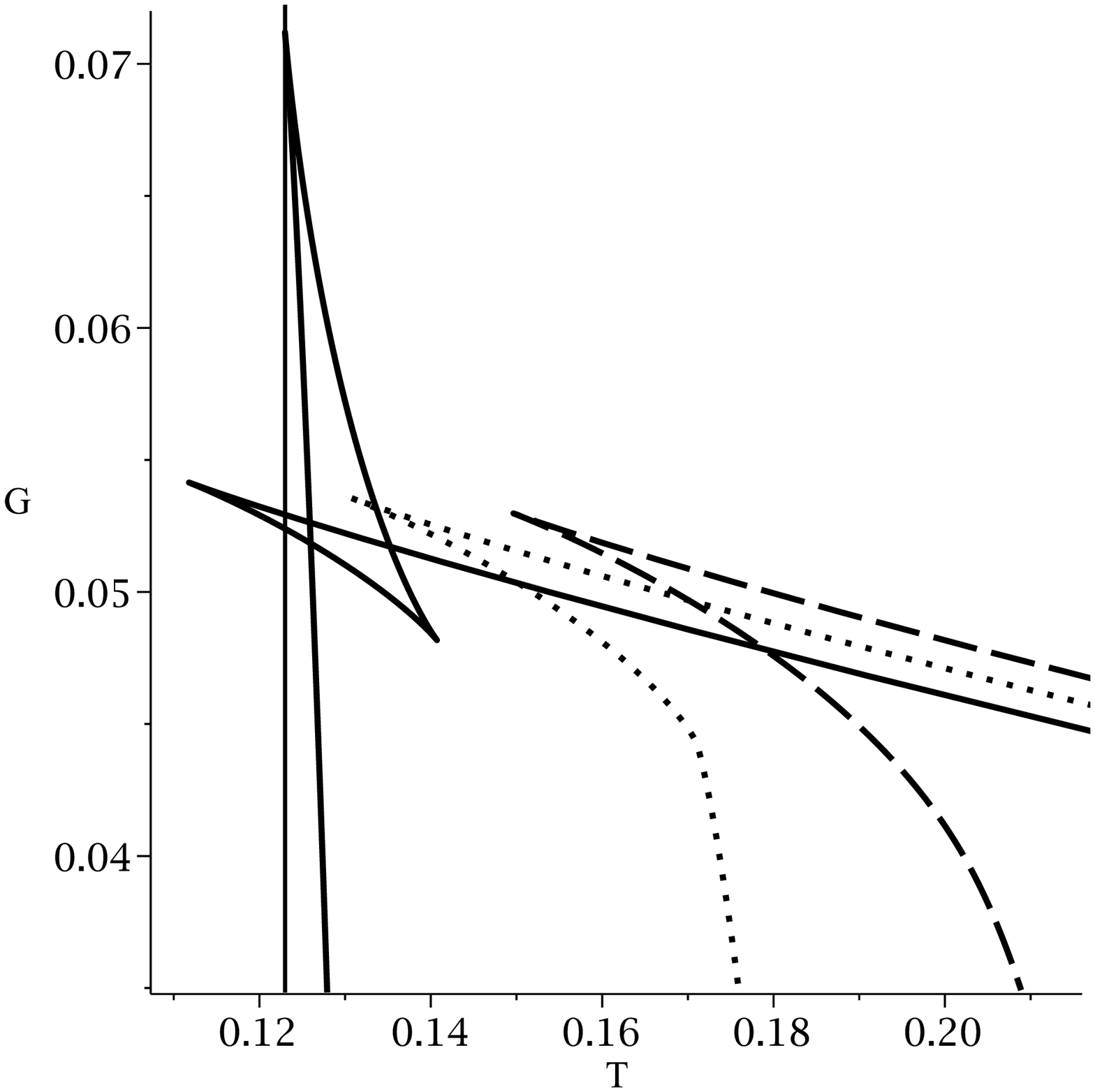}%
\end{array}
$%
\caption{\textbf{\emph{Einstein solutions:}} $P-r_{+}$ (Left), $T-r_{+}$
(Middle) and $G-T$ (Right) diagrams for $k=1$, $n=4$, $q=1$ and $\protect%
\beta =0.045$. \newline
$P-r_{+}$ diagram: $T=0.8T_{c}$ (continuous line), $T=T_{c}$ (dotted line)
and $T=1.2T_{c}$ (dashed line). \newline
$T-r_{+}$ diagram: $P=0.5P_{c}$ (continuous line), $P=P_{c}$ (dotted line)
and $P=1.5P_{c}$ (dashed line). \newline
$G-T$ diagram: $P=0.5P_{c}$ (continuous line), $P=P_{c}$ (dotted line) and $%
P=1.5P_{c}$ (dashed line).}
\label{Ebeta045n4}
\end{figure}
\begin{figure}[tbp]
$%
\begin{array}{ccc}
\epsfxsize=5cm \epsffile{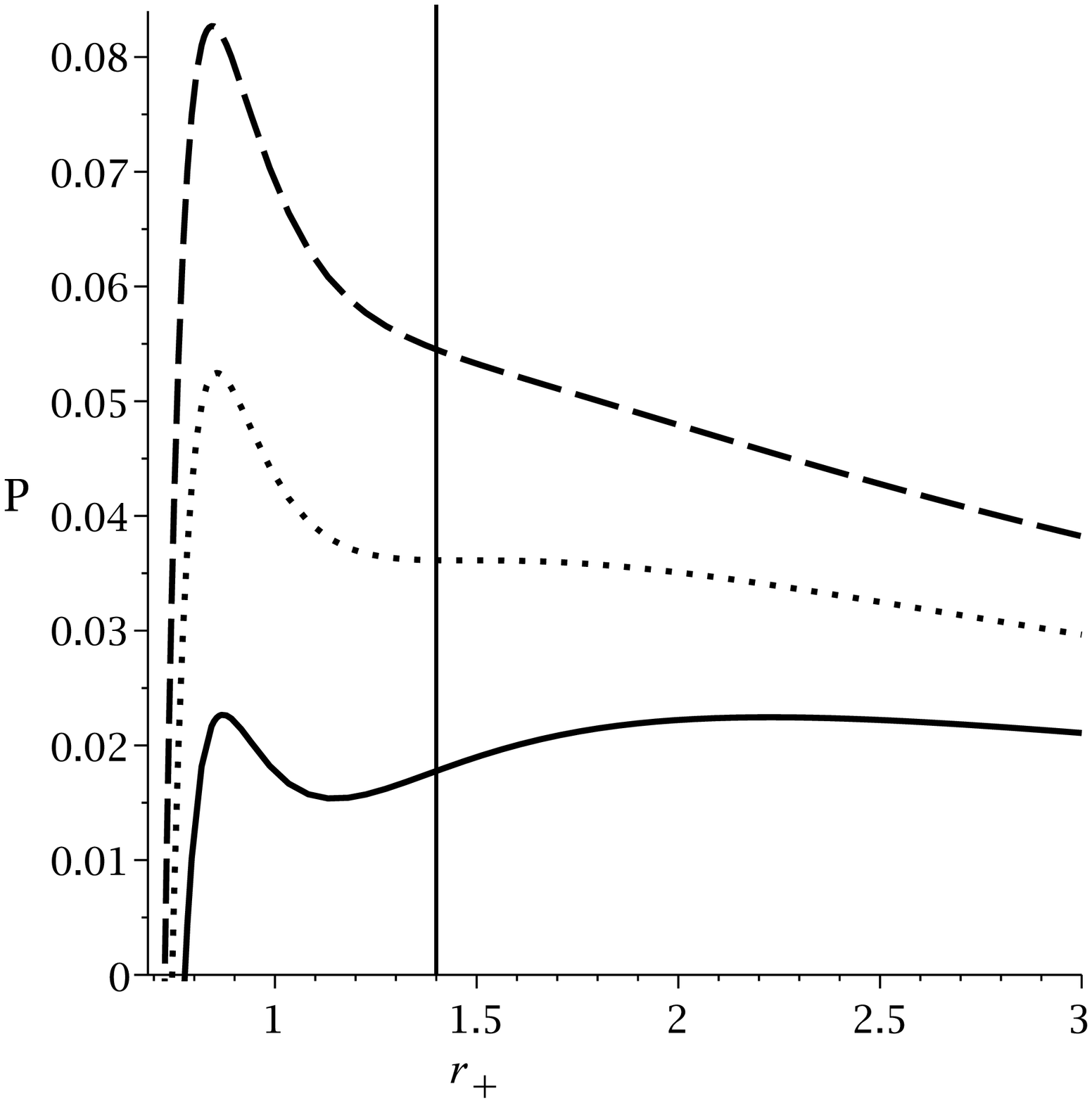} & \epsfxsize=5cm %
\epsffile{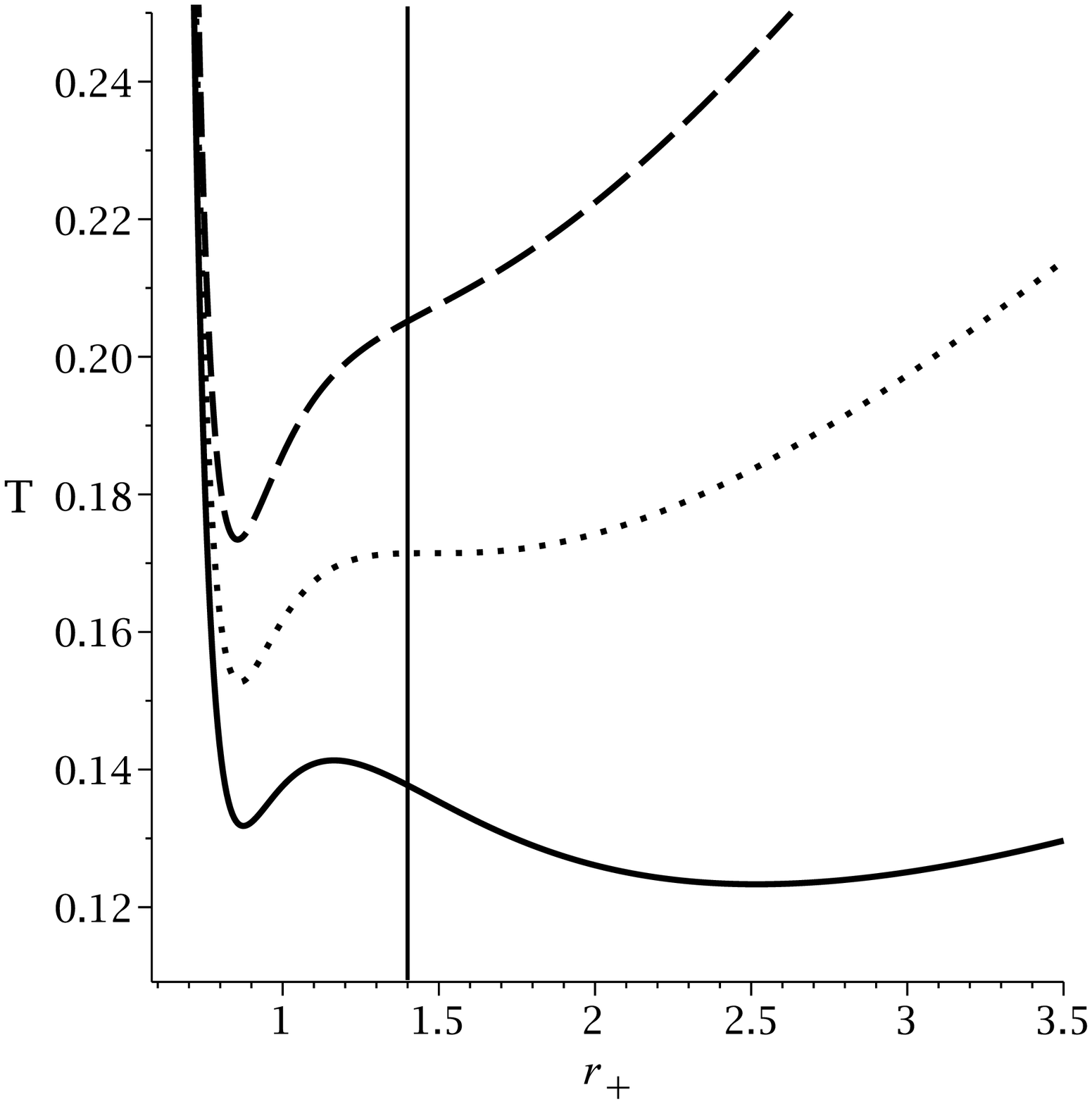} & \epsfxsize=5cm \epsffile{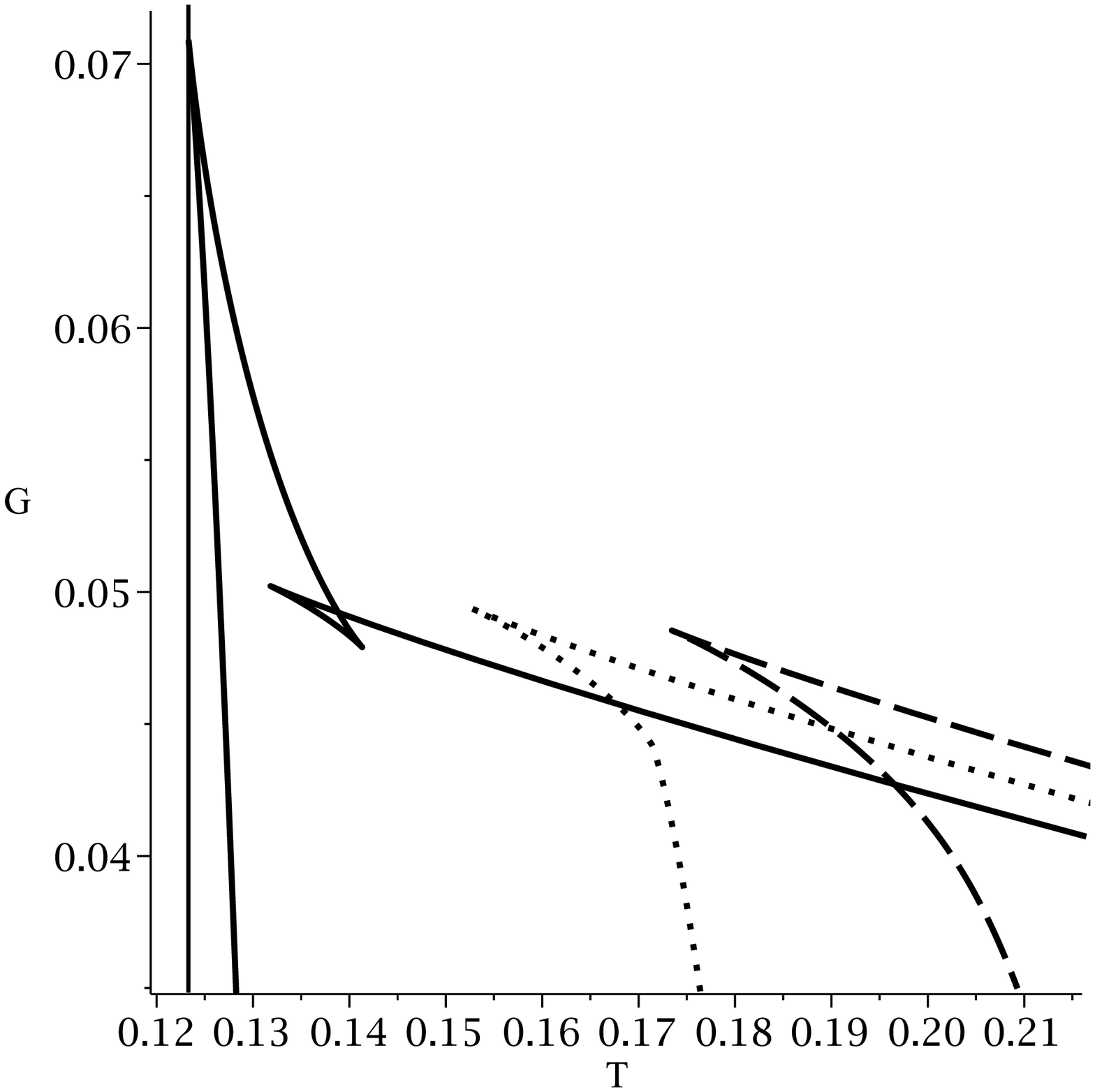}%
\end{array}
$%
\caption{\textbf{\emph{Einstein solutions:}} $P-r_{+}$ (Left), $T-r_{+}$
(Middle) and $G-T$ (Right) diagrams for $k=1$, $n=4$, $q=1$ and $\protect%
\beta =0.07$. \newline
$P-r_{+}$ diagram: $T=0.8T_{c}$ (continuous line), $T=T_{c}$ (dotted line)
and $T=1.2T_{c}$ (dashed line). \newline
$T-r_{+}$ diagram: $P=0.5P_{c}$ (continuous line), $P=P_{c}$ (dotted line)
and $P=1.5P_{c}$ (dashed line). \newline
$G-T$ diagram: $P=0.5P_{c}$ (continuous line), $P=P_{c}$ (dotted line) and $%
P=1.5P_{c}$ (dashed line).}
\label{Ebeta07n4}
\end{figure}
\begin{figure}[tbp]
$%
\begin{array}{ccc}
\epsfxsize=5cm \epsffile{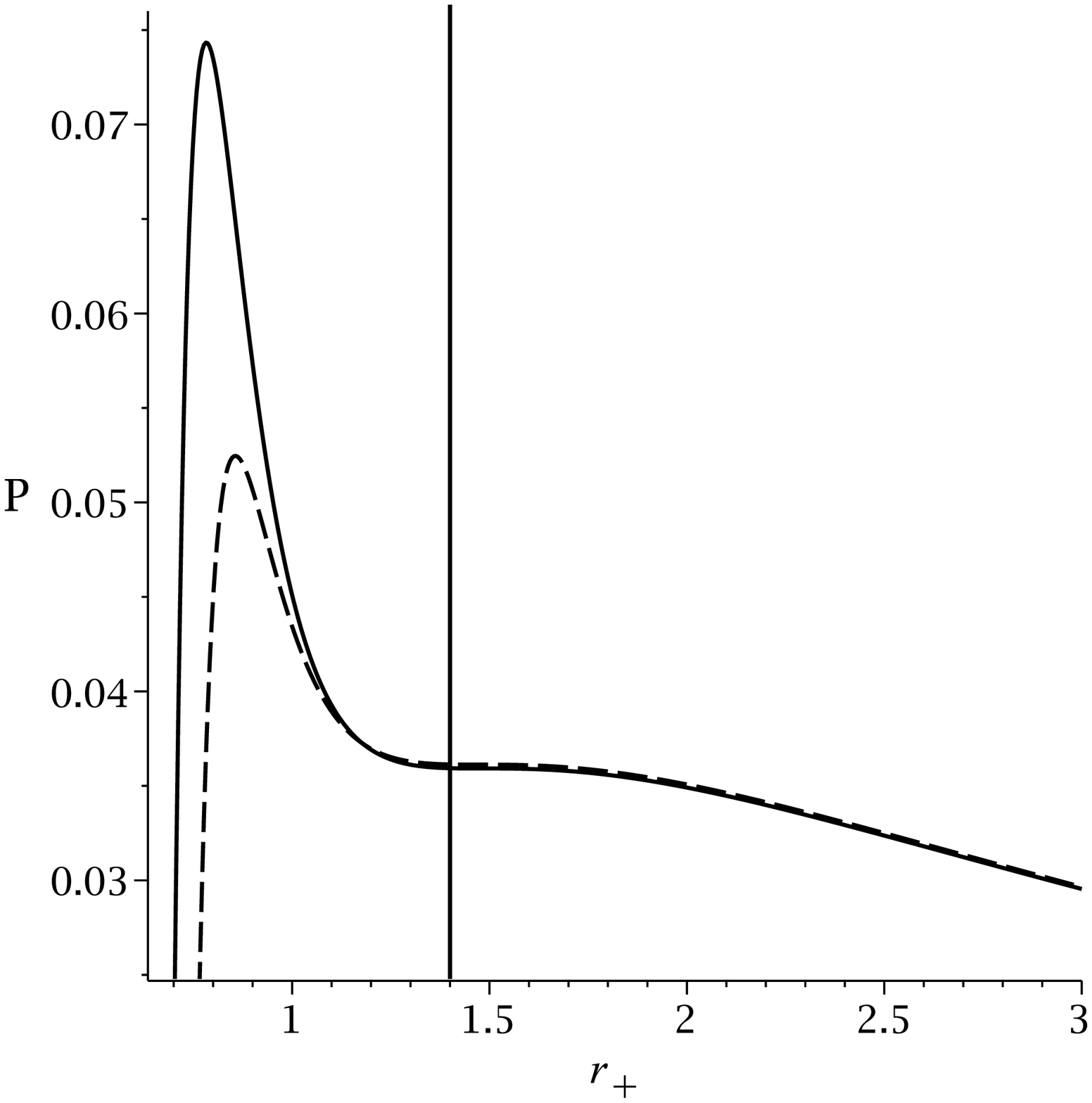} & \epsfxsize=5cm %
\epsffile{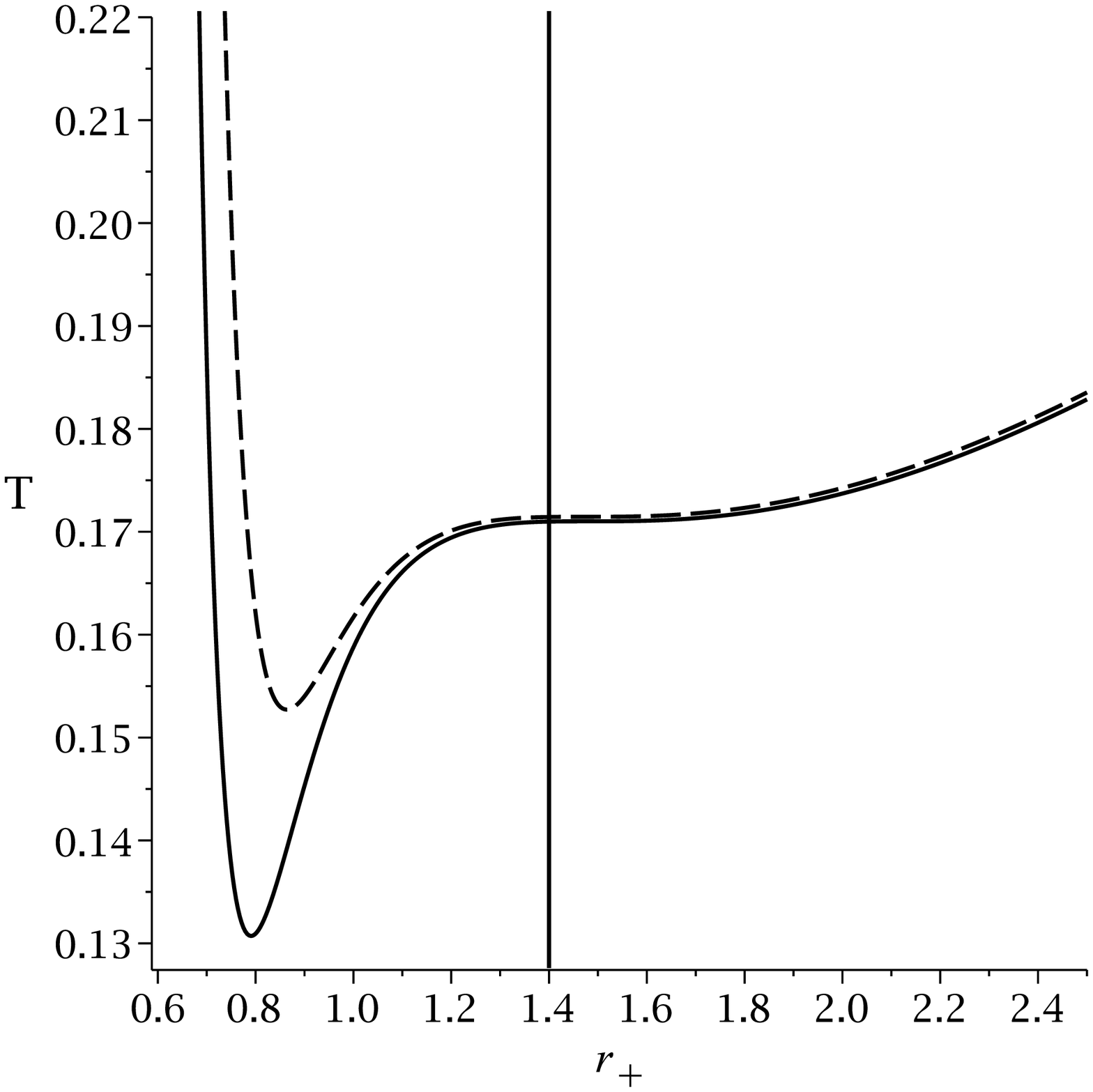} & \epsfxsize=5cm %
\epsffile{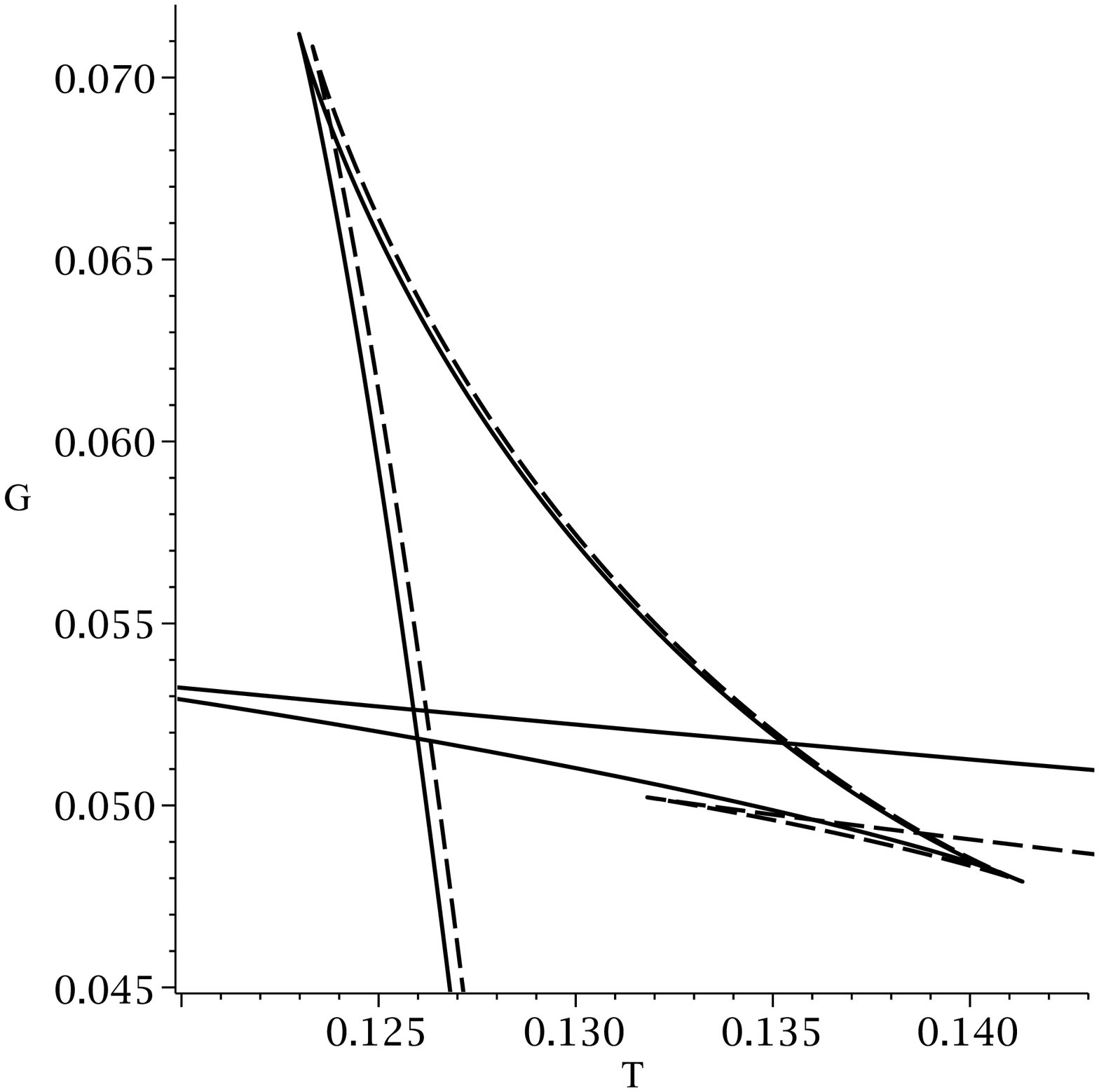}%
\end{array}
$%
\caption{\textbf{\emph{Einstein solutions:}} $P-r_{+}$ for $T=T_{c}$ (Left),
$T-r_{+}$ for $P=P_{c}$ (Middle) and $G-T$ for $P=0.5P_{c}$ (Right) diagrams
for $k=1$, $n=4$, $q=1$, $\protect\beta =0.045$ (continuous line) and $%
\protect\beta =0.07$ (dashed line).}
\label{comparEbeta04507n4}
\end{figure}
\begin{figure}[tbp]
$%
\begin{array}{ccc}
\epsfxsize=5cm \epsffile{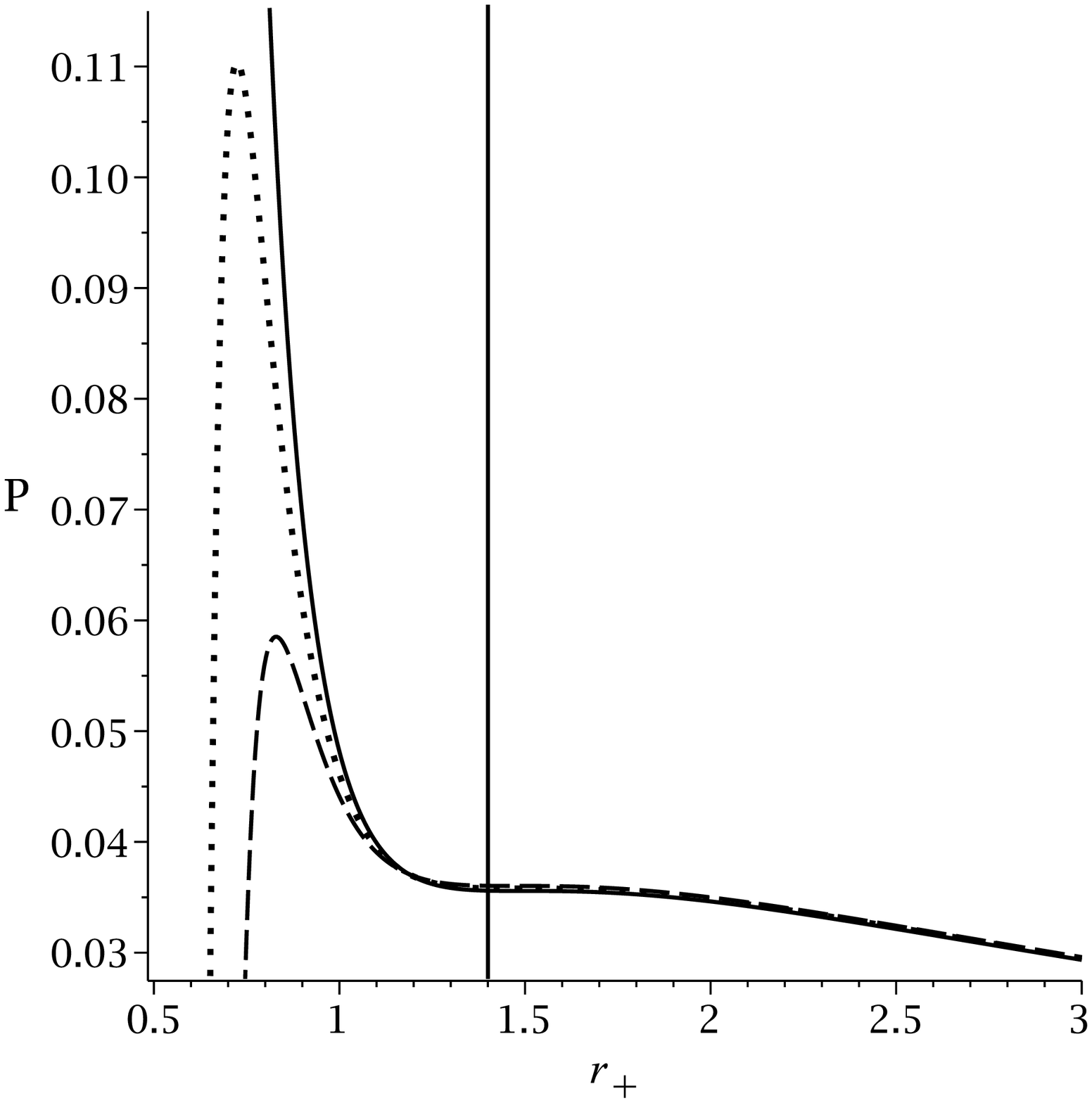} & \epsfxsize=5cm %
\epsffile{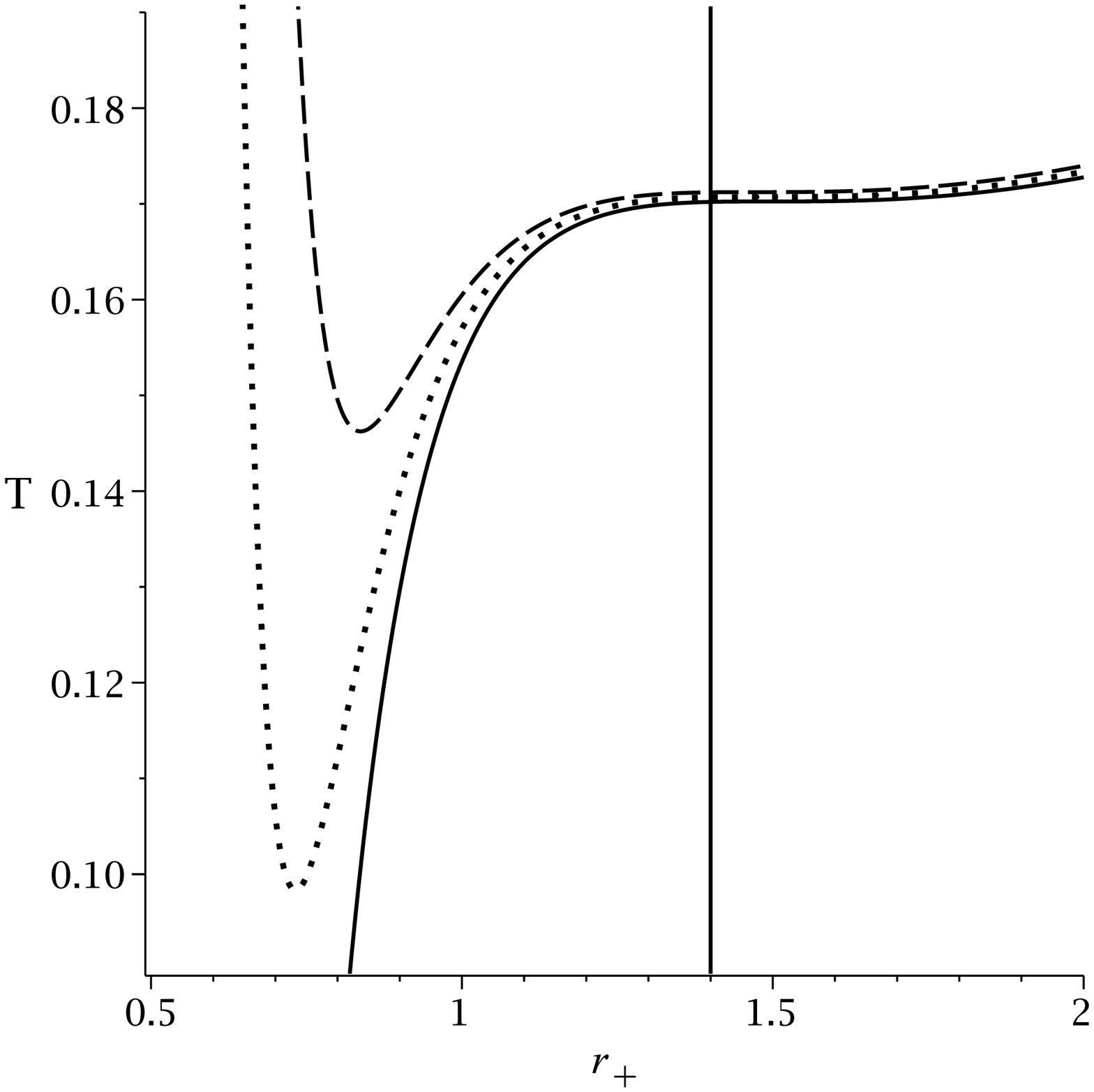} & \epsfxsize=5cm %
\epsffile{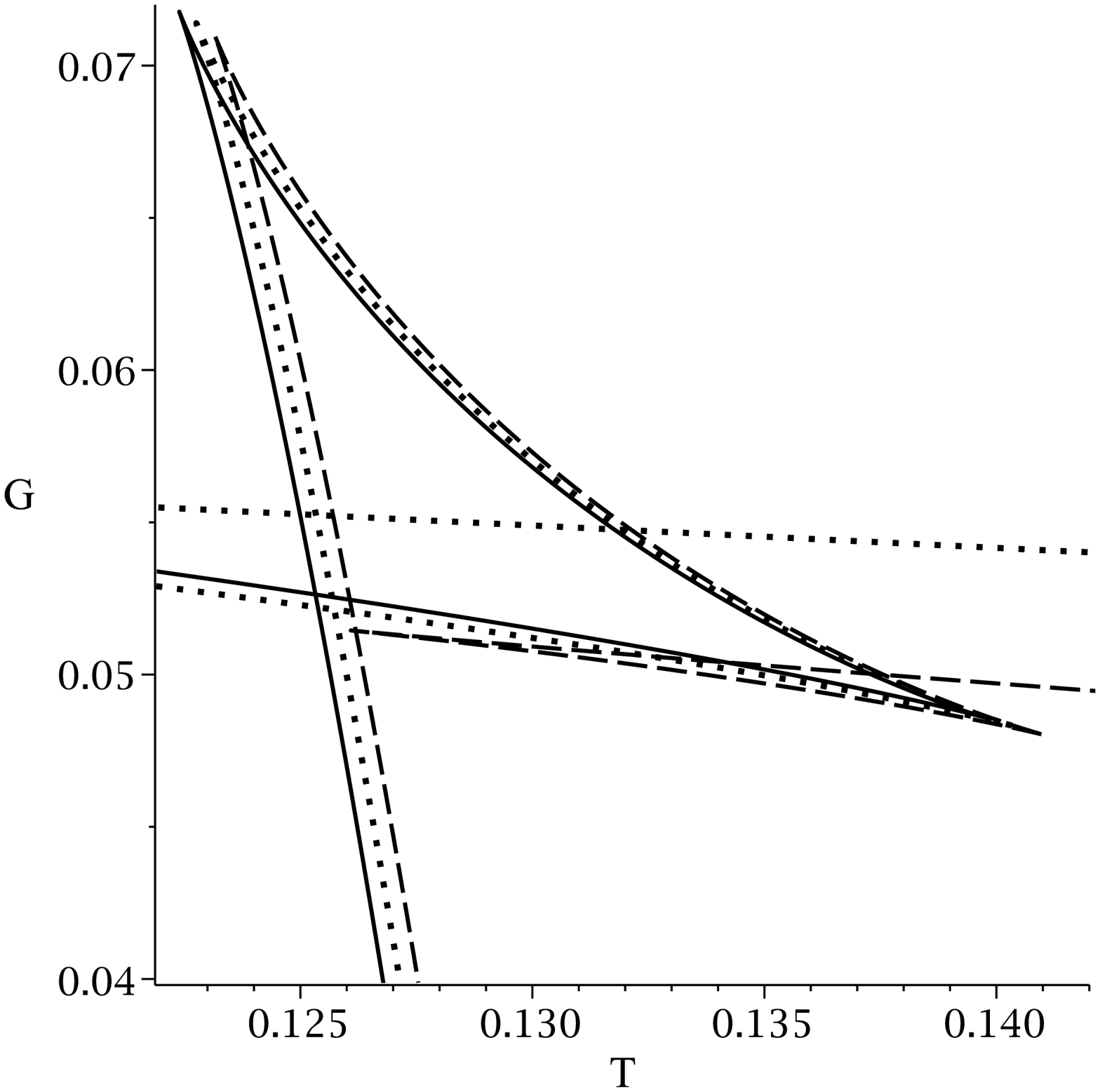}%
\end{array}
$%
\caption{\textbf{\emph{GB solutions:}} $P-r_{+}$ for $T=T_{c}$ (Left), $%
T-r_{+}$ for $P=P_{c}$ (Middle) and $G-T$ for $P=0.5P_{c}$ (Right) diagrams
for $k=1$, $n=4$, $q=1$, $\protect\alpha ^{\prime }=10^{-4}$, and $\protect%
\beta =0$ (continuous line), $\protect\beta =0.03$ (dotted line) and $%
\protect\beta =0.06$ (dashed line).}
\label{comparGBbeta010507alpha0001n4}
\end{figure}
\begin{figure}[tbp]
$%
\begin{array}{ccc}
\epsfxsize=5cm \epsffile{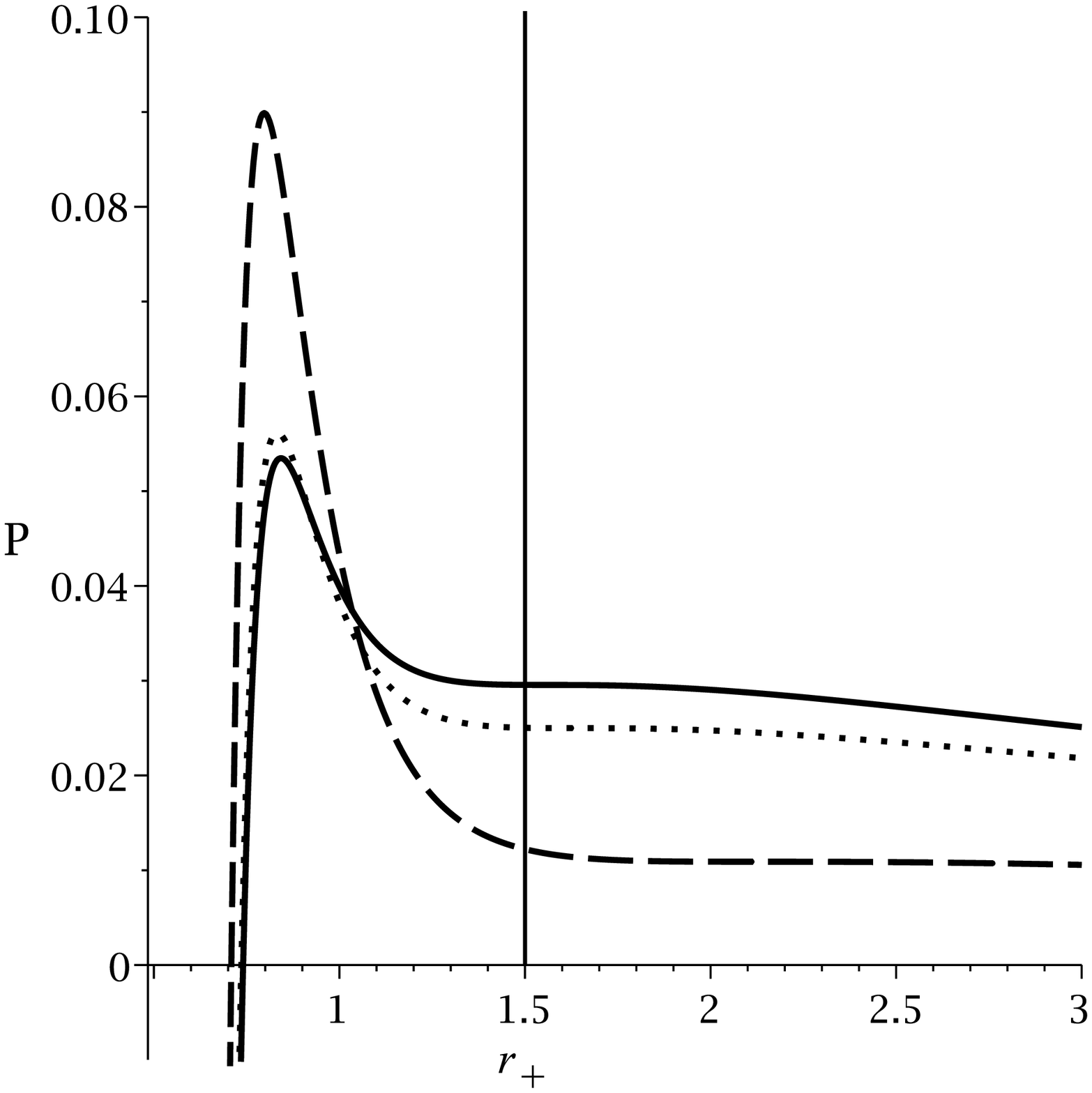} & \epsfxsize=5cm %
\epsffile{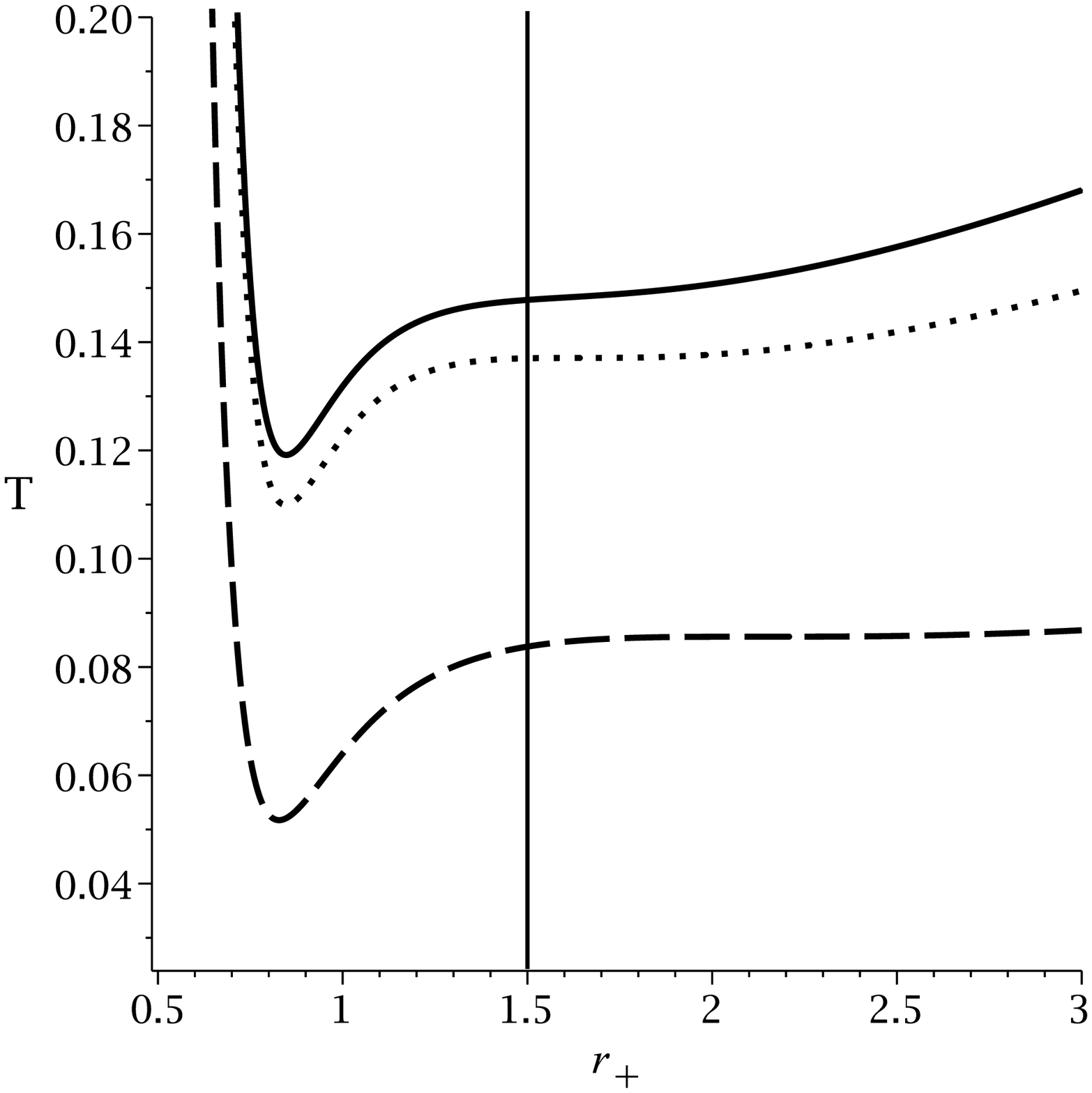} & \epsfxsize=5cm %
\epsffile{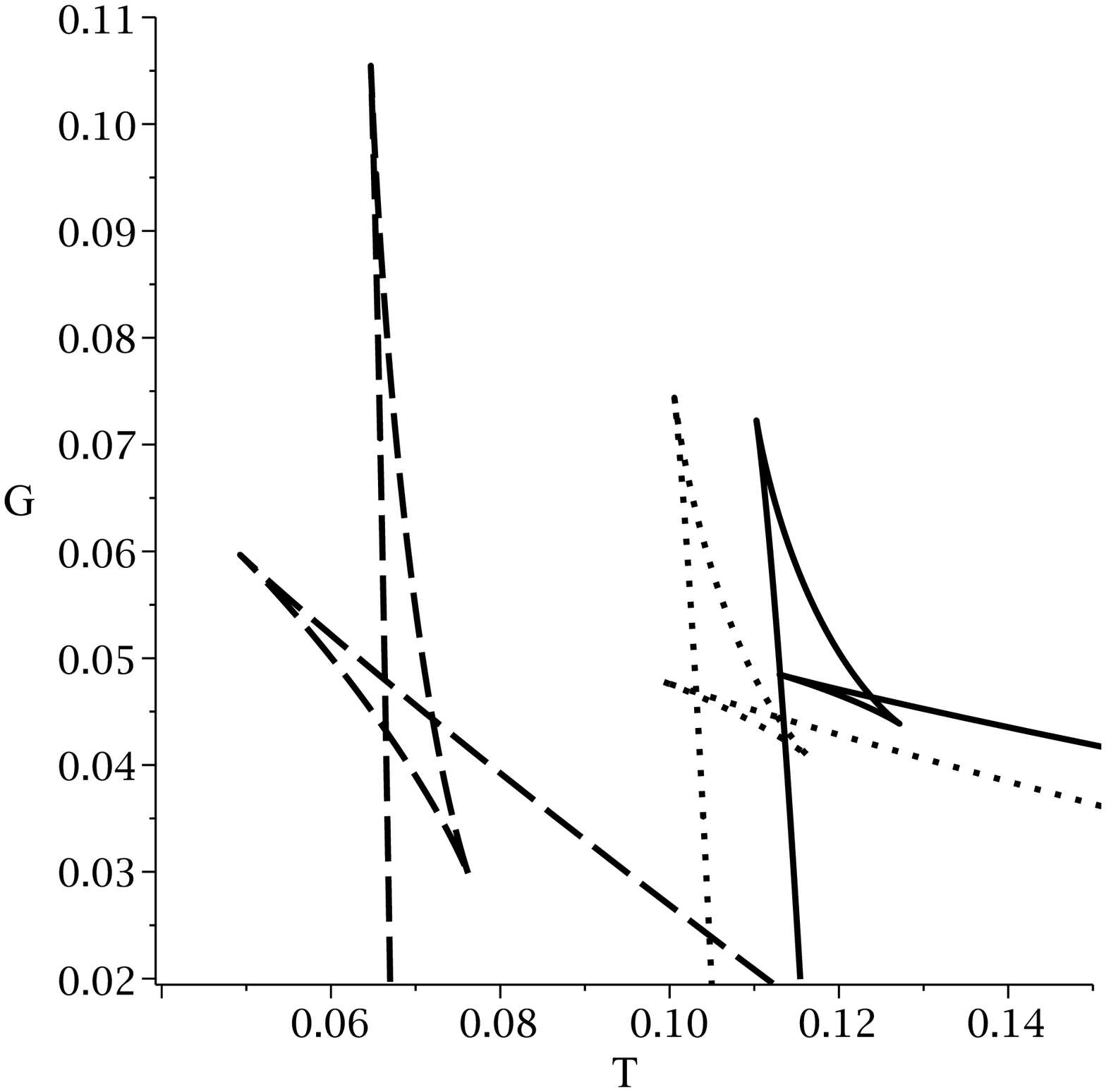}%
\end{array}
$%
\caption{\textbf{\emph{GB solutions:}} $P-r_{+}$ (Left), $T-r_{+}$ (Middle)
and $G-T$ (Right) diagrams for $k=1$, $n=4$, $q=1$ and $\protect\beta =0.07$%
. \newline
$P-r_{+}$ diagram: $T=T_{c}$ and $\protect\alpha ^{\prime }=0.05$
(continuous line), $\protect\alpha ^{\prime }=0.1$ (dotted line) and $%
\protect\alpha ^{\prime }=0.5$ (dashed line). \newline
$T-r_{+}$ diagram: $P=P_{c}$ and $\protect\alpha ^{\prime }=0.05$
(continuous line), $\protect\alpha ^{\prime }=0.1$ (dotted line) and $%
\protect\alpha ^{\prime }=0.5$ (dashed line). \newline
$G-T$ diagram: $P=0.5P_{c}$ and $\protect\alpha ^{\prime }=0.05$ (continuous
line), $\protect\alpha ^{\prime }=0.1$ (dotted line) and $\protect\alpha %
^{\prime }=0.5$ (dashed line).}
\label{comparGBalpha010515beta07n4}
\end{figure}
\begin{figure}[tbp]
$%
\begin{array}{ccc}
\epsfxsize=5cm \epsffile{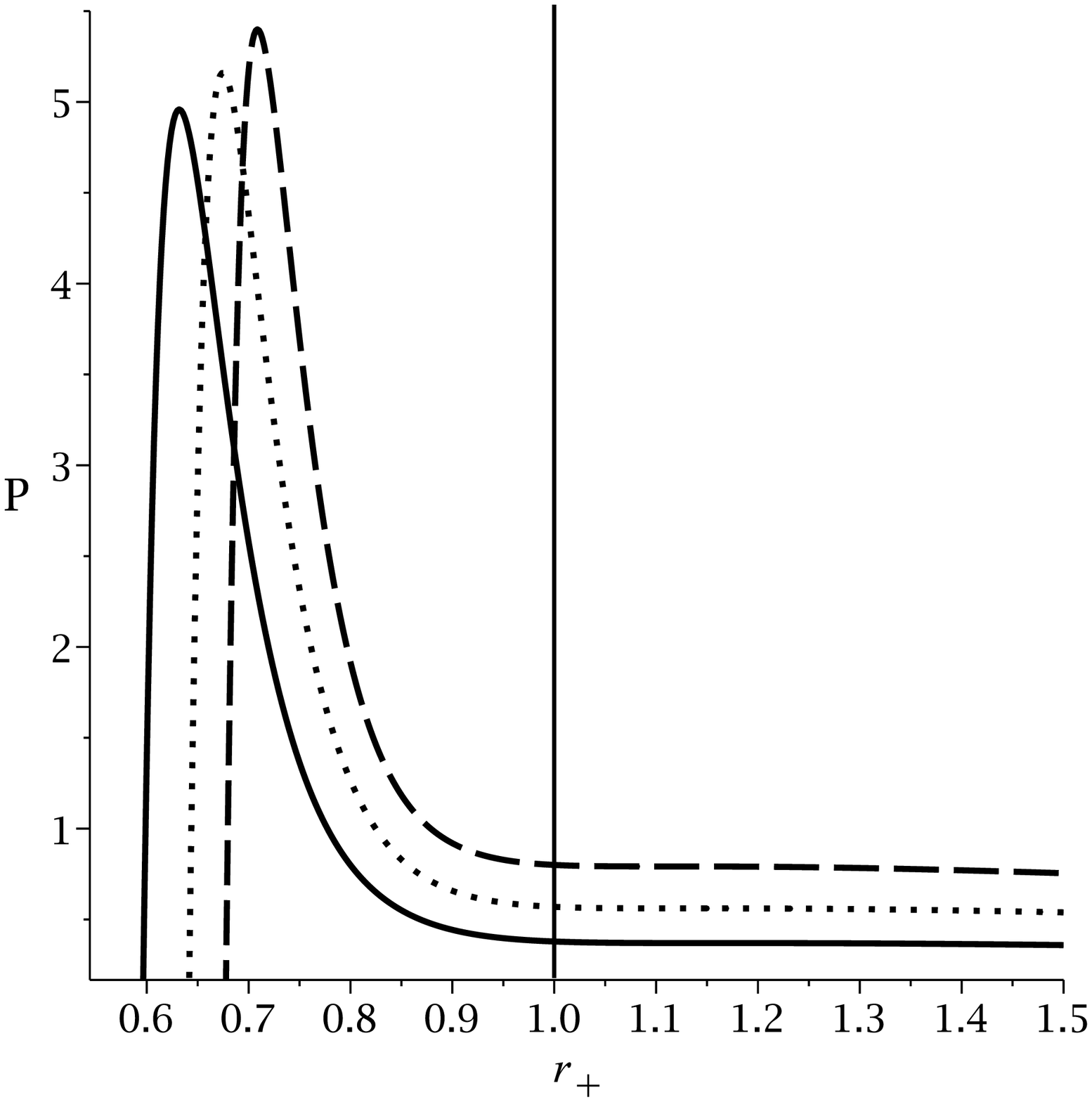} & \epsfxsize=5cm \epsffile{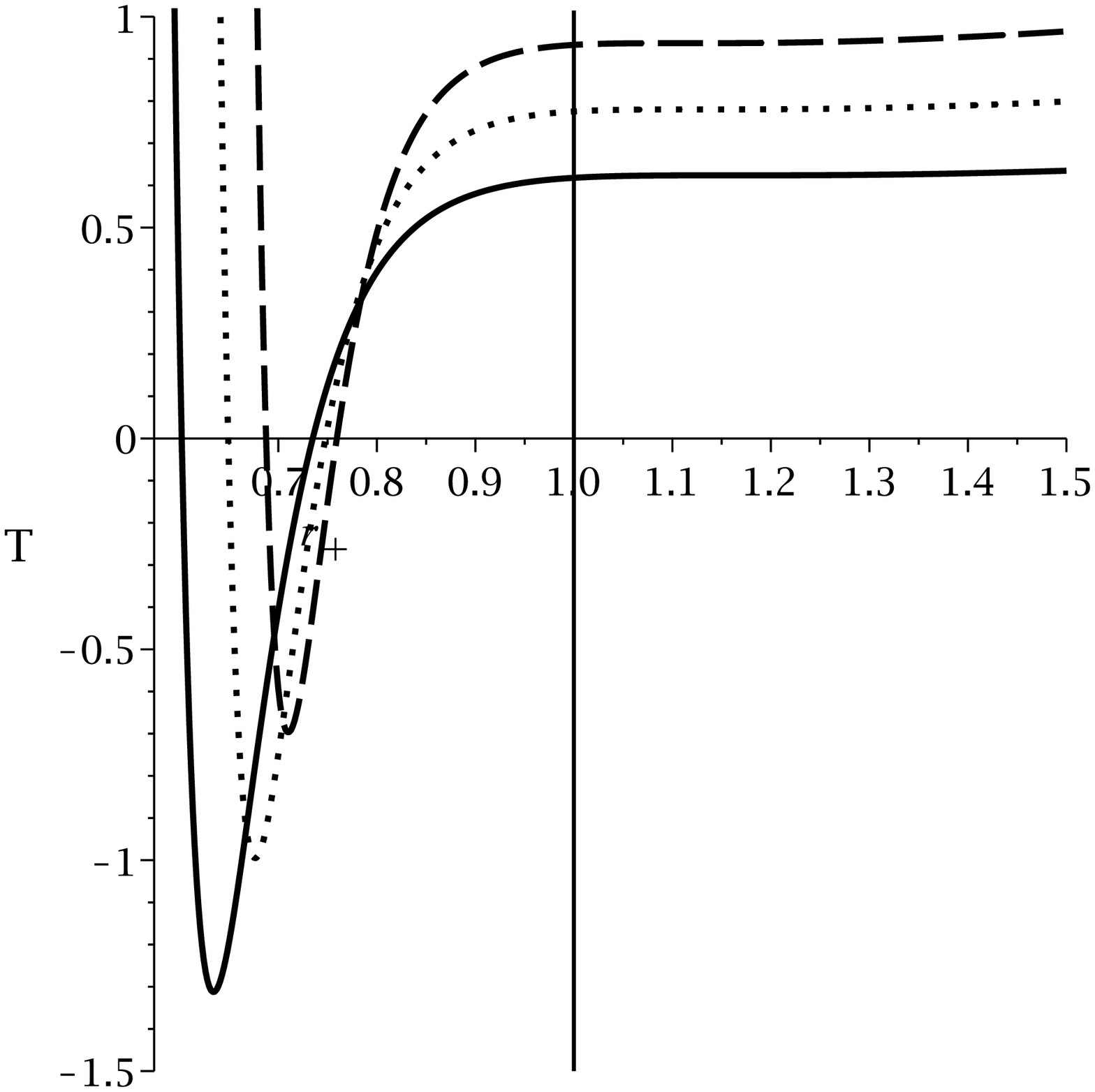}
& \epsfxsize=5cm \epsffile{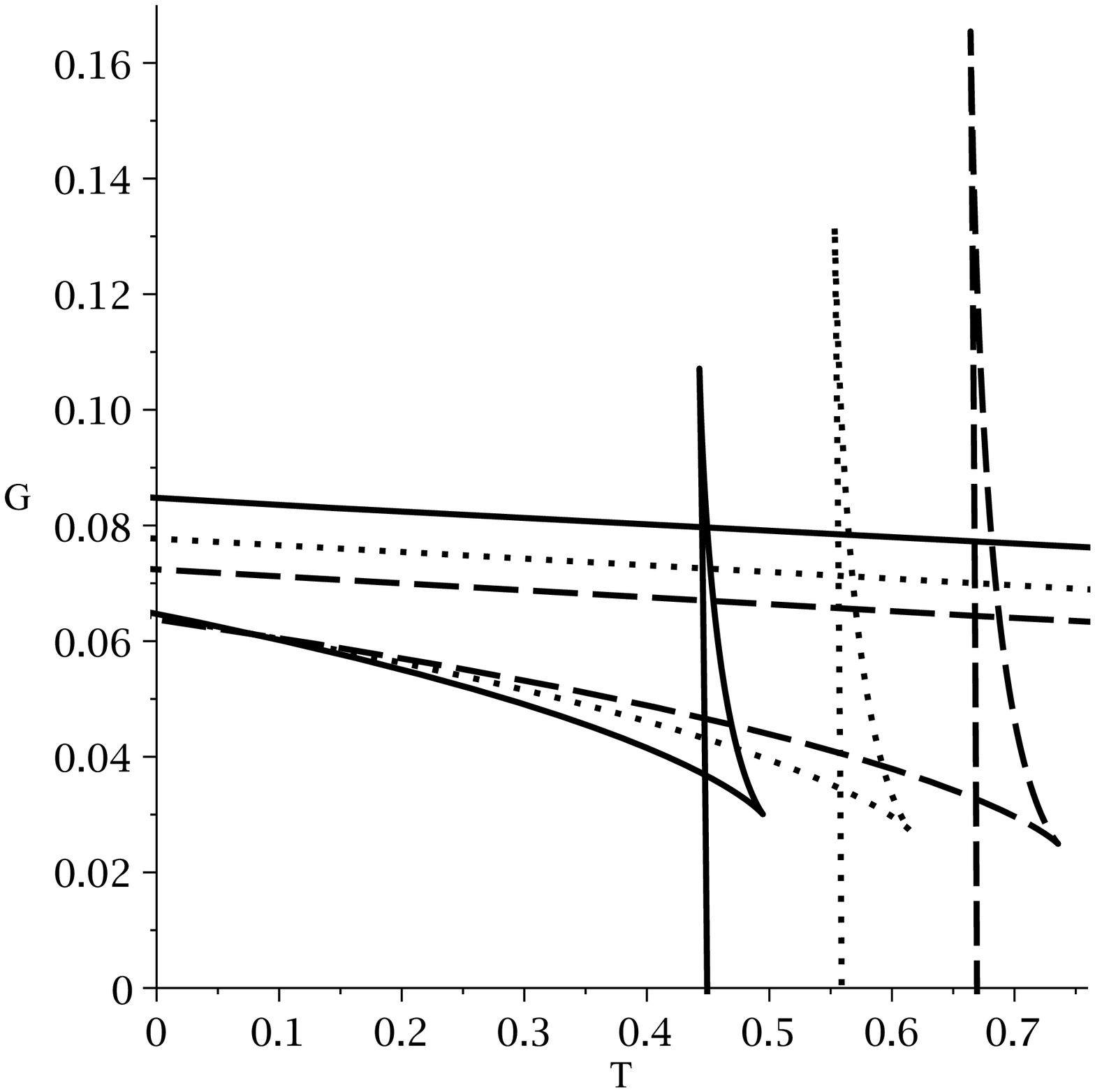}%
\end{array}
$%
\caption{\textbf{\emph{Einstein solutions:}} $P-r_{+}$ for $T=T_{c}$ (Left),
$T-r_{+}$ for $P=P_{c}$ (Middle) and $G-T$ for $P=0.5P_{c}$ (Right) diagrams
for $k=1$, $q=1$, $\protect\beta =0.001$, $n=7$ (continuous line), $n=8$
(dotted line) and $n=9$ (dashed line).}
\label{En789}
\end{figure}
\begin{figure}[tbp]
$%
\begin{array}{ccc}
\epsfxsize=5cm \epsffile{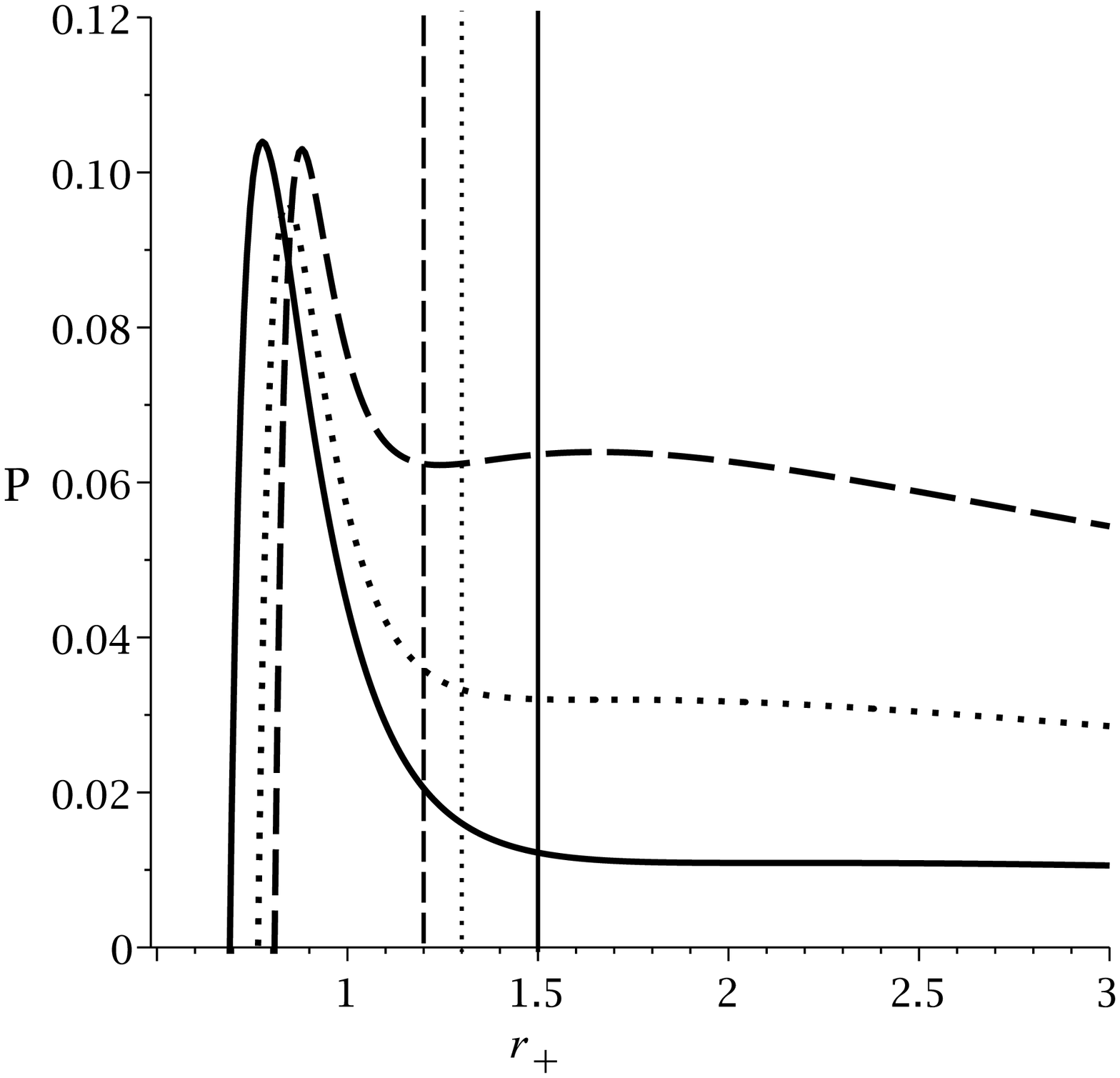} & \epsfxsize=5cm %
\epsffile{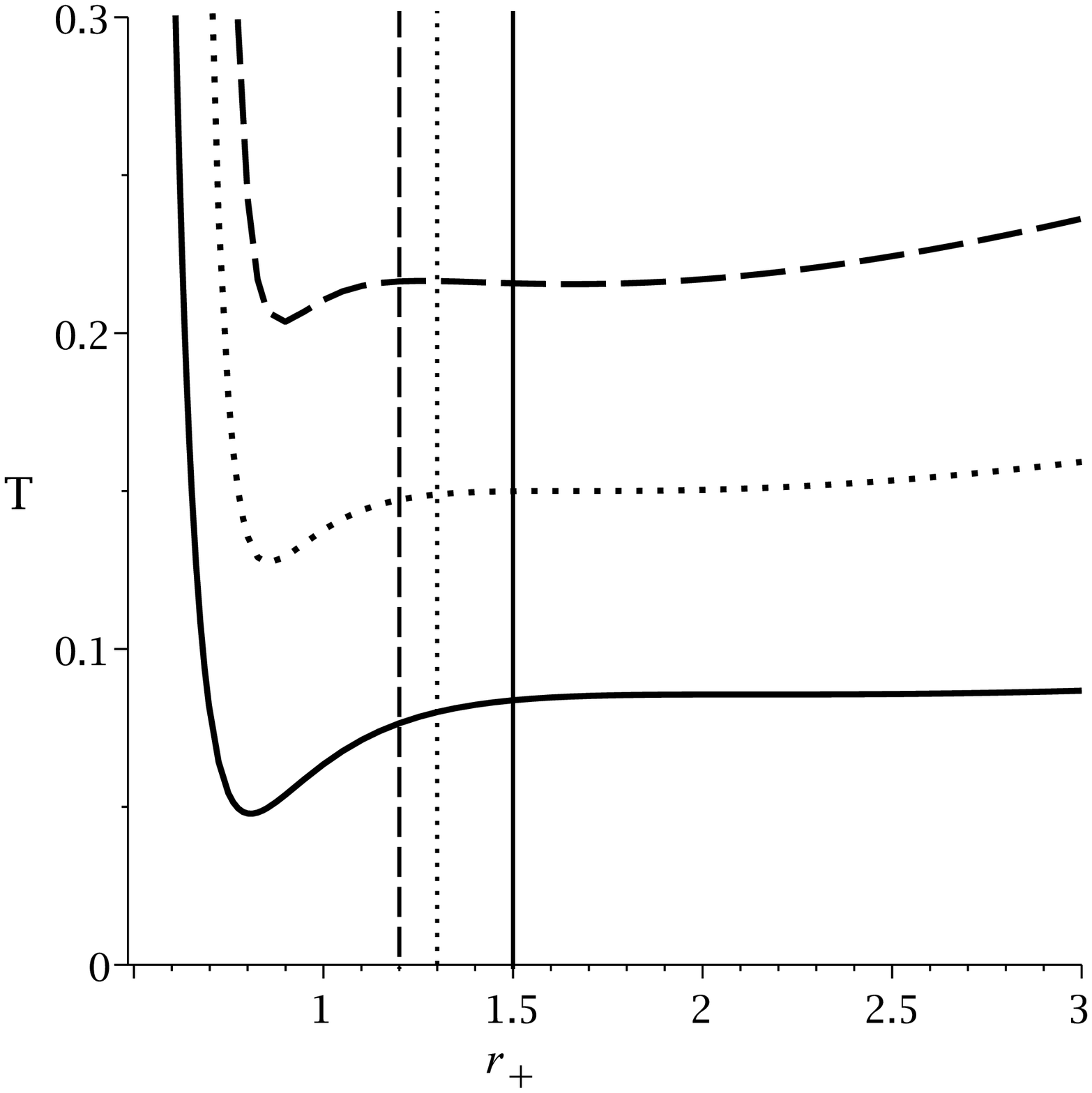} & \epsfxsize=5cm \epsffile{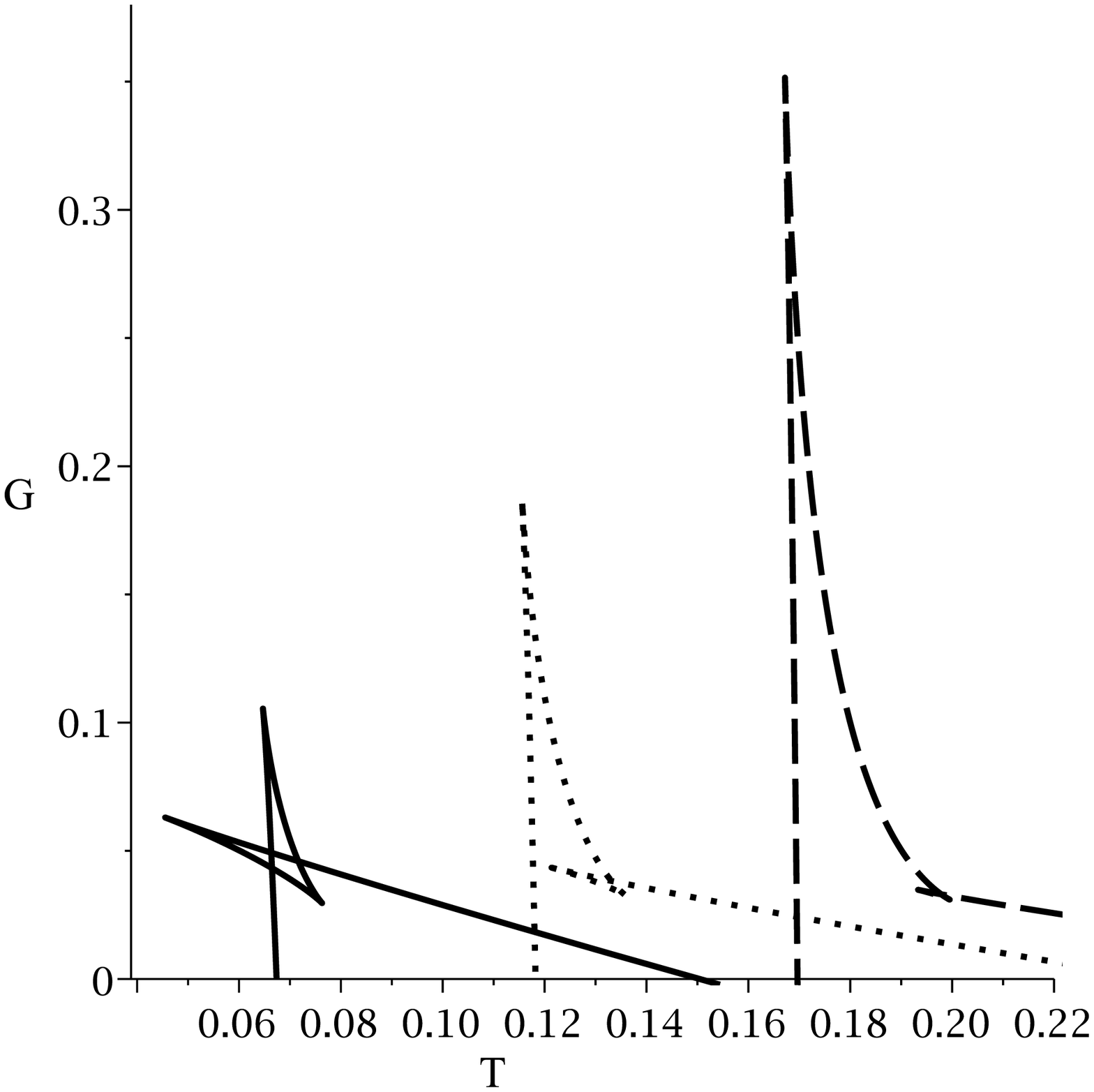}%
\end{array}
$%
\caption{\textbf{\emph{GB solutions:}} $P-r_{+}$ for $T=T_{c}$ (Left), $%
T-r_{+}$ for $P=P_{c}$ (Middle) and $G-T$ for $P=0.5P_{c}$ (Right) diagrams
for $k=1$, $q=1$, $\protect\beta =0.06$, $\protect\alpha ^{\prime }=0.5$, $%
n=4$ (continuous line), $n=5$ (dotted line) and $n=6$ (dashed line).}
\label{GBn456}
\end{figure}

\begin{figure}[tbp]
$%
\begin{array}{ccc}
\epsfxsize=7cm \epsffile{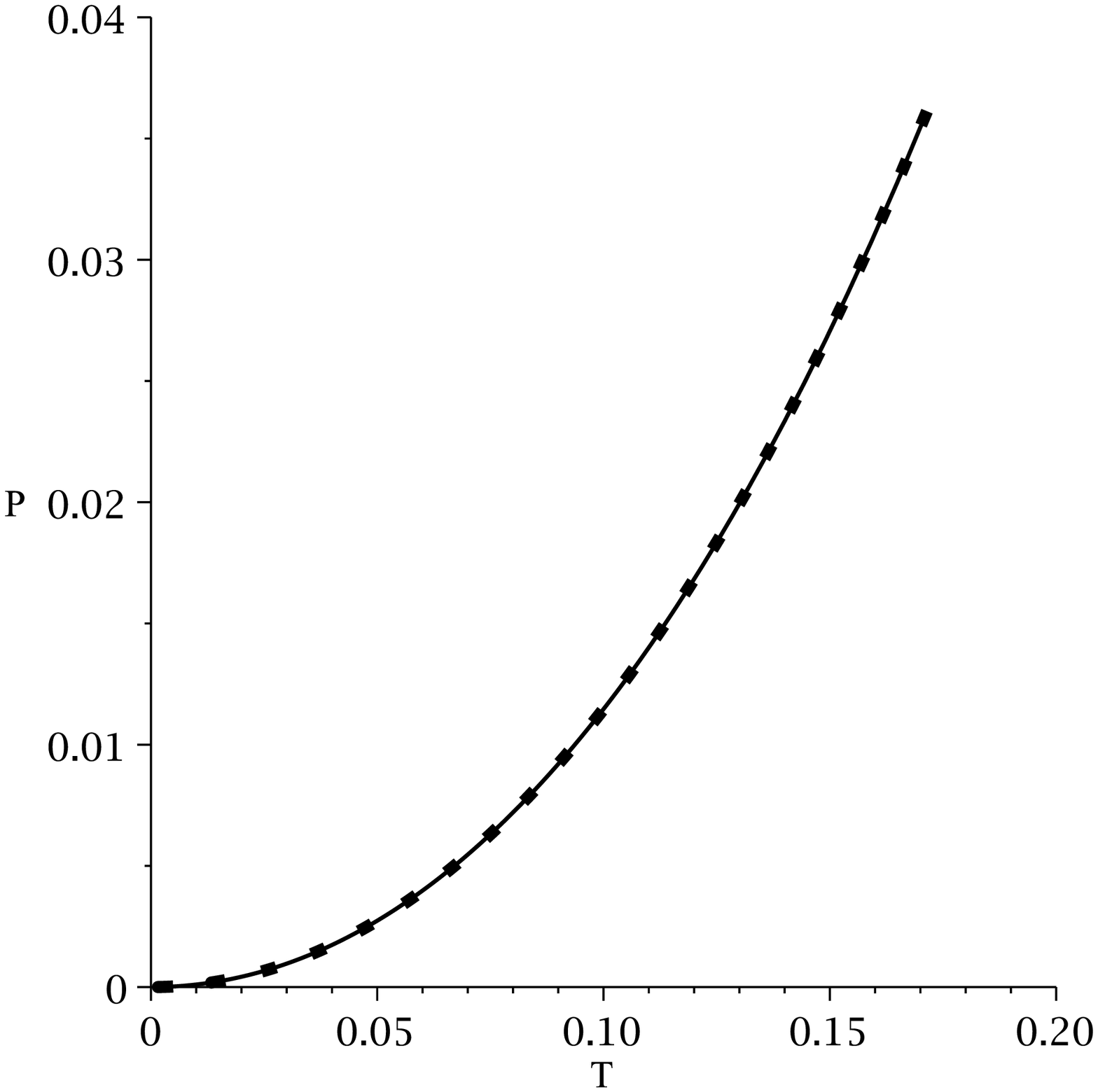} & \epsfxsize=7cm %
\epsffile{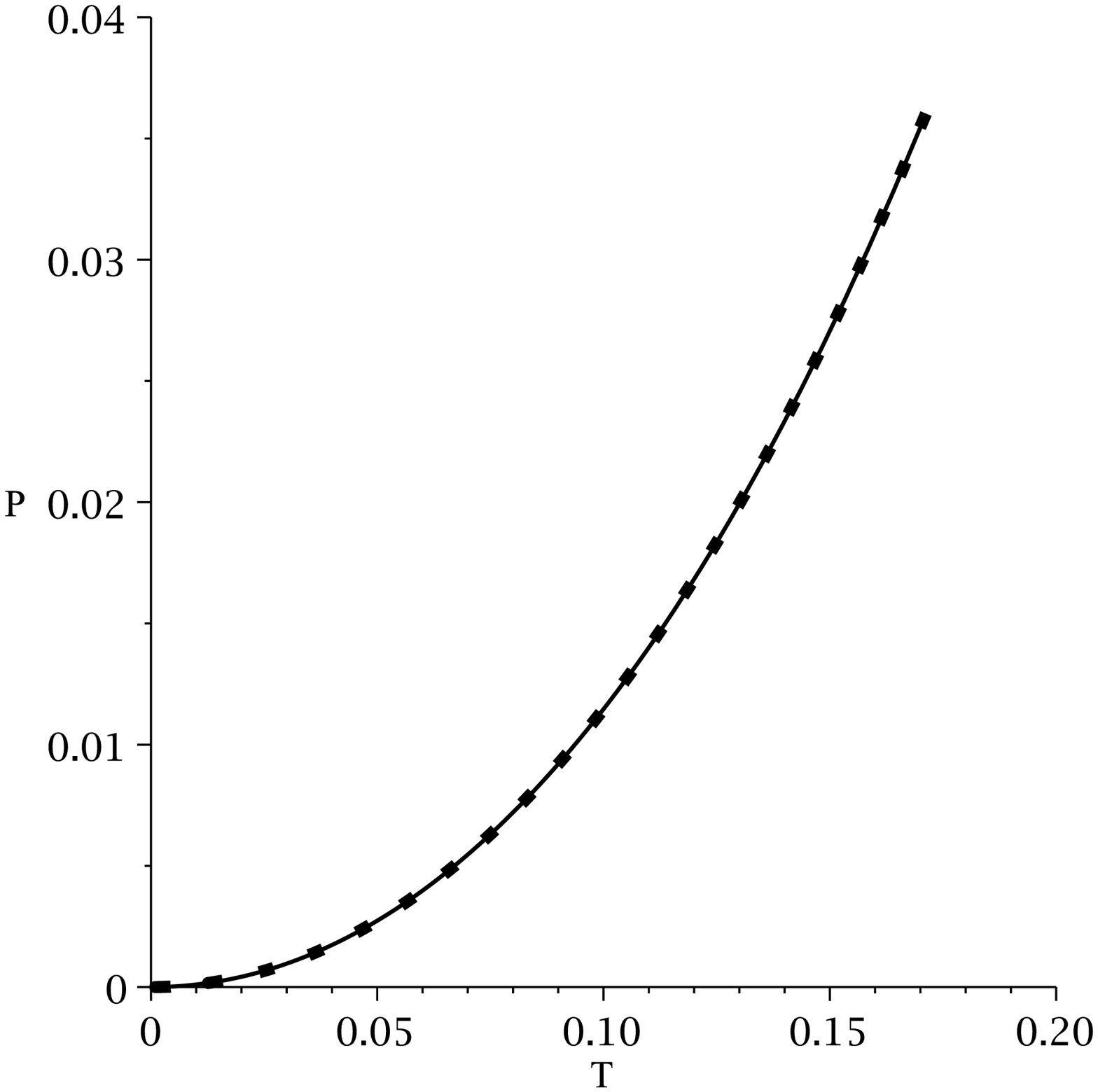} &
\end{array}
$%
\caption{\textbf{\emph{Coexistence line for Einstein (Left) and GB (Right)
gravities:}} $P-T$ for $k=1$, $n=4$ and $q=1$.\newline
Left diagram: $\protect\beta =0.045$ (continuous line) and $\protect\beta %
=0.07$ (dotted line). \newline
Right diagram: $\protect\alpha^{\prime }=10^{-4}$ and $\protect\beta =0$
(continuous line) and $\protect\beta =0.06$ (dotted line).}
\label{PTbeta}
\end{figure}
\begin{figure}[tbp]
$%
\begin{array}{ccc}
\epsfxsize=7cm \epsffile{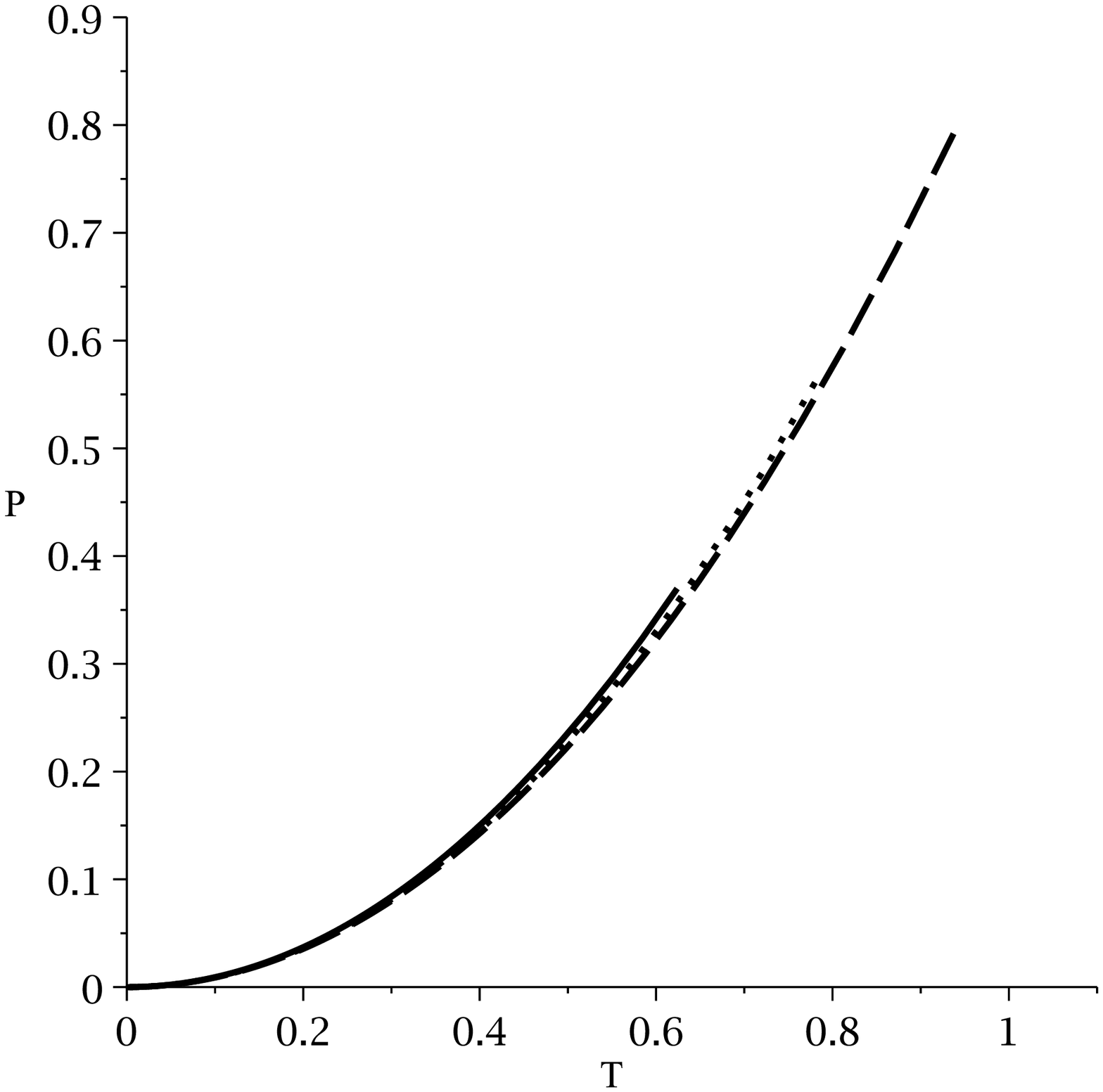} & \epsfxsize=7cm %
\epsffile{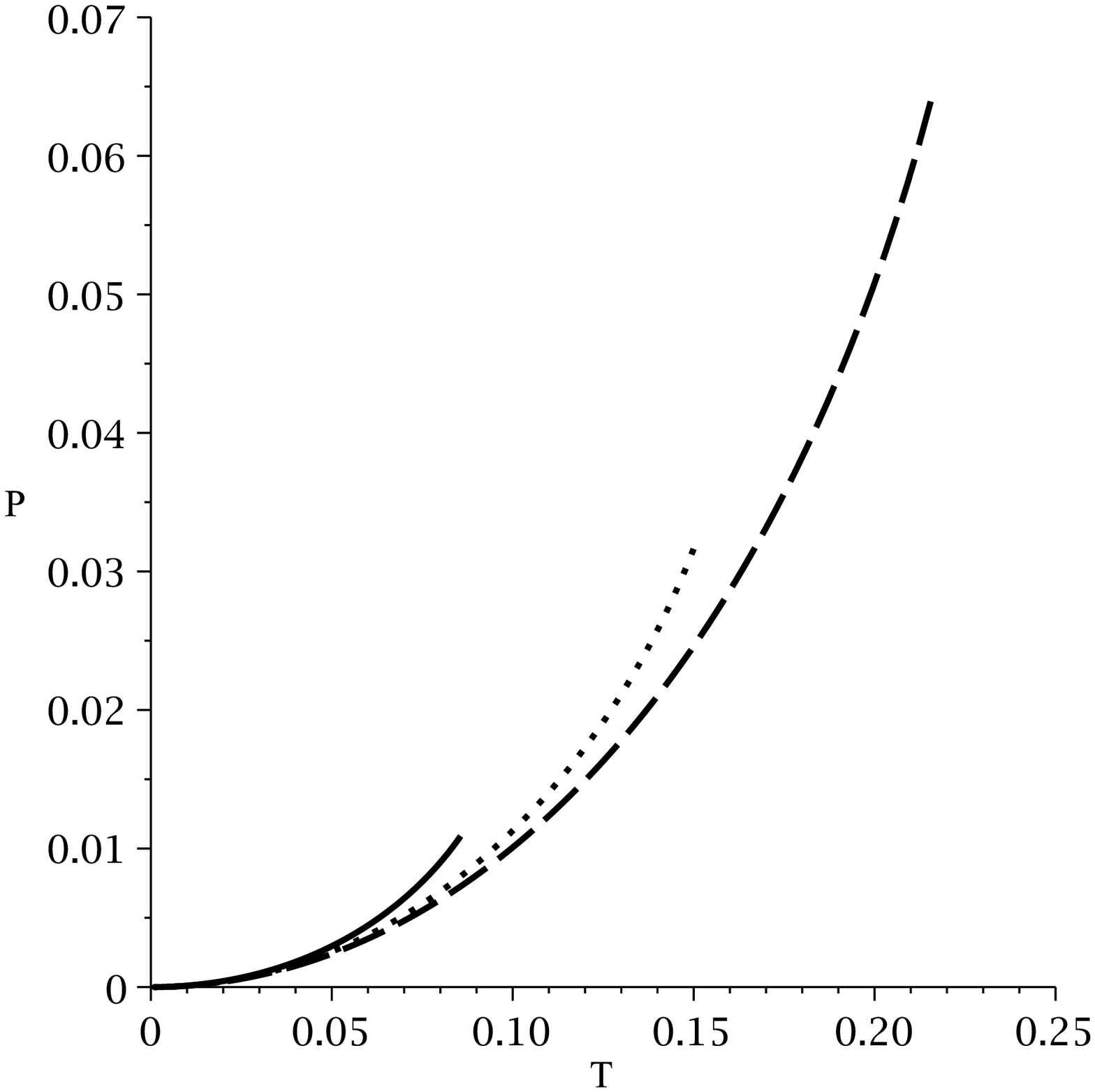} &
\end{array}
$%
\caption{\textbf{\emph{Coexistence line for Einstein (Left) and GB (Right)
gravities:}} $P-T$ for $k=1$, $n=4$ and $q=1$.\newline
Left diagram: $\protect\beta =0.001$ and $n=7$ (continuous line), $n=8$
(dotted line) and $n=9$ (dashed line). \newline
Right diagram: $\protect\beta =0.06$, $\protect\alpha ^{\prime }=0.5$ and $%
n=4$ (continuous line), $n=5$ (dotted line) and $n=6$ (dashed line).}
\label{PTn}
\end{figure}
\begin{figure}[tbp]
$%
\begin{array}{ccc}
\epsfxsize=7cm \epsffile{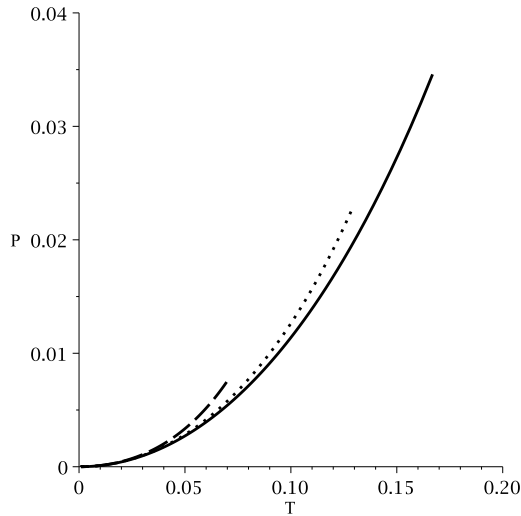} &  &
\end{array}
$%
\caption{\textbf{\emph{Coexistence line for GB gravity:}} $P-T$ for $k=1$, $%
n=4$, $q=1$, $\protect\beta =0.07$ and $\protect\alpha ^{\prime }=0.05$
(continuous line), $\protect\alpha ^{\prime }=0.1$ (dotted line) and $%
\protect\alpha ^{\prime }=0.5$ (dashed line).}
\label{PTalpha}
\end{figure}

\begin{center}
\begin{tabular}{|c|c|c|}
\hline
\begin{tabular}{ccccc}
\hline\hline
$\beta $ & $r_{c}$ & $T_{c}$ & $P_{c}$ & $\frac{P_{c}v_{c}}{T_{c}}$ \\
\hline\hline
$0$ & $2.44948$ & $0.04330$ & $0.00331$ & $0.18750$ \\ \hline
$0.00100$ & $%
\begin{array}{c}
0.31209 \\
2.44917%
\end{array}%
$ & $0.04331$ & $0.00332$ & $0.18748$ \\ \hline
$0.05000$ & $%
\begin{array}{c}
0.85371 \\
2.43312%
\end{array}%
$ & $0.04343$ & $0.00333$ & $0.18569$ \\ \hline
$0.10000$ & $%
\begin{array}{c}
1.03212 \\
2.41568%
\end{array}%
$ & $0.04356$ & $0.00336$ & $0.18635$ \\ \hline
$0.15000$ & $%
\begin{array}{c}
1.15889 \\
2.39698%
\end{array}%
$ & $0.04370$ & $0.00338$ & $0.18569$ \\ \hline
\end{tabular}
& \;\;\;\;\;\;\;\;\; &
\begin{tabular}{ccccc}
\hline\hline
$\beta $ & $r_{c}$ & $T_{c}$ & $P_{c}$ & $\frac{P_{c}v_{c}}{T_{c}}$ \\
\hline\hline
$0$ & $1.49534$ & $0.17029$ & $0.03558$ & $0.31250$ \\ \hline
$0.00100$ & $%
\begin{array}{c}
0.45502 \\
1.49505%
\end{array}%
$ & $0.17030$ & $0.03559$ & $0.31247$ \\ \hline
$0.05000$ & $%
\begin{array}{c}
0.89202 \\
1.47940%
\end{array}%
$ & $0.17110$ & $0.03596$ & $0.31093$ \\ \hline
$0.10000$ & $%
\begin{array}{c}
1.01944 \\
1.46018%
\end{array}%
$ & $0.17201$ & $0.03639$ & $0.30890$ \\ \hline
$0.15000$ & $%
\begin{array}{c}
1.11349 \\
1.43543%
\end{array}%
$ & $0.17310$ & $0.03690$ & $0.30603$ \\ \hline
\end{tabular}
\\ \hline
\end{tabular}
\\[0pt]
Table $1$ (left): Einstein gravity for $q=1$ and $n=3$. Table $2$ (right):
Einstein gravity for $q=1$ and $n=4$.
\end{center}


\begin{center}
\begin{tabular}{|c|c|c|}
\hline
\begin{tabular}{ccccc}
\hline\hline
$\beta $ & $r_{c}$ & $T_{c}$ & $P_{c}$ & $\frac{P_{c}v_{c}}{T_{c}}$ \\
\hline\hline
$0$ & $1.29271$ & $0.31658$ & $0.10714$ & $0.43750$ \\ \hline
$0.00100$ & $%
\begin{array}{c}
0.55202 \\
1.29247%
\end{array}%
$ & $0.31661$ & $0.10716$ & $0.43746$ \\ \hline
$0.05000$ & $%
\begin{array}{c}
0.91471 \\
1.27952%
\end{array}%
$ & $0.31804$ & $0.10820$ & $0.43532$ \\ \hline
$0.10000$ & $%
\begin{array}{c}
1.01396 \\
1.26233%
\end{array}%
$ & $0.31978$ & $0.10949$ & $0.43221$ \\ \hline
$0.15000$ & $%
\begin{array}{c}
1.09170 \\
1.23597%
\end{array}%
$ & $0.32202$ & $0.11117$ & $0.42668$ \\ \hline
\end{tabular}
& \;\;\;\;\;\; &
\begin{tabular}{ccccc}
\hline\hline
$\beta $ & $r_{c}$ & $T_{c}$ & $P_{c}$ & $\frac{P_{c}v_{c}}{T_{c}}$ \\
\hline\hline
$0$ & $1.49550$ & $0.17024$ & $0.03557$ & $0.31248$ \\ \hline
$0.01000$ & $%
\begin{array}{c}
0.67158 \\
1.49252%
\end{array}%
$ & $0.17040$ & $0.03564$ & $0.31219$ \\ \hline
$0.03000$ & $%
\begin{array}{c}
0.81326 \\
1.48626%
\end{array}%
$ & $0.17071$ & $0.03579$ & $0.31158$ \\ \hline
$0.05000$ & $%
\begin{array}{c}
0.89202 \\
1.47957%
\end{array}%
$ & $0.17105$ & $0.03594$ & $0.31091$ \\ \hline
$0.07000$ & $%
\begin{array}{c}
0.95018 \\
1.47237%
\end{array}%
$ & $0.17140$ & $0.03610$ & $0.31017$ \\ \hline
\end{tabular}
\\ \hline
\end{tabular}
\\[0pt]
Table $3$ (left): Einstein gravity for $q=1$ and $n=5$. Table $4$ (right):
GB gravity for $q=1$, $\alpha =10^{-4}$ and $n=4$.
\end{center}


\begin{center}
\begin{tabular}{|c|c|c|}
\hline
\begin{tabular}{ccccc}
\hline\hline
$\alpha $ & $r_{c}$ & $T_{c}$ & $P_{c}$ & $\frac{P_{c}v_{c}}{T_{c}}$ \\
\hline\hline
$0.01000$ & $%
\begin{array}{c}
0.94914 \\
1.48981%
\end{array}%
$ & $0.16688$ & $0.03458$ & $0.30873$ \\ \hline
$0.05000$ & $%
\begin{array}{c}
0.94608 \\
1.55579%
\end{array}%
$ & $0.15157$ & $0.02955$ & $0.30333$ \\ \hline
$0.10000$ & $%
\begin{array}{c}
0.94374 \\
1.63116%
\end{array}%
$ & $0.13707$ & $0.02499$ & $0.29747$ \\ \hline
$0.50000$ & $%
\begin{array}{c}
0.93836 \\
2.12855%
\end{array}%
$ & $0.08560$ & $0.01089$ & $0.27103$ \\ \hline
$1.00000$ & $%
\begin{array}{c}
0.89005 \\
2.65829%
\end{array}%
$ & $0.06346$ & $0.00619$ & $0.25942$ \\ \hline
\end{tabular}
& \;\;\;\;\;\; &
\begin{tabular}{ccccc}
\hline\hline
$\beta $ & $r_{c}$ & $T_{c}$ & $P_{c}$ & $\frac{P_{c}v_{c}}{T_{c}}$ \\
\hline\hline
$0$ & $1.49550$ & $0.17024$ & $0.03557$ & $0.31248$ \\ \hline
$0.01000$ & $%
\begin{array}{c}
0.67158 \\
1.49252%
\end{array}%
$ & $0.17040$ & $0.03564$ & $0.31219$ \\ \hline
$0.03000$ & $%
\begin{array}{c}
0.81326 \\
1.48626%
\end{array}%
$ & $0.17071$ & $0.03579$ & $0.31158$ \\ \hline
$0.05000$ & $%
\begin{array}{c}
0.89202 \\
1.47957%
\end{array}%
$ & $0.17105$ & $0.03594$ & $0.31091$ \\ \hline
$0.07000$ & $%
\begin{array}{c}
0.95018 \\
1.47237%
\end{array}%
$ & $0.17140$ & $0.03610$ & $0.31017$ \\ \hline
\end{tabular}
\\ \hline
\end{tabular}
\\[0pt]
Table $5$ (left): GB gravity for $q=1$, $\beta =0.07$ and $n=4$. Table $6$
(right): GB gravity for $q=1$, $\alpha =10^{-4}$ and $n=5$.
\end{center}


\begin{center}
\begin{tabular}{ccccc}
\hline\hline
$\alpha $ & $r_{c}$ & $T_{c}$ & $P_{c}$ & $\frac{P_{c}v_{c}}{T_{c}}$ \\
\hline\hline
$0.01000$ & $%
\begin{array}{c}
0.95884 \\
1.28530%
\end{array}%
$ & $0.30889$ & $0.10367$ & $0.43139$ \\ \hline
$0.05000$ & $%
\begin{array}{c}
0.95598 \\
1.32895%
\end{array}%
$ & $0.27681$ & $0.08766$ & $0.42089$ \\ \hline
$0.10000$ & $%
\begin{array}{c}
0.95414 \\
1.37662%
\end{array}%
$ & $0.24745$ & $0.07358$ & $0.40937$ \\ \hline
$0.50000$ & $%
\begin{array}{c}
0.95469 \\
1.65361%
\end{array}%
$ & $0.15000$ & $0.03196$ & $0.35241$ \\ \hline
$1.00000$ & $%
\begin{array}{c}
0.96227 \\
1.91323%
\end{array}%
$ & $0.11049$ & $0.01837$ & $0.31815$ \\ \hline
\end{tabular}
\\[0pt]
\vspace{0.1cm} Table ($7$): GB gravity for $q=1$, $\beta =0.07$ and $n=5$.
\end{center}


It is a well-known fact that inclusion of higher order polynomial
terms (e.g. $r_{+}^d$, $d\in\mathbb{N}$) in the Van der Waals
equation of state changes the location of critical point in
thermodynamic space ($P, V, T$), but does not change the
universality class and therefore leads to the same exponents as
the Van der Waals fluid (see \cite{Vahidinia,GBMaxwell} for more
details). This is different from the case of pure Lovelock gravity
solutions \cite{DolanMann}, which the equation of state is no
longer a polynomial form.

\section{Discussion on the results of diagrams}

In order to study the behavior of phase transition for these black
holes in more details, we have plotted $P-r_{+}$, $T-r_{+}$ and
$G-T$ diagrams. For having better insight regarding the effects of
correction on critical behavior of the system, we have also
plotted some diagrams for the case of $\beta =0$ which is the
Maxwell theory.

like Van der Waals system, the usual characteristic swallow tail
is seen in black holes only in the presence of linear Maxwell
field (right panels of Figs. \ref{EMbeta0dim} and
\ref{GBbeta0dim}). In other words, considering Maxwell
electromagnetic field leads to a Van der Waals like behavior and
usual phase transition. The existence and usual behavior of phase
transition are also evident from studying $P-r_{+}$ and $T-r_{+}$
diagrams. In case of the absence of nonlinearity parameter, for
$P-r_{+}$ (left panels of Figs. \ref{EMbeta0dim} and
\ref{GBbeta0dim}), pressure is a decreasing function of horizon
radius but for a range of horizon radius, it is an increasing
function of it which is not a physical behavior. Before and after
this region there are two values of horizon with same pressure.
The phase transition takes place between these two points which in
case of black holes, it is small/large black hole phase
transition. It is crucial to mention that, this region is only
seen for case of $T\leq T_{c}$ whereas for $T>T_{c}$ pressure is
only a decreasing function of horizon radius. On the other hand,
for $\beta =0$ in case of $T-r_{+}$ diagrams (middle panels of
Figs. \ref{EMbeta0dim} and \ref{GBbeta0dim}), if $P=P_{c}$ the
temperature is an increasing function of horizon radius and for a
region of horizon radius, temperature is fixed. This place is the
region where phase transition takes place and known as subcritical
isobar.

As for the effects of dimensions on the critical behavior of the system in
the presence of linear Maxwell field Figs. \ref{EMbeta0dim} and \ref%
{GBbeta0dim} are plotted. As one can see, the swallow tail (right
panels of Figs. \ref{EMbeta0dim} and \ref{GBbeta0dim}), critical
pressure (left panels of Figs. \ref{EMbeta0dim} and
\ref{GBbeta0dim}) and temperature (middle panels of Figs.
\ref{EMbeta0dim} and \ref{GBbeta0dim}) are increasing functions of
dimension while critical horizon radius and subcritical isobars
are decreasing functions of it.

In addition, for the case of absence of nonlinearity parameter, we compare
critical behavior of these two gravities with each other (Fig. \ref%
{comparEandGBbeta0n4}). As one can see considering GB gravity leads to
increasing the size of swallow tail (right panel of Fig. \ref%
{comparEandGBbeta0n4}). The place of swallow tail is also shifted to lower
values of temperature. On the other hand, pressure (left panel of Fig. \ref%
{comparEandGBbeta0n4}) and temperature (middle panel of Fig. \ref%
{comparEandGBbeta0n4}) of the critical point are greater in Einstein gravity
whereas the length of subcritical isobar and the critical horizon radius are
greater in GB gravity. Also, the needed energy for the phase transition in
Einstein gravity is more than GB case.

Next, we are considering $\beta \neq 0$ and plot one set of graphs for
Einstein gravity for variation of nonlinearity parameter (Figs. \ref%
{Ebeta045n4}-\ref{comparEbeta04507n4}) and Figs. \ref%
{comparGBbeta010507alpha0001n4} and \ref{comparGBalpha010515beta07n4}\ for
variation of nonlinearity and GB parameters for GB gravity. Also, we plot
Figs. \ref{En789} and \ref{GBn456} to investigate the effects of dimensions
on the solutions of Einstein and GB gravities, respectively. It is notable
that in these figures physical solutions exist only after the vertical line.
This means that before these vertical lines the second term in Eq. (\ref{Ftr}%
) is not small enough (with respect to Maxwell term). The vertical lines in $%
P-r_{+}$ ($T-r_{+}$) and $G-T$ diagrams are interpreted as the minimum value
of the authorized horizon radius and temperature, respectively. In comparing
figures there are more than one vertical line. Because of overlapping of
these vertical lines with the vertical line of $G-T$ diagrams, these lines
are not presented.

In case of variation of nonlinearity parameter same behavior is
observed for Einstein and GB gravities. For this case the
following results are obtained. Interestingly, contrary to Maxwell
theory, in this case ($\beta \neq 0$), the Van der Waals like
behavior is not preserved. The plotted graphs for Gibbs free
energy versus temperature show the existence of a phase
transition, a turning point. In other words, the characteristic
swallow tail of phase transition in this nonlinear theory is
modified and its shape is different from the usual thermodynamical
systems. For small values of nonlinearity parameter (right panel
of Fig. \ref{Ebeta045n4}), the usual swallow tail is observed with
a turning point and a minimum temperature (vertical line) which
are located before swallow tail. It is evident that the minimum
energy and the distance between turning point and swallow tail,
are decreasing functions of $\beta$ whereas the minimum
temperature is an increasing function(see Figs.
\ref{comparEbeta04507n4} and \ref{comparGBbeta010507alpha0001n4}).
It is worthwhile to mention that minimum temperature and critical
temperature are two different quantities.

As for the $P-r_{+}$ , it is evident that the related graphs are modified
like $G-T$ diagrams. First, pressure is an increasing function of horizon
radius, then after a turning point, it changes into being a decreasing
function of $r_{+}$. In this case, a part of the graphs shows the usual
behavior of phase transition whereas there is another part which is
irregular (left panels of Figs. \ref{Ebeta045n4} and \ref{Ebeta07n4}). In
case of $T=T_{c}$, one can find two horizon radii for critical pressure. In
this case the critical horizon radius is a decreasing function of
nonlinearity parameter whereas the turning point and critical pressure are
increasing functions of it (Tables $1-4$ and $6$).

Moreover, in studying $T-r_{+},$ same abnormal behavior is
observed (middle panels of Figs. \ref{Ebeta045n4} and
\ref{Ebeta07n4}). Usually, we are expecting temperature to be a
decreasing function of horizon radius except in place of phase
transition in which temperature is fixed and horizon radius
increases. This region is know as subcritical isobars. But in this
case, first temperature is a decreasing function of $r_{+}$ then
it becomes an increasing function of it. As one can see in case of
$P=P_{c}$ the subcritical isobar is observed but another value of
horizon radius exists which has the same temperature as
subcritical isobar. This may show that in this place phase
transition takes place. The critical temperature is an increasing
function of nonlinearity parameter.

Comparing the variational effects of GB and nonlinearity parameters, we find
that they have opposite effects (Figs. \ref{comparGBbeta010507alpha0001n4}
and \ref{comparGBalpha010515beta07n4}). Thus, we leave out discussions of GB
parameter for reasons of economy.

It is worthwhile to mention a few characteristic behavior of graphs. As one
can see, in case of small values of nonlinearity parameter, the distance
between the critical point and turning point is large (right panels of Figs. %
\ref{Ebeta045n4} and \ref{Ebeta07n4}). Similarly, in case of $P-r_{+}$ ($%
T-r_{+}$) the region of $r_{+}$ in which pressure being increasing
(decreasing) function of $r_{+}$ is large too. As nonlinearity
parameter increases, the distance between these points decreases
in $G-T$ diagrams, and interestingly, in case of $P-r_{+}$
($T-r_{+}$) the region of r+, distance between the critical
horizon radius, and turning point decrease (Figs.
\ref{comparEbeta04507n4} and \ref{comparGBbeta010507alpha0001n4}).
In the end, we mention that $P_{c}r_{c}/T_{c}$ is a decreasing
function of nonlinearity and GB parameters.

Next, we have considered the effects of dimensions on the critical
behavior. As one can see, the temperature of swallow tail
formation, energy gap between two states and minimum of
temperature (turning point), critical horizon (middle panels of
Figs. \ref{En789} and \ref{GBn456}), pressure (left panels of Figs. \ref%
{En789} and \ref{GBn456}) and $P_{c}r_{c}/T_{c}$ are increasing
functions of dimensions. An abnormal behavior for $4$-dimension is
observed in $G-T$ diagrams for Einstein gravity. This behavior is
due to power of $r_{+}$ in the last term of Gibbs free energy. The
same abnormality could be obtained for the case of GB gravity in
$5$-dimension which is due to structure of Gibbs free energy,
temperature and pressure. In other words, there are terms in these
equations that vanish in case of $n=4$.

Finally, we have plotted the coexistence line in which along this
curve, small and large black holes have alike temperature and
pressure (Figs. \ref{PTbeta}, \ref{PTn}, and \ref{PTalpha}).
Critical points are located at the end of the coexistence line
where above these points the phase transition does not occur. In
addition, Fig. \ref{PTbeta} indicates that nonlinearity parameter
does not significantly affect the coexistence line. For both
Einstein and GB gravities as dimension increases the critical
pressure and critical temperature increase too (Fig. \ref{PTn}).
In case of GB parameter, the critical pressure and critical
temperature are decreasing functions of GB parameter (Fig.
\ref{PTalpha} and table 5).

\section{Phase transition points through heat capacity}

Since we have observed an abnormal behavior in plotted phase
diagrams, it will be worthwhile to test the existence of reentrant
phase transitions through another method. In Ref. \cite{HendiIJMD}
a new method for studying critical behaviors was introduced. The
method is based on obtaining a relation for thermodynamical
pressure by using the denominator of the heat capacity. In other
words, by replacing the cosmological constant with its
corresponding thermodynamical pressure and solving the denominator
of the heat capacity with respect to this pressure a relation is
obtained. This relation is different from the pressure which is
obtainable from Eq. (\ref{T}). The maximums of this relation are
representing places in which phase transitions take place.
Therefore, the maximums in plotting a $P-r_{+}$ diagram for this
relation will give us critical pressure and horizon radius. This
method was employed to obtain critical pressure in several papers
which has proven to be an successful one
\cite{HendiIJMD,HendiJHEP}.

The heat capacity is obtained by
\begin{equation}
C_{Q}=\left( \frac{\partial M}{\partial S}\right) _{Q}\left( \frac{\partial
^{2}M}{\partial S^{2}}\right) _{Q}^{-1}.  \label{CQ}
\end{equation}

\begin{figure}[tbp]
$%
\begin{array}{ccc}
\epsfxsize=5.5cm \epsffile{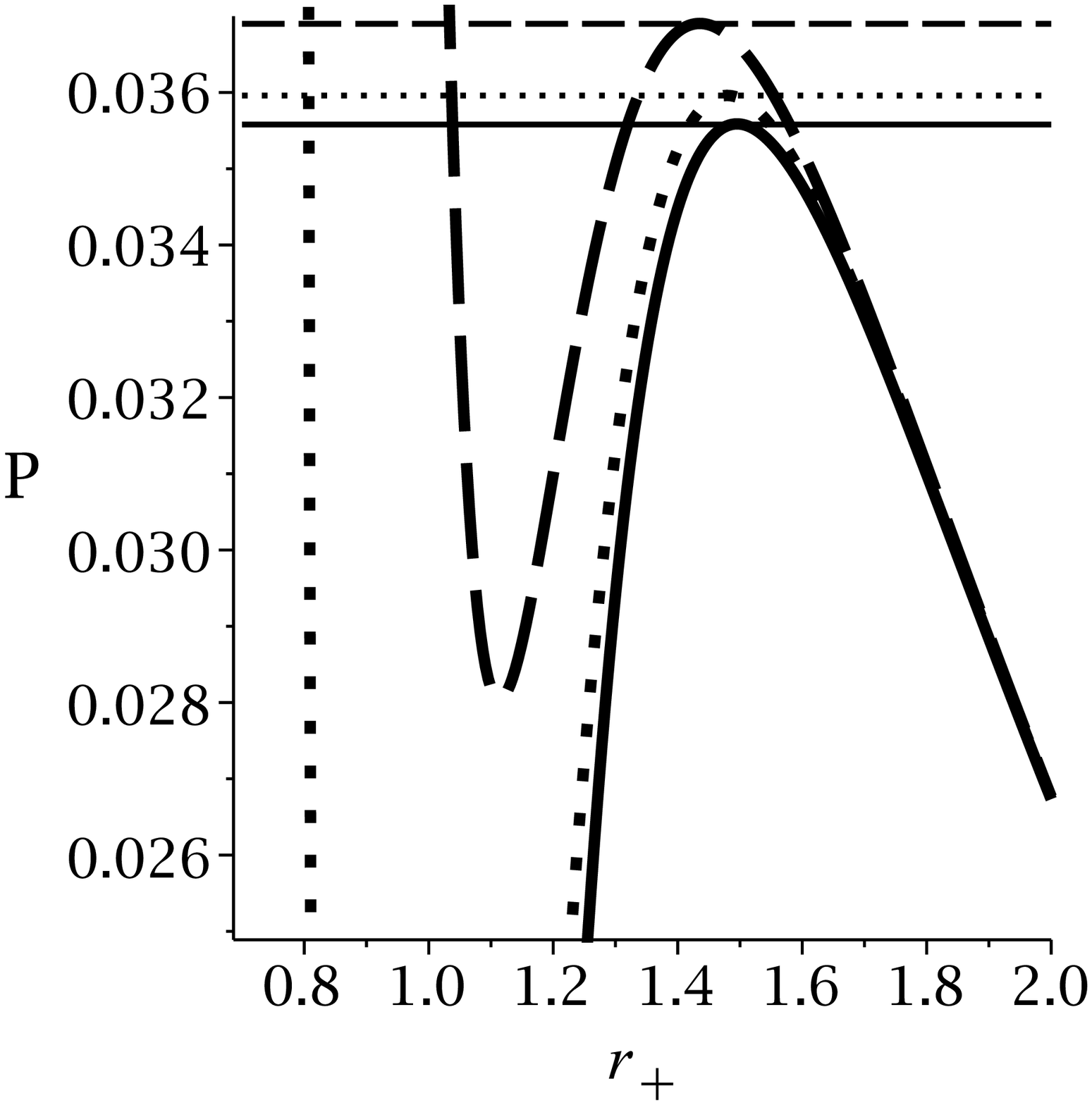} & \epsfxsize=5.5cm %
\epsffile{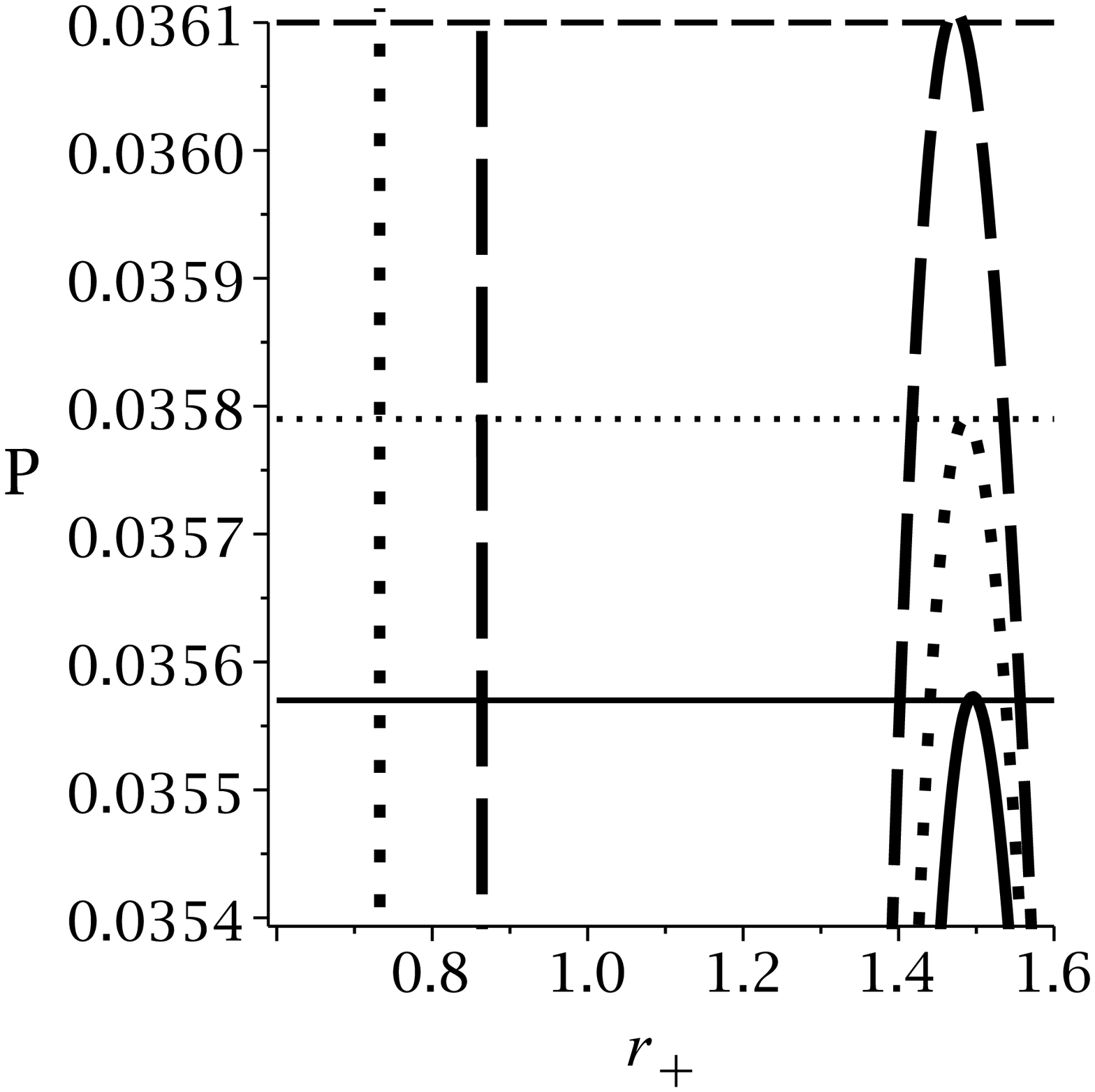} & \epsfxsize=5.5cm \epsffile{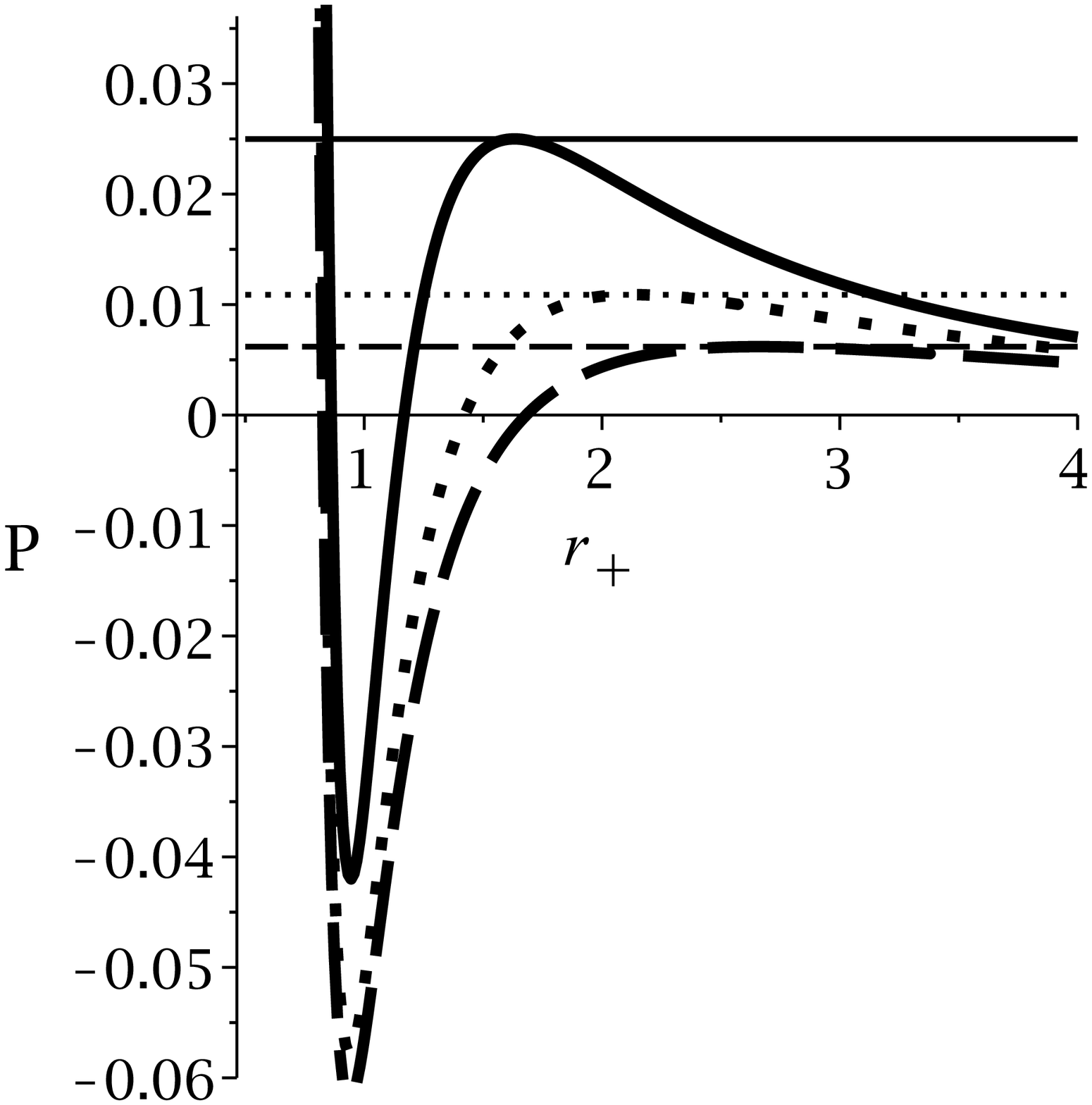}%
\end{array}
$%
\caption{$P$ versus $r_{+}$ diagrams for $q=1$ and $k=1$. \newline
left panel: $\protect\alpha=0$ and $\protect\beta=0$ (bold continues line), $%
P=0.03558$ (continues line), $\protect\beta=0.05$ (bold dotted line), $%
P=0.03596$ (dotted line), $\protect\beta=0.15$ (bold dashed line), $P=0.03690
$ (dashed line). \newline
middle panel: $\protect\alpha=10^{-4}$ and $\protect\beta=0$ (bold continues
line), $P=0.03557$ (continues line), $\protect\beta=0.03$ (bold dotted
line), $P=0.03579$ (dotted line), $\protect\beta=0.07$ (bold dashed line), $%
P=0.03610$ (dashed line). \newline
right panel: $\protect\beta=0.07$ and $\protect\alpha=0.1$ (bold continues
line), $P=0.02499$ (continues line), $\protect\alpha=0.5$ (bold dotted
line), $P=0.01089$ (dotted line), $\protect\alpha=1$ (bold dashed line), $%
P=0.00619$ (dashed line).}
\label{NEW}
\end{figure}

Using Eqs. (\ref{T}) and (\ref{S}) and replacing cosmological
constant with its corresponding relation with pressure and solving
its denominator with respect to pressure will lead to a relation
for pressure which for economical reasons we will not bring it. We
will present the results of this relation by considering mentioned
values in different tables in following diagrams (Fig. \ref{NEW}).

First of all, in absence of the nonlinearity, the plotted diagrams
shows existence of only one maximum. The pressure and horizon
radius of this maximum is exactly located where phase transition
takes place. Interestingly, by adding the nonlinearity to the
system, two extrema are observed: a minimum and a maximum. The
place of this maximum is exactly where phase transition occurs for
mentioned vales for different parameters. But as one can see, no
other maximum exists which indicates that, at least no other
second order phase transition is observed. In other words,
although for specific critical pressure in these diagrams two
critical horizons are observed, the type of these critical points
are not the same. Therefore, one can conclude that no reentrant
phase transitions happens which is consistent with what was
observed in coexistence diagrams.

\section{Conclusions}

In this paper we have considered a quadratic Maxwell invariant as
a correction term to the Maxwell Lagrangian. We studied the
thermodynamic behavior of these solutions in Einstein and GB
gravities. We considered cosmological constant as thermodynamic
pressure and related conjugated quantity as volume of the black
hole. By doing so, the interpretation of the mass as internal
energy was changed into Enthalpy of the system. Therefore, not
only the interpretation of the mass of the black hole was changed,
but also we extended phase space.

It is worthwhile to make some discussion regarding mass of the
black hole in this case. Usually, the interpretation of mass is
internal energy. In this point of view mass is a conserved
quantity which is representing only the total internal energy. But
in case of the new interpretation, mass of the black hole is a
combination of internal energy and pressure. In other words, not
only the mass of the black holes determines the internal energy
and the shape of the black hole, but also we are expecting it to
have information regarding interaction of constituents of black
holes which is known as pressure of the system. Although, one may
state that due to natural properties of black holes, it is
impossible to study the interaction of constituents of black
holes, one must take the approach of string theory to the matter
into consideration. In other words, in context of string theory
some attempts were made to study microstates of the black holes
and their interpretation which can be extended by considering this
point of view.

The volume of the black hole is determined by topological structure of
metric. Therefore, one expects that the calculated value of this quantity
and the topological structure of the metric be in agreement. This result was
obtained in calculation of volume.

In linear electromagnetic field, the critical behavior of the system for
both gravities were usual ones. One critical horizon radius was found, hence
one phase transition was expected. The formation of the characteristic
swallow tail was observed in $G-T$ diagrams. In $T-r_{+}$ and $P-r_{+}$
diagrams the properties of phase transition were seen. But amazingly, in
consideration of additional term the behavior of the system differed
completely.

By considering this nonlinear electromagnetic field, the structure of the
phase diagrams and the thermodynamic behavior of the system were highly
modified. Calculations regarding critical horizon radius lead to \textbf{%
existence of two critical horizon radii}. It is not unusual to find two
critical horizon radii for black holes \cite{7}, but in our case the
presence of the second critical horizon radius was observed in plotted
graphs. In other words, in case of this nonlinear electromagnetic field,
contrary to other cases of nonlinear theories, a phase transition and a
turning point were observed. The existence of these two points is related to
existence of number of critical horizon radius. This result is evident by
comparing Maxwell theory with this nonlinear theory. In case of Maxwell
theory only one positive critical horizon radius was found which resulted
into existence of one phase transition whereas for this case of nonlinear
electromagnetic field, two critical horizon radii were found which resulted
into existence of the phase transition and turning point.

In case of $G-T$ , three points were observed in which the behavior of the
system changed in them. These points are representing different phases. In $%
P-r_{+},$ for $T=T_{c}$, three horizon radii were found with same pressure
(critical pressure) which indicates the existence of three phases and one
turning point. For $T-r_{+},$ in case of $P=P_{c}$, two horizon radii were
found with same temperature (critical temperature). One of these horizons
was located on subcritical isobar which is the usual phase transition point
and the other one was out of the subcritical isobar.

Another interesting issue was the effects of nonlinearity and GB
parameters were opposite of each other. In case of nonlinearity
parameter, the smaller critical horizon radius was an increasing
function of nonlinearity parameter whereas the larger critical
horizon radius was a decreasing function of $\beta $. In this case
the smaller critical horizon radius was highly function of
nonlinearity parameter comparing to larger critical horizon
radius. Therefore, in case of increasing nonlinearity parameter,
the distance between these two critical horizon radii decreased.
The critical temperature and pressure were increasing functions of
nonlinearity parameter whereas $P_{c}r_{c}/T_{c}$ was a decreasing
function of it. Interestingly, in case of GB parameter, the
smaller critical horizon radius was a decreasing function of GB
parameter and the larger critical horizon radius was an increasing
function of it. The value of larger critical horizon radius was
highly function of variation of GB parameter. Therefore, the
distance between two critical horizon radii was an increasing
function of GB parameter. As for the critical temperature and
pressure, they were decreasing functions of GB parameter. In this
case, $P_{c}r_{c}/T_{c}$ was also a decreasing function of GB
parameter.

These two parameters ($\alpha $ and $\beta $) are describing two aspects of
the black holes; gravitational and matter fields. In case of increasing $%
\beta $, the nonlinearity behavior of the system increases. In context of
electromagnetic tensor, increasing this parameter leads to decreasing value
of this tensor (\ref{Ftr}). Therefore, the values of the electromagnetic
tensor are decreasing functions of nonlinearity parameter. This fact leads
to a conclusion that increasing the power of nonlinearity leads to
decreasing power of electromagnetic field. But in case of GB parameter, by
increasing its value, the power of the gravitational field increases. In
case of this generalization due to considering higher orders of curvature
scalar, we are increasing the gravitational force. As one can see,
increasing gravitational force causes the critical temperature and pressure
decrease which indicating that in this case phase transition is taking place
in lower temperature, hence lower energy. Therefore, increasing the power of
the gravitational force causes the system to have higher value of internal
energy. This can also be seen in studying entropy of the GB gravity. As one
can see, in case of spherical symmetric, entropy is an increasing function
of GB parameter. In comparing GB and Einstein gravities, it is evident that
GB gravity has higher entropy. Having higher entropy means that system has
higher internal energy which causes the system to have phase transition in
lower temperature. This result was found in studying phase diagrams of the
GB and Einstein diagrams.

By what was mentioned in last paragraph one can conclude following
results: first of all generalization of GB gravity causes the
internal energy of the system increases and the black hole has
phase transition faster comparing to Einstein gravity. Therefore,
in this theory black holes need to absorb less mass to have phase
transition. Second, the entropy may be a function of the
complexity of the black holes structure. Considering higher orders
of curvature scalar causes the system to have more complicated
structure. Therefore, it is arguable that the entropy and the
complexity of the structure of the system are related to each
other. In other words, complexity of the structure is a
measurement for entropy of the system. Nonlinearity of the system
and gravitational force are two opposing factors. They are
decreasing each others effects and at some points they may cancel
each others effects. This argument is stating that it may be
possible to fix parameters in a way which system has critical
values same as ones in Einstein-Maxwell theory.

Due to existence of the abnormal behavior in phase diagrams a
question regarding the existence of reentrant phase transitions
could rise which was answered through two methods. The coexistence
diagrams and a new method showed that only one of the critical
horizons are representing second order phase transition, while the
other one could be another type of the phase transition.

As for the effects of the dimensions, the smaller critical
horizon, pressure, temperature and $P_{c}r_{c}/T_{c}$ were
increasing functions of dimensions whereas, the subcritical
isobars and larger critical horizon radius were decreasing
functions of it. This behavior is indicating that the higher
dimensional black holes need to absorb more mass to have phase
transition. It is worthwhile to mention that case $n=3$ for
Einstein gravity and $n=4$ for GB gravity are considered to be
special cases in these gravities. As for the GB gravity, in this
specific dimension, some terms in Eqs. (\ref{T}) and (\ref{rc})
vanish. In case of Einstein gravity, it is evident from Eq.
(\ref{rc}) that for case of $n=3$ the power of the horizon radius
is different from other dimensions in a way which causes the Gibbs
free energy to have abnormal behavior which was observed in
studying phase diagrams.

As it was pointed out, considering whether we are working in
context of stringy correction or classical Born-Infeld theories,
the results may be different. As it was seen, in case of later
correction, there was a limiting point which for specific values
makes the system never acquire non-Van der Waals behavior, whereas
in case of stringy corrected electromagnetic field such limitation
did not exist and the system would have non-Van der Waals like
behavior. This points out that in stringy corrected version of the
theory, different terms have different dominant regions which
correspondingly modify phase structure and phase transitions of
the solutions.

Finally, it is worthwhile to mention that regarding string theory,
the quadratic Maxwell invariant arises independent of Born-Infeld
electromagnetic field. If one conducts the study that was done in
this paper by motivations of string theory (not correction to
Maxwell theory), there will be no limitation for considered
values. Therefore, one can omit vertical limiting lines in plotted
graphs. In addition, we should note that modification in phase
diagrams, previously was obtained by changing the gravitational
part of action. Here in this paper, the modification of matter
field caused the abnormal behavior of the phase diagrams.

\begin{acknowledgements}
We also thank the Shiraz University Research Council. This work
has been supported financially by the Research Institute for
Astronomy and Astrophysics of Maragha, Iran.
\end{acknowledgements}


\begin{thebibliography}{999}
\bibitem{1} S. W. Hawking, Phys. Rev. Lett. \textbf{26} (1971) 1344;

J. D. Bekenstein, {Phys. Rev.} D \textbf{7} (1973) 2333;

J. M. Bardeen, B. Carter and S. W. Hawking, {Commun. Math. Phys.} \textbf{31}
(1973) 161.

\bibitem{2} P. Hut, {Mon. Not. Roy. Astron. Soc.} \textbf{180} (1977) 379;

P. C. W. Davies, {Rep. Prog. Phys.} \textbf{41} (1978) 8.

\bibitem{3} M. W. Zemansky and R. H. Dittman, \textit{Heat and
thermodynamics: an intermediate textbook}, McGraw-Hill, U.S.A. (1997).

\bibitem{4} R. Banerjee, S. Ghosh and D. Roychowdhury, {Phys. Lett. }B
\textbf{696} (2011) 156.

\bibitem{5} G. Ruppeiner, {Phys. Rev.} A \textbf{20} (1979) 1608.

\bibitem{6} M. Cvetic and S. S. Gubser, {JHEP} \textbf{04} (1999) 024;

M. M. Caldarelli, G. Cognola and D. Klemm, {Class. Quantum
Gravit.} \textbf{17} (2000) 399.

\bibitem{ADSCFT} C. V. Johnson, Class. Quantum Gravit. \textbf{31} (2014)
205002;

B. P. Dolan, Mod. Phys. Lett. A, \textbf{30}, (2015) 1540002;

B. P. Dolan, JHEP \textbf{10} (2014) 179.

\bibitem{9} D. Kastor, S. Ray and J. Traschen, {Class. Quantum Gravit.} \textbf{%
26} (2009) 195011;

M. Urano, A. Tomimatsu and H. Saida, {Class. Quantum Gravit.}
\textbf{26} (2009) 105010;

B. Dolan, {Class. Quantum Gravit. }\textbf{28} (2011) 125020;

B. P. Dolan, {Class. Quantum Gravit.} \textbf{28} (2011) 235017;

B. P. Dolan,{\ Phys. Rev.} D \textbf{84} (2011) 127503.

\bibitem{Chamblin} A. Chamblin, R. Emparan, C. Johnson and R. Myers, {Phys.
Rev.} D \textbf{60} (1999) 064018;

A. Chamblin, R. Emparan, C. Johnson and R. Myers, {Phys. Rev.} D \textbf{60}
(1999) 104026.

\bibitem{Goldenfeld} N. Goldenfeld, \textit{Lectures on Phase Transitions
and the Renormalization Group}. Westview Press, New York, (1992).

\bibitem{7} C. Peca and J. Lemos, {Phys. Rev.} D \textbf{59} (1999) 124007;

G. Gibbons, M. Perry and C. Pope, {Class. Quantum Gravit.}
\textbf{22} (2005) 1503;

S. Wang, S. Q. Wu, F. Xie and L. Dan, {Chin. Phys. Lett.} \textbf{23} (2006)
1096;

Y. Sekiwa, {Phys. Rev.} D \textbf{73} (2006) 084009;

G. L. Cardoso and V. Grass, {Nucl. Phys.} B \textbf{803} (2008) 209;

H. Elvang, R. Emparan and P. Figueras, {JHEP} \textbf{05} (2007) 056;

R. Emparan, T. Harmark, V. Niarchos, N. A. Obers and M. Rodriguez, {JHEP}
\textbf{10} (2007) 110;

O. Dias, P. Figueras, R. Monteiro, J. E. Santos and R. Emparan, {Phys. Rev.}
D \textbf{80} (2009) 111701;

R. Banerjee, S. K. Modak and S. Samanta, {Phys. Rev.} D \textbf{84} (2011)
064024;

C. Niu, Y. Tian and X. N. Wu, {Phys. Rev.} D \textbf{85} (2012) 024017;

Y. D. Tsai, X. Wu and Y. Yang, {Phys. Rev.} D \textbf{85} (2012) 044005;

M. S. Ma, H. H. Zhao, L. C. Zhang and R. Zhao, {Int. J. Mod. Phys. }A
\textbf{29} (2014) 1450050;

S. Chen, X. Liu, C. Liu and J. Jing, {Chin. Phys. Lett.} \textbf{30} (2013)
060401;

R. Zhao, H. H. Zhao, M. S. Ma and L. C. Zhang, {Eur. Phys. J}. C \textbf{73}
(2013) 2645;

J. X. Mo, X. X. Zeng, G. Q. Li, X. Jiang and W. B. Liu, {JHEP} \textbf{10}
(2013) 056;

M. B. J. Poshteh, B. Mirza and Z. Sherkatghanad, {Phys. Rev.} D \textbf{88}
(2013) 024005;

S. A. Hosseini Mansoori and B. Mirza, {Eur. Phys. J.} C \textbf{74} (2014)
2681;

B. Mirza and Z. Sherkatghanad, {Phys. Rev.} D \textbf{90} (2014) 084006.

\bibitem{Vahidinia} S. Hendi and M. Vahidinia, {Phys. Rev.} D \textbf{88}
(2013) 084045.

\bibitem{8} S. W. Hawking and D. N. Page, {Commun. Math. Phys.} \textbf{87}
(1983) 577.

\bibitem{KubiznakMann} D. Kubiznak and R. B. Mann, {JHEP} \textbf{07} (2012)
033;

S. Gunasekaran, D. Kubiznak and R. Mann, {JHEP} \textbf{11} (2012)
110.

\bibitem{10} S. W. Wei and Y. X. Liu, {Phys. Rev. }D \textbf{90} (2014)
044057;

S. W. Wei and Y. X. Liu, {Phys. Rev.} D \textbf{87} (2013) 044014;

D. Zou, Y. Liu and B. Wang, {Phys. Rev.} D \textbf{87} (2013) 044014;

W. Xua, H. Xub and L. Zhaoc, {Eur. Phys. J.} C \textbf{74} (2014) 2970;

S. H. Hendi, S. Panahiyan and B. Eslam Panah, Prog. Theor. Exp. Phys.
\textbf{2015}, 103E01 (2015);

\bibitem{GBMaxwell} R. G. Cai, L. M. Cao, L. Li and R. Q. Yang, {JHEP}
\textbf{09} (2013) 005.

\bibitem{maximal pressure} A. M. Frassino, D. Kubiznak, R. B. Mann and F.
Simovic, JHEP \textbf{09} (2014) 080;

R. A. Hennigar, W. G. Brenna and R. B. Mann, JHEP \textbf{07}
(2015) 077.

\bibitem{11} M. Born and L. Infeld, {Proc. Roy. Soc. Lond}. A \textbf{143}
(1934) 410;

M. Born and L. Infeld, {Proc. Roy. Soc. Lond}. A \textbf{144} (1934) 425.

\bibitem{12} B. Hoffmann,{\ Phys. Rev}. \textbf{47} (1935) 877.

\bibitem{13} W. Yao and J. Jing, {JHEP} \textbf{05} (2014) 058;

S. Gangopadhyay, {Mod. Phys. Lett.} A \textbf{29} (2014) 1450088;

S. Gangopadhyay and D. Roychowdhury, {JHEP} \textbf{05} (2012) 156;

S. Gangopadhyay and D. Roychowdhury, {JHEP} \textbf{05} (2012) 002;

J. Jing, L. Wang, Q. Pan and S. Chen, {Phys. Rev.} D \textbf{83} (2011)
066010;

J. Jing and S. Chen, {Phys. Lett. }B \textbf{686} (2010) 68.

\bibitem{14} H. Q. Lu, L. M. Shen, P. Ji, G. F. Ji and N. J. Sun, {Int. J.
Theor. Phys.} \textbf{42} (2003) 837;

M. H. Dehghani and S. H. Hendi, {Gen. Relativ. Gravit.}
\textbf{41} (2009) 1853;

E. F. Eiroa and G. F. Aguirre, {Eur. Phys. J.} C \textbf{72} (2012) 2240.

\bibitem{15} S. H. Hendi, {Adv. High Energy Phys.} \textbf{2014} (2014)
697863.

\bibitem{16} M. H. Dehghani, N. Alinejadi and S. H. Hendi, {Phys. Rev.} D
\textbf{77} (2008) 104025;

M. H. Dehghani and S. H. Hendi, {Phys. Rev}. D \textbf{73} (2006) 084021;

M. Allahverdizadeh, S. H. Hendi, J. P. S. Lemos and A. Sheykhi, {Int. J.
Mod. Phys.} D \textbf{23} (2014) 1450032;

R. Banerjee and D. Roychowdhury, {Phys. Rev.} D \textbf{85} (2012) 104043;

A. Lala and D. Roychowdhury, {Phys. Rev.} D \textbf{86} (2012) 084027;

R. Banerjee and D. Roychowdhury, {Phys. Rev.} D \textbf{85} (2012) 044040;

P. Li, R. H. Yue and D. C. Zou, {Commun. Theor. Phys.} \textbf{56} (2011)
845;

D. C. Zou, Z. Y. Yang, R. H. Yue and P. Li, {Mod. Phys. Lett}. A \textbf{26}
(2011) 515;

A. Ghodsi and D. M. Yekta, {Phys. Rev.} D \textbf{83} (2011) 104004;

R. G. Cai and Y. W. Sun, {JHEP} \textbf{09} (2008) 115;

S. H. Mazharimousavi, M. Halilsoy and Z. Amirabi, {Phys. Rev.} D \textbf{78}
(2008) 064050;

W. A. Chemissany, M. de Roo and S. Panda, {Class. Quantum Gravit.}
\textbf{25} (2008) 225009;

Y. S. Myung, Y. W. Kim and Y. J. Park, {Phys. Rev.} D \textbf{78} (2008)
084002;

Y. S. Myung, Y. W. Kim and Y. J. Park, {Phys. Rev.} D \textbf{78} (2008)
044020;

O. Miskovic and R. Olea, {Phys. Rev.} D \textbf{77} (2008) 124048;

I. Zh. Stefanov, S. S. Yazadjiev and M. D. Todorov, {Phys. Rev.} D \textbf{75%
} (2007) 084036;

S. Fernando, {Phys. Rev.} D \textbf{74} (2006) 104032;

R. G. Cai, D. W. Pang and A. Wang, {Phys. Rev.} D \textbf{70} (2004) 124034;

M. Aiello, R. Ferraro and G. Giribet, {Phys. Rev}. D \textbf{70} (2004)
104014;

T. K. Dey, {Phys. Lett.} B \textbf{595} (2004) 484;

T. Tamaki, {JCAP} \textbf{05} (2004) 004;

S. Fernando and D. Krug, {Gen. Relativ. Gravit.} \textbf{35}
(2003) 129;

M. Wirschins, A. Sood and J. Kunz, {Phys. Rev.} D \textbf{63} (2001) 084002;

M. Cataldo and A. Garcia, {Phys. Lett.} B \textbf{456} (1999) 28.

\bibitem{17} S. H. Hendi, {J. Math. Phys.} \textbf{49} (2008) 082501;

M. H. Dehghani, S. H. Hendi, A. Sheykhi and H. Rastegar Sedehi, {JCAP}
\textbf{02} (2007) 020;

M. H. Dehghani and S. H. Hendi, {Int. J. Mod. Phys.} D \textbf{16} (2007)
1829;

M. H. Dehghani and H. Rastegar Sedehi, {Phys. Rev}. D \textbf{74} (2006)
124018;

S. H. Hendi, {Phys. Rev.} D \textbf{82} (2010) 064040;

D. J. Cirilo-Lombardo, {Gen. Relativ. Gravit.} \textbf{37} (2005)
847;

V. Ferrari, L. Gualtieri, J. A. Pons and A. Stavridis, {Mon. Not. Roy.
Astron. Soc.} \textbf{350} (2004) 763.

\bibitem{18} E. Fradkin and A. Tseytlin, {Phys. Lett.} B \textbf{163} (1985)
123;

R. Matsaev, M. Rahmanov and A. Tseytlin, {Phys. Lett.} B \textbf{193} (1987)
205;

C. Callan, C. Lovelace, C. Nappi and S. Yost, {Nucl. Phys.} B \textbf{308}
(1988) 221;

O. Andreev and A. Tseytlin, {Nucl. Phys.} B \textbf{311} (1988) 221;

R. Leigh, {Mod. Phys. Lett.} A \textbf{04} (1989) 2767.

\bibitem{19} H. H. Soleng, {Phys. Rev.} D \textbf{52} (1995) 6178.

\bibitem{20} S. H. Hendi, {JHEP} \textbf{03} (2012) 065.

\bibitem{21} S. H. Hendi, Ann. Phys. (N.Y.) \textbf{333} (2013) 282;

S. H. Hendi and A. Sheykhi, {Phys. Rev.} D \textbf{88} (2013) 044044;

S. H. Hendi, Ann. Phys. (N.Y.) \textbf{346} (2014) 42;

S. H. Hendi, {Adv. High Energy Phys.} \textbf{2014} (2014) 697914.

\bibitem{22} D. H. Delphenich, \textit{Nonlinear electrodynamics and QED},
[arXiv: hep-th/0309108];

D. H. Delphenich, \textit{Nonlinear optical analogies in quantum
electrodynamics}, [arXiv: hep-th/0610088];

W. Heisenberg and H. Euler, {Z. Phys.} \textbf{98} (1936) 714. {Translation
by}: W. Korolevski and H. Kleinert, \textit{Consequences of Dirac's Theory
of the Positron}, [physics/0605038];

J. Schwinger, Phys. Rev. \textbf{82} (1951) 664;

P. Stehle and P. G. DeBaryshe, {Phys. Rev.} \textbf{152} (1966) 1135.

\bibitem{Lorenci} V. A. De Lorenci and M. A. Souza, Phys. Lett. B \textbf{512%
} (2001) 417;

V. A. De Lorenci and R. Klippert, Phys. Rev. D \textbf{65} (2002) 064027;

M. Novello and E. Bittencourt, Phys. Rev. D \textbf{86} (2012) 124024;

M. Novello et al., Class. Quantum Gravit. \textbf{20} (2003) 859.

\bibitem{AyonBeato1} E. Ayon-Beato and A. Garcia, Gen. Relativ. Gravit.
\textbf{31} (1999) 629.

\bibitem{AyonBeato2} E. Ayon-Beato and A. Garcia, Phys. Lett. B \textbf{464}
(1999) 25;

H. P. Oliveira, Class. Quantum Gravit. \textbf{11} (1994) 1469;

D. Palatnik, Phys. Lett. B \textbf{432} (1998) 287;

E. Ayon--Beato and A. Garcia, Phys. Rev. Lett. \textbf{80} (1998) 5056.

\bibitem{Corda} V. A. De Lorenci, R. Klippert, M. Novello and J. M. Salim,
Phys. Rev. D \textbf{65} (2002) 063501;

I. Dymnikova, Class. Quantum Gravit. \textbf{21} (2004) 4417;

C. Corda and H. J. Mosquera Cuesta, Mod. Phys. Lett A \texttt{25} (2010)
2423;

C. Corda and H. J. Mosquera Cuesta, Astropart. Phys. \textbf{34} (2011) 587.

\bibitem{Mosquera} H. J. Mosquera Cuesta and J. M. Salim, Mon. Not. Roy.
Astron. Soc. \textbf{354} (2004) L55;

H. J. Mosquera Cuesta and J. M. Salim, Astrophys. J. \textbf{608} (2004) 925;

Z. Bialynicka-Birula and I. Bialynicka-Birula, Phys. Rev. D \textbf{2}
(1970) 2341.

\bibitem{Gopakumar} R. Gopakumar, S. Minwalla, N. Seiberg, A. Strominger,
JHEP \textbf{08} (2000) 008;

T. Tamaki and K. Maida, Phys. Rev. D \textbf{62} (2000) 084041;

S. Kar and S. Majumdar, Int. J. Mod. Phys. A \textbf{21} (2006) 6087.

\bibitem{Brigante} M. Brigante, H. Liu, R. C. Myers, S. Shenker and S.
Yaida, Phys. Rev. D \textbf{77} (2008) 126006;

Y. Kats and P. Petrov, JHEP \textbf{01} (2009) 044;

P. Kovtun, D. T. Son and A. O. Starinets, JHEP \textbf{10} (2003) 064;

X. H. Ge, Y. Matsuo, F. W. Shu, S. J. Sin and T. Tsukioka, JHEP \textbf{10}
(2008) 009.

\bibitem{23} A. Tseytlin, {Nucl. Phys.} B \textbf{276} (1985) 391;

D. J. Gross and J. H. Sloan, {Nucl. Phys.} B \textbf{291} (1987) 41;

Y. Kats, L. Motl and M. Padi, {JHEP} \textbf{12} (2007) 068;

D. Anninos and G. Pastras, {JHEP} \textbf{07} (2009) 030;

R. G. Cai, Z. Y. Nie and Y. W. Sun, {Phys. Rev}. D \textbf{78} (2008) 126007;

N. Seiberg and E. Witten, {JHEP} \textbf{09} (1999) 032;

E. Bergshoeff, E. Sezgin, C. Pope and P. Townsend, {Phys. Lett.} B \textbf{%
188} (1987) 70.

\bibitem{HendiMomen} S. H. Hendi and M. Momennia, {Eur. Phys. J.} C \textbf{%
75} (2015) 54.

\bibitem{ThreeMag} S. H. Hendi, B. Eslam Panah, M. Momennia and S.
Panahiyan, Eur. Phys. J. C \textbf{75} (2015) 457.

\bibitem{HendiEPJC} S. H. Hendi, {Eur. Phys. J.} C \textbf{73} (2013) 2634;

S. H. Hendi, S. Panahiyan and B. Eslam Panah, Eur. Phys. J. C \textbf{75}
(2015) 296.

\bibitem{HendiPanah} S. H. Hendi and S. Panahiyan, {Phys. Rev.} D \textbf{90}
(2014) 124008.

\bibitem{HendiIJMPD} S. H. Hendi Int. J. Mod. Phys. D \textbf{24} (2015)
1550040.

\bibitem{SL} Y. C. Bruhat, J. Math. Phys. \textbf{29} (1988) 1891;

X. O. Camanho and J. D. Edelstein, JHEP \textbf{04} (2010) 007;

X. O. Camanho and J. D. Edelstein, JHEP \textbf{06} (2010) 099;

K. Izumi, Phys. Rev. D \textbf{90} (2014) 044037;

X. O. Camanho, J. D. Edelstein, J. Maldacena and A. Zhiboedov,
[arXiv:1407.5597];

G. D'Appollonio, P. Di Vecchia, R. Russo and G. Veneziano, JHEP \textbf{05}
(2015) 144.

\bibitem{dS} M. A. Ainou, Eur. Phys. J. C \textbf{75}, (2015) 34;

M. A. Ainou, Phys. Rev. D \textbf{91} (2015) 064049.

\bibitem{HendiIJMD} S. H. Hendi, S. Panahiyan and B. Eslam Panah, Int. J.
Mod. Phys. D \textbf{25} (2016) 1650010.

\bibitem{HendiJHEP} S. H. Hendi, B. Eslam Panah and S. Panahiyan, JHEP
\textbf{11} (2015) 157;

S. H. Hendi, S. Panahiyan and B. Eslam Panah, JHEP \textbf{01}
(2016) 129.

\bibitem{Cosm} R. Aldrovandi, J. P. Beltran Almeida and J. G. Pereira, Int.
J. Mod. Phys. D \textbf{13} (2004) 2241.

\bibitem{Smarr} D. Kastor, S. Ray, and J. Traschen, Class. Quantum Gravit.
\textbf{27} (2010) 235014;

D. C. Zou, S. J. Zhang and B. Wang, Phys. Rev. D \textbf{89} (2014) 044002;

Z. Sherkatghanad, B. Mirza, Z. Mirzaeyan and S. A. H. Mansoori,
[arXiv:1412.5028].

\bibitem{DolanMann} B. P. Dolan, A. Kostouki, D. Kubiznak and R. B. Mann,
[arxiv:1407.4783].
\end{thebibliography}
\end{document}